\documentclass[a4paper,11pt]{article}
\usepackage{jheppub,xcolor} 
\usepackage{float}
\usepackage{graphicx,subcaption,caption,braket,dsfont}
\usepackage[normalem]{ulem}
\usepackage{orcidlink}
\usepackage{bbm}
\usepackage{hyperref}

\usepackage{listings}

\newcommand{\nn}{\nonumber}

\title{\boldmath Probing weak chaos in $\mathcal N=4$ super Yang-Mills and long-range spin chains}

\author[a,b,c]{Pawel Caputa,}
\author[a]{Brian Creed,}
\author[d,e,f]{Rathindra Nath Das,}
\author[g]{Saskia Demulder,}
\author[h]{Tristan McLoughlin}

\affiliation[a]{
The Oskar Klein Centre and Department of Physics, Stockholm University, AlbaNova, 106 91 Stockholm, Sweden}
\affiliation[b]{Yukawa Institute for Theoretical Physics, Kyoto University, Kitashirakawa Oiwakecho, Sakyo-ku, Kyoto 606-8502, Japan}
\affiliation[c]{Faculty of Physics, University of Warsaw, Pasteura 5, 02-093 Warsaw, Poland}
\affiliation[d]{Department of Particle Physics and Astrophysics, Weizmann Institute of Science,\\
Rehovot 7610001, Israel}
\affiliation[e]{Center for Theoretical Physics — a Leinweber Institute, MIT, Cambridge, MA 02139, USA}
\affiliation[f]{Institute for Theoretical Physics and Astrophysics, and\\ W\"urzburg-Dresden Cluster of Excellence ctd.qmat,\\
Julius-Maximilians-Universit\"at W\"urzburg, Am Hubland, 97074 W\"urzburg, Germany}
\affiliation[g]{Department of Quantitative Methods, CUNEF Universidad,\\ Calle Almansa 101,
28040 Madrid, Spain}
\affiliation[h]{School of Mathematics and Hamilton Mathematics Institute, Trinity College, \\Dublin 2, Ireland}

\preprint{YITP-26-69}

\abstract{We study signatures of quantum chaos in finite-loop truncations of the planar dilatation operator in the $\mathfrak{su}(2)$ sector of $\mathcal N=4$ super Yang-Mills and its $\beta$-deformation. These truncations define holographically motivated long-range deformations of the nearest-neighbour XXX spin chain. At one-loop the model is integrable, while the all-loop planar theory is expected to again be integrable. Finite-loop truncations therefore provide a natural setting for investigating how chaotic behaviour emerges between these two integrable limits. We analyse this question using spectral statistics, eigenvector diagnostics and spread complexity. We find that the two- and four-loop truncations develop GOE-like level statistics at sufficiently large coupling but with features characteristic of weak integrability breaking. The integrability breaking at four-loops is weaker than at two-loops and the critical coupling at which chaos occurs is larger, at least for long spin chains. The three-loop truncation does not show the same onset of chaos in the range studied. Eigenvector diagnostics show that the corresponding eigenstates remain less random than GOE vectors, indicating weak ergodicity and multifractality. Finally, we can identify signatures of the eigenvalue and eigenvector chaos in the Krylov-space data. Namely, we demonstrate a correlation of the level spacing statistics with the peak of spread complexity and disorder on the Krylov chain. The delocalisation of the initial state in the Hamiltonian eigenbasis is shown to strongly affect the saturation of complexity. Our results suggest that finite-loop dilatation operators are not generic long-range spin chain Hamiltonians, but already display patterns consistent with the restoration of integrability in the all-loop planar theory.}

\begin{document}
\maketitle
\flushbottom

\section{Introduction}

Chaos in the context of holography is most often studied in its strongest form: maximal chaos, as defined by maximal Lyapunov growth. This has become a central theme as it provides a bridge between quantum many-body dynamics and black hole physics with effective descriptions of near-horizon behaviour \cite{Sekino:2008he,Shenker:2013pqa,Maldacena:2015waa,Cotler:2016fpe,Altland:2026tog}. However, holographic gauge theories exhibit a much broader range of dynamics beyond the maximally chaotic regime, corresponding to different manifestations of many-body chaos. One example is the weak chaos observed in BPS sectors of holographic gauge theories \cite{Chen:2024oqv, Chen:2026vml}. Another is arithmetic quantum chaos \cite{Benjamin:2021ygh,Haehl:2023tkr}, which may be connected to black-hole interiors \cite{DeClerck:2023fax}. The focus of the present work is yet another regime, namely weak integrability breaking in the spectrum of perturbative planar anomalous dimensions described by long-range spin chains.

Our starting point is the planar dilatation operator of $\mathcal N=4$ super Yang--Mills theory (SYM) and its deformations, in particular the $\mathcal N=1$ $\beta$-deformed theory \cite{Leigh:1995ep}. A remarkable fact about this system is that the one-loop dilatation operator acting on single-trace operators can be mapped to an integrable spin chain Hamiltonian with nearest-neighbour interactions \cite{Minahan:2002ve, Beisert:2003tq, Beisert:2003yb}. At higher orders in perturbation theory the Hamiltonian becomes long-ranged and it is believed that the planar spectral problem admits an integrable all-order completion with a finite radius of convergence set by $|\lambda|<\pi^2$ in terms of the 't Hooft coupling. While the all-order Hamiltonian is not known, the spectrum of anomalous dimensions is encoded in the $\mathcal N=4$ SYM Quantum Spectral Curve \cite{Gromov:2014caa} which is valid at finite coupling and smoothly interpolates between known results at weak and strong coupling. This integrable structure underlies an impressive range of exact and all-loop results; see \cite{Beisert:2010jr, Arutyunov:2009ga, Bombardelli:2016rwb, Gromov:2017blm} for reviews.  Integrability is highly exceptional, both in spin chains and in the string sigma-models arising in AdS/CFT \cite{Basu:2011fw,Stepanchuk:2012xi}. Classical strings in less symmetric backgrounds often display chaotic dynamics. Precisely because planar integrability is so constraining, controlled departures from it are of particular interest and understanding how such non-integrable behaviour emerges from the gauge-theory spectral problem remains an important question.

While the single trace sector of the planar theory is believed to be integrable for finite values of the 't Hooft coupling, the dilatation operator when truncated at a perturbative order beyond one-loop is not integrable. Such a truncated Hamiltonian treated as an exact Hamiltonian will break the integrability of the one-loop term and generate chaotic dynamics. However, the integrability breaking terms are not of a generic form but are fixed by the gauge theory in a highly structured fashion. At loop order $\ell$, the corresponding spin chain Hamiltonian contains interactions involving at most $\ell+1$ adjacent spins. Moreover the interaction terms can be generated as a deformation of the one-loop result by a recursive method \cite{Bargheer:2008jt, Bargheer:2009xy} and arise from the spin chain version of $T\bar T$-type deformations \cite{Pozsgay:2019ekd,Marchetto:2019yyt}. As a result these models permit quasi-conserved charges which commute with the Hamiltonian up to corrections of higher order in the 't Hooft parameter. The planar gauge theory thus provides a non-trivial example of integrability breaking which is weak in a very specific sense and is of broader interest, e.g. \cite{Pozsgay:2019xak, Kurlov_2022,Surace:2023wqq,Vanovac:2024tkj,Adans:2026jue}, due to its dynamical properties such as anomalously long thermalisation times. Higher-loop truncations of the dilatation operator therefore provide a particularly natural setting in which to study weak integrability breaking and the onset of chaos. 

Our interest is not simply in whether integrability is broken, but what kind of non-integrable and chaotic features exist and how they emerge as a function of the integrability breaking parameter. A characteristic feature of weak integrability breaking models occurs in the system-size dependence of the critical coupling, that is the coupling at which the spectral statistics cross from Poisson-like behaviour to random-matrix behaviour. For generic perturbations of gapless integrable models this critical coupling is expected to scale as a power of the volume, with an exponent close to $-3$ \cite{modak2014finite, modak2014universal}. In models with weak integrability breaking, including for the two-loop $\mathcal N=4$ SYM dilatation operator and its $\beta$-deformation, it has been shown that the critical coupling scales more slowly, with a less negative exponent \cite{Szasz-Schagrin:2021pqg, McLoughlin:2022jyt}. In this work we extend this analysis to the long-range spin chain Hamiltonians arising from the three- and four-loop planar dilatation operator. We find that the exponent becomes even less negative at higher loop order, so that the onset of random-matrix behaviour is further delayed in the thermodynamic limit. This points towards the emergence of finite-coupling integrability in the (as yet unknown) all-order Hamiltonian.

More generally, non-integrability and chaos can manifest in many different ways, and distinct diagnostics need not yield the same conclusions. Spectral observables, such as level-spacing statistics and spectral form factors (SFFs), probe correlations between eigenvalues. However, it was already recognized in the early literature on quantum chaos that the eigenvectors themselves contain independent and equally important information. In \cite{berry1977regular}, Berry conjectured that eigenstates of chaotic systems resemble random superpositions. In many-body systems, this perspective is closely tied to ergodicity and thermalisation. More broadly, chaotic systems may exhibit a variety of eigenstate structures, including multifractality, quantum scars, and other forms of non-ergodic behaviour, with no single diagnostic of chaos necessarily implying another. Consequently, asking whether a system is chaotic is only part of the story; the more precise question is in what sense it is chaotic. 

This is precisely one of the points we will address in this work: are the novel features seen in the planar dilatation operator spin chain merely spectral, or do they extend to eigenstate ergodicity.
We numerically compute the (multi-)fractal dimensions, as well as the structural entropy, for the eigenstates of the two-, three- and four-loop Hamiltonians. We find that the onset of GOE-like spectral correlations is accompanied only by weak ergodicity at the level of the eigenvectors: in both the spin-product and one-loop integrable basis, the states remain systematically less random than GOE-vectors. Nonetheless, the loop-order hierarchy seen in the eigenvalue statistics persists in the eigenvector diagnostics, with the two-loop truncations coming closest to ergodic behaviour, the thee-loop truncation remaining closest to the integrable regime, and the four-loop case again lying in a regime in between.

Recently, the Recursion Method \cite{ViswanathMuller1994} has made a resurgence in the many-body quantum systems and holography literature due to the increasing popularity of Krylov operator complexity \cite{Parker_2019} and spread complexity for states \cite{Balasubramanian_2022}. This framework provides an efficiently computable way of organising dynamics in a Krylov basis fixed by the generator of time evolution and by a chosen initial operator or state. It has been used to study questions about operator growth, quantum chaos and late-time dynamics, including questions motivated by black hole physics, see reviews \cite{Nandy_2025,Baiguera_2026,Rabinovici:2025otw}. It has also reconnected with its original purpose of probing the dynamics of many-body systems, in particular with regard to their classification as integrable or chaotic \cite{Rabinovici:2022beu,Erdmenger_2023,Balasubramanian:2023kwd,Balasubramanian:2024ghv,kristjansen2026blackholestatesquantum,Das:2026hbw,Bhattacharjee:2022vlt,Huh:2023jxt,Camargo:2026szl,Das:2024tnw}. To explore these applications one needs to take into account both the generator of the system's dynamics and a particular state or observable. Therefore, this approach is naturally sensitive to aspects of both the spectrum of the generator and the structure of the chosen initial state in its eigenbasis.

For the purpose of relating the specific measures of chaos that we will consider in this paper, we focus on spread complexity \cite{Balasubramanian_2022}. The eigenvalues and eigenvectors of the Hamiltonian will be analysed directly as measures of chaos, as is typical for Hermitian quantum systems such as spin chains. It is therefore natural to study the dynamics of states generated by the Hamiltonian itself, rather than the dynamics of operators generated by the Liouvillian. Although the Liouvillian data of a closed finite-dimensional system are determined by the Hamiltonian spectrum and eigenstates, operator and state Krylov complexities depend on different choices of initial state and probe different spaces. Since our eigenvector diagnostics are formulated in the Hamiltonian Hilbert space, spread complexity provides the more direct comparison. However, it is worth emphasising that the spin chain eigenstates of the Hamiltonian correspond to local, single-trace operators in the gauge theory.

The paper is organised as follows. In section \ref{sec:bckgrd}, we review the planar $\mathcal N=4$ SYM dilatation operator and associated spin chain picture, its higher-loop truncations and deformation considered in this work. In section \ref{sec:eigenvals}, we analyse eigenvalue-based diagnostics of chaos, including level-spacing statistics, the $r$-ratio and the spectral form factor. In section \ref{sec:eigenvecs}, we study eigenvector diagnostics and the extent to which the onset of spectral chaos is accompanied by eigenstate ergodicity, both in the spin-product basis and in the one-loop integrable basis.  In section \ref{sec:Krylov}, we turn to Krylov-space observables and compare them with the spectral and eigenvector diagnostics. We conclude in section \ref{sec:disc}.

\section{$\mathcal{N}=4$ SYM spin chain}
\label{sec:bckgrd}
We start by reviewing the conventions and background for the $\mathcal N=4$ SYM spin chains and its marginal deformations. More pedagogical expositions and details can be found in e.g. \cite{Beisert:2010jr, Arutyunov:2009ga, Bombardelli:2016rwb, Gromov:2017blm}. 
\subsection{Scaling dimensions and spin chains }
The field content of $\mathcal{N}=4$ SYM includes, in addition to the gauge field $A_\mu$, four chiral fermionic fields $\Psi_\alpha$, their anti-chiral conjugates, and three complex scalar fields all transforming in the adjoint representation of the gauge group. The $\mathfrak{su}(2)$ subsector of the planar theory comprises the single-trace gauge-invariant operators composed of two of the complex scalars, $X$ and $Z$, e.g. Tr$(ZXZZX\dots)$. As in any conformal field theory, we are generally interested in computing the spectrum of scaling dimensions, $\Delta$, of the conformal primary operators. Such operators have definite weight under coordinate rescalings or equivalently satisfy the transformation law
 \begin{align}
     [\mathcal{D}, \mathcal{O}]=-i \Delta \mathcal{O}\,,
 \end{align}
 where $\mathcal{D}$ is the dilatation operator and we take $\mathcal{O}$ to be at the origin. A generic operator in the $\mathfrak{su}(2)$ sector will however not be an eigenstate of $\mathcal{D}$ but instead will mix with other operators in the same sector with the mixing matrix depending on the 't Hooft coupling, $\lambda=g_{\text{YM}}^2 N_c$ where $N_c$ is the rank of the gauge group. 
 
The problem of computing the conformal primary eigenstates and their eigenvalues in $\mathcal{N}=4$ SYM can be mapped onto the problem of diagonalising a Hamiltonian acting on a one-dimensional lattice, see \cite{Minahan:2010js} for details. Restricting to the $\mathfrak{su}(2)$ subsector, a basis at each lattice site is given by the eigenvectors of the $z$-component of spin, $S_z=\sigma_z/2$, i.e. $\ket{\uparrow}, \ket{\downarrow}$. The mapping from gauge operators is done by identifying each $Z$ inside the trace with a $\ket{\uparrow}$ state and each $X$ field with a $\ket{\downarrow}$ state. The spin chain Hilbert space is a tensor product over each of the lattice sites $(\mathbb{C}^2)^{\otimes L}$, 
 and so a basis for the full Hilbert space is given by the states
\begin{align}
    \ket{\uparrow\uparrow\dots \uparrow}~, ~~~\ket{\downarrow\uparrow\dots \uparrow}~,~~~\ket{\uparrow\downarrow\dots \uparrow}~, \dots, ~~~\ket{\downarrow\downarrow\dots \downarrow}~.
\end{align}
However the space of gauge theory operators is smaller as the cyclicity of the trace corresponds to restricting to states that are invariant under cyclic shifts of the lattice sites, and therefore restricting to states in the zero lattice-momentum sector.
 
The mixing matrix of anomalous dimensions is then given by a Hamiltonian acting on these spin chain states. This Hamiltonian can be perturbatively computed from the gauge theory as an expansion in the 't Hooft coupling \
\begin{align} \label{eq:HamiltonianExpansion}
	H&=\sum_{\ell=1}^{\infty} g^{2(\ell-1)} H_\ell\,,~~~\text{with}~~~g^2=\frac{\lambda}{16\pi^2}\,,
\end{align}
where the contribution at each loop-order is a sum over local densities, $H_\ell$, of range $\leq \ell+1$. 
The leading term, corresponding to one-loop anomalous dimensions, is the Heisenberg XXX-Hamiltonian with nearest-neighbour (NN) interactions which is well known to be integrable. The higher-loop corrections can then be viewed as long-range deformations of this model, and at each order we can define a truncated Hamiltonian as a function of the coupling
\begin{align}
\label{eq:HamTrunc}
    H^{(l)}(g^2)=\sum_{\ell=1}^l g^{2(\ell-1)} H_\ell~.
\end{align}
At two-loop order, we can write the range-3 Hamiltonian explicitly in terms of Pauli matrices acting on each lattice site $p$ as 
\begin{align}
    H^{(2)}{(g^2)}=\sum_{p=1}^L (\mathbbm{1}_{p,p+1}-\vec{\sigma}_p\cdot \vec{\sigma}_{p+1})-g^2 \sum_{p=1}^L (3\mathbbm{1}_{p,p+1}-4 \vec{\sigma}_p\cdot \vec{\sigma}_{p+1}+\vec{\sigma}_p\cdot \vec{\sigma}_{p+2})~.\label{H2g2trunc}
\end{align}
It can be easily seen that this Hamiltonian has a cyclically invariant ground state $\otimes_{p=1}^L \ket{\uparrow}_p $ corresponding to the BPS operator Tr$(ZZZ\dots Z)$ (we could equally take the BPS state Tr$(XXX\dots X)$ as the ground state). As the Hamiltonian commutes with the total spin operator $\vec{S}=\sum_{p=1}^L \vec{\sigma}_p$, the impurity number $M=L/2-S^z$, corresponding to the number of $X$ fields in the operator, is conserved and is a good quantum number to label eigenstates. This Hamiltonian is not integrable but it is perturbatively integrable such that there exist a family of higher charges which commute with $H^{(2)}$ up to corrections of $\mathcal{O}(g^{4})$. Moreover the spectrum can be found from the asymptotic Bethe ansatz again up to corrections at $\mathcal{O}(g^{4})$. As we will see, this perturbative integrability leaves a trace in the properties of the spectrum and eigenstates, even at finite values of the coupling.

One interesting feature of the two-loop Hamiltonian is a simplification at the point $g^2=1/4$ where the NN interactions cancel. Thus the even and odd lattice sites decouple and the model simplifies into two integrable nearest-neighbour spin chains. As this point is beyond the radius of convergence of the perturbative gauge theory, $g^2 \leq 1/16$, its significance for the spectrum of anomalous dimensions is unclear. However from spin chain perspective it is a novel feature which non-trivially affects the properties of the system.  More generally, it is only when $g^2$ is taken small, that we should view the system as a deformed NN spin chain.  At sufficiently large $g^2$, where the higher-order terms dominate, the model is better viewed as an exotic long-range system perturbed by short-range interactions. It is the small $g^2$ region that corresponds to perturbative gauge theory but the large $g^2$ is still of interest from a purely quantum spin chain perspective.

As the spin chain Hamiltonian densities become progressively more complicated at higher orders it is convenient to introduce the
 interaction symbols $\{a, b, c, \dots \}$ representing 
 \begin{equation}
\{a, b, c, \dots \}=\sum_{p=1}^L P_{p+a}P_{p+b}P_{p+c}\dots~,
 \end{equation}
 where $P_{p}$ is the nearest-neighbour permutation which interchanges the spin-states on lattice site $p$ and $p+1$. In this notation we have the one-, two- and three-loop contributions\footnote{ In appendix \ref{app:Ham_Pauli}, we show and workout how the loop contributions to the spin chain Hamiltonian can be expressed in terms of Pauli-matrices.}
\begin{eqnarray}\label{eq:123loops}
	H_1&=&2\{\}-2\{1\}\,,\\
	H_2&=&-8\{\}+12\{1\}-2(\{1,2\}+\{2,1\})\,,
    \\
H_3&=& 
+60\{\}
-104\{1\}
+4\{1,3\}
+24\big(\{1,2\}+\{2,1\}\big)
\nn\\
&&-4 \epsilon_2\{1,3,2\}
+4 \epsilon_2\{2,1,3\}
-4\big(\{1,2,3\}+\{3,2,1\}\big)\,,
\end{eqnarray}
where $\epsilon_2$ is a parameter, equal to -1 in gauge theory, that can be removed by making a similarity transform and so doesn't affect the spectrum.

These expressions are essentially fixed by the symmetries of the theory, however starting at four-loops there is a one parameter family of models consistent with the symmetries and asymptotic integrability. The four-loop term,
\begin{align}
\label{eq:4-loops}
H_4=&
+\big(-560-4\beta_{2,3}\big) \{\} 
+\big(+1072+12\beta_{2,3}+8\epsilon_{3a}\big) \{1\} 
+\big(-84-6\beta_{2,3}-4\epsilon_{3a}\big) \{1,3\} 
\nn\\
&
-4\{1,4\}
+\big(-302-4\beta_{2,3}-8\epsilon_{3a}\big) \big( \{1,2\} + \{2,1\} \big) \nn\\
&
+\big(+4\beta_{2,3}+4\epsilon_{3a}+2i\epsilon_{3c}-4i\epsilon_{3d}\big) \{1,3,2\} 
+\big(+4\beta_{2,3}+4\epsilon_{3a}-2i\epsilon_{3c}+4i\epsilon_{3d}\big) \{2,1,3\}
\nn\\
&
+\big(4-2i\epsilon_{3c}\big) \big( \{1,2,4\} + \{1,4,3\} \big)
+\big(4+2i\epsilon_{3c}\big) \big( \{1,3,4\} + \{2,1,4\} \big)
\nn\\
&
+\big(+96+4\epsilon_{3a}\big) \big( \{1,2,3\} + \{3,2,1\} \big)
\nn\\
&+\big(-12-2\beta_{2,3}-4\epsilon_{3a}\big) \{2,1,3,2\} 
+\big(+18+4\epsilon_{3a}\big) \big( \{1,3,2,4\} + \{2,1,4,3\} \big)
\nn\\
&+\big(-8 - 2\epsilon_{3a}-2i\epsilon_{3b}\big) \big( \{1,2,4,3\} + \{1,4,3,2\} \big)
\nn\\
&
+\big(-8 - 2\epsilon_{3a}+2i\epsilon_{3b}\big) \big( \{2,1,3,4\} + \{3,2,1,4\} \big) 
\nn\\
&-10\big( \{1,2,3,4\} + \{4,3,2,1\} \big)\,,
\end{align}
was given in \cite{Beisert:2007hz} with the parameter $\beta_{2,3}=4 \zeta(3)$ consistent with the predictions of the BES dressing phase \cite{Beisert:2006ez} ($\epsilon_{3a-d}$ are additional parameters related to similarity transformations).
 For spin chains of length greater than four, the Hamiltonian $H^{(4)}$ can be diagonalized by the asymptotic Bethe Ansatz and reproduces the planar spectrum of anomalous dimensions up to corrections of $\mathcal{O}(g^{10})$. Moreover, this long-range spin chain is of the type considered in \cite{Bargheer:2009xy} and can be viewed as a deformation of a nearest neighbour spin chain with the generators of the deformations of boost and bi-local form. 

While $\mathcal{N}=4$ SYM is the best understood holographic gauge theory, it is naturally interesting to generalise to different theories particularly those with less supersymmetry. Moreover, when we turn to the numerical computations it is useful to consider Hamiltonians with reduced symmetry as it can lead to improved statistics. To this end we introduce a twist deformation $\beta$, by defining a twisted permutation operator
\begin{align}
	P\ket{\uparrow\uparrow}=\ket{\uparrow\uparrow}~,~~~P\ket{\uparrow\downarrow}=e^{i\beta} \ket{\downarrow\uparrow}~,~~~P\ket{\downarrow\uparrow}=e^{-i\beta} \ket{\uparrow\downarrow}~,~~~P\ket{\downarrow\downarrow}= \ket{\downarrow\downarrow}\,,
\end{align}
and use the same expression for the Hamiltonian. If we use the same twist parameter for all permutation operators the Hamiltonian will remain perturbatively integrable and diagonalisable by the asymptotic Bethe Ansatz with twisted boundary conditions \cite{Beisert:2005if}. For real values, this is the planar dilatation operator for the $\beta$-deformed $\mathcal{N}=4$ SYM theory which is a marginal deformation of Leigh-Strassler type \cite{Leigh:1995ep} preserving $\mathcal{N}=1$ supersymmetry. While this deformation breaks the SO$(6)$ $R$-symmetry of $\mathcal{N}=4$ SYM to U$(1)^3$, the operators corresponding to the ground states of the spin chain, Tr$(ZZ\dots Z)$, are still BPS. Alternatively, if we use different parameters - $\beta_2$, in the two-loop $H_2$ term, $\beta_3$ in $H_3$, $\beta_4$ in $H_4$, etc. - the Hamiltonian will no longer be even perturbatively integrable but rather generically chaotic and so provides a useful comparison to the gauge theory Hamiltonian. 

\section{Eigenvalue diagnostics}\label{sec:eigenvals}
A standard characterization of quantum chaos follows from the BGS conjecture \cite{Bohigas:1983er}, which states that the spectral fluctuations of quantum systems whose classical dynamics are chaotic coincide with those of random matrix ensembles. Random Matrix Theory (RMT) captures the correlations between energy eigenvalues and makes predictions for correlation functions which can be analytically computed using well known methods, see e.g. \cite{Guhr:1997ve, mehta2004random}. In this section we compare these predictions with the results of direct numerical computations for the gauge theory spin chains.

To make this comparison one must consider a sector with fixed global quantum numbers, that is one must desymmetrise the spectrum. For the $\mathfrak{su}(2)$ invariant theory, in addition to focusing on states with lattice momentum zero, we would need to restrict to highest weight states, i.e. those annihilated by $S^+=\sum_i \sigma_i^+$, and having definite lattice parity.  On the other hand, for the $\beta$-deformed long-range spin chain this SU$(2)$ symmetry is broken, as is lattice parity.  As the $z$-component of the total spin $S^z=\tfrac{1}{2} \sum_i \sigma_i^z$ is still conserved, we restrict to states of fixed length $L$ and impurity number $M$. For a given $L$ and $M$ this results in larger sectors with better statistics compared to the $\mathfrak{su}(2)$ invariant theory, and so we generally consider the case of $\beta\neq 0$\footnote{In general one should consider $\beta/\pi$ irrational, as a rational value has additional unwanted symmetry relating to an orbifold \cite{Lunin_2005}.}. Similarly, while at fixed $L$ the largest sectors occur for impurity number close to half-filling, $M=L/2$, at exactly half-filling there is an additional spin-flip symmetry and so we avoid this sector. Where we are interested in the spectrum as a function of system size we, increase $L$ and $M$ while keeping the ratio $L/M\simeq 16/7$ fixed. Additionally, as the universal predictions of RMT are for the spectral fluctuations we must remove the mean eigenvalue density by either unfolding the spectrum (see appendix \ref{sec:unfolding} for further details) or by considering observables which are independent of these global properties; we will do both below. Finally, it is not the case that the entire spectrum becomes chaotic even in generic spin chains. States with energies near the edges of the spectrum, that is near the ground state or close to the maximum energy, are known to behave in a non-generic fashion and for diagnosing the onset of chaos it is conventional to remove, or clip, such states and focus instead on states in the bulk of the spectrum. Where we follow this,  we generally remove 15\% of the eigenvalues from each edge of the spectrum. 

\subsection{Level-spacing distribution}
A common characterisation of the spectral statistics is given by the level spacing distribution for the unfolded eigenvalues, $\xi_a$. After ordering the values, $\xi_1\leq \xi_2 \leq \dots $, the level spacings are given by the differences $s_a\equiv\xi_{a+1}-\xi_a$. This is thus sensitive to short range correlations within the spectrum. We can numerically approximate the distribution, $P(s)$, by computing the normalised histogram of values. For an integrable system, where the eigenvalues are uncorrelated, the distribution of spacings is Poisson, $P(s)\sim e^{-s}$, while for a chaotic system the distribution is well described by the Wigner surmise $P(s)=\frac{\pi s}{2}e^{-\pi s^2/4}$. This distribution reflects the well-known fact that in the chaotic case energy levels repel which results in the distribution vanishing at small $s$. For finite dimensional systems with an integrability breaking perturbation, the distribution will smoothly interpolate between the integrable and chaotic distributions for increasing values of the deformation parameter. A useful description of the distribution for generic cases is given by the Brody distribution:
\begin{align}
	P_{\text{Brody}}(s;\omega_B)=
		\frac{
			\Gamma\left(\frac{2+ \omega_B}{1+\omega_B}\right)^{1+\omega_B}(1+\omega_B)
		}{
		\Gamma\left(\frac{1+ \omega_B}{1+\omega_B}\right)^{2+ \omega_B}
	}
		s^{\omega_B}
		\text{exp}
		\left[-\left(
	\tfrac{
		\Gamma\left(\frac{2+ \omega_B}{1+\omega_B}\right)}{
		\Gamma\left(\frac{1+ \omega_B}{1+\omega_B}\right)}
		\right)^{1+ \omega_B}
	s^{1+\omega_B}
\right]\,.
\end{align}
This distribution depends on a parameter $\omega_B\in [0,1]$ which is $0$ for the integrable case and $1$ for the chaotic case described by the GOE Wigner distribution. 
\begin{figure}
	\vspace{0.3cm}
	\begin{center}
		\includegraphics[height=0.31\linewidth]{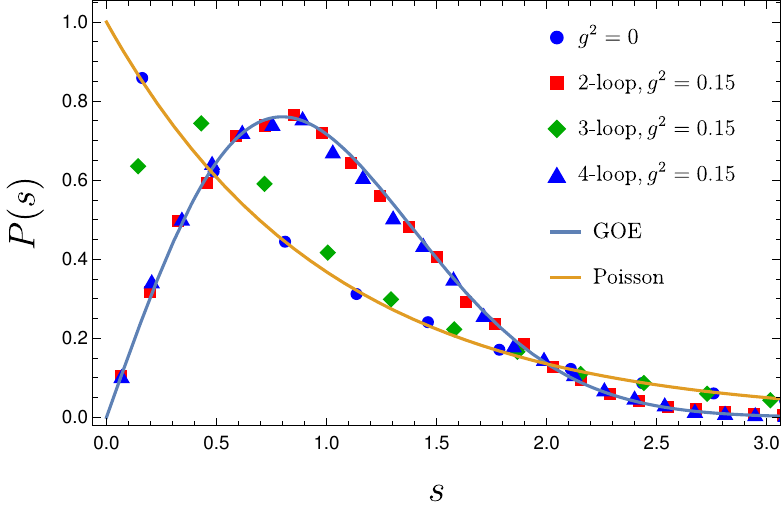}
        \hspace{3mm}
        \includegraphics[height=0.31\linewidth]{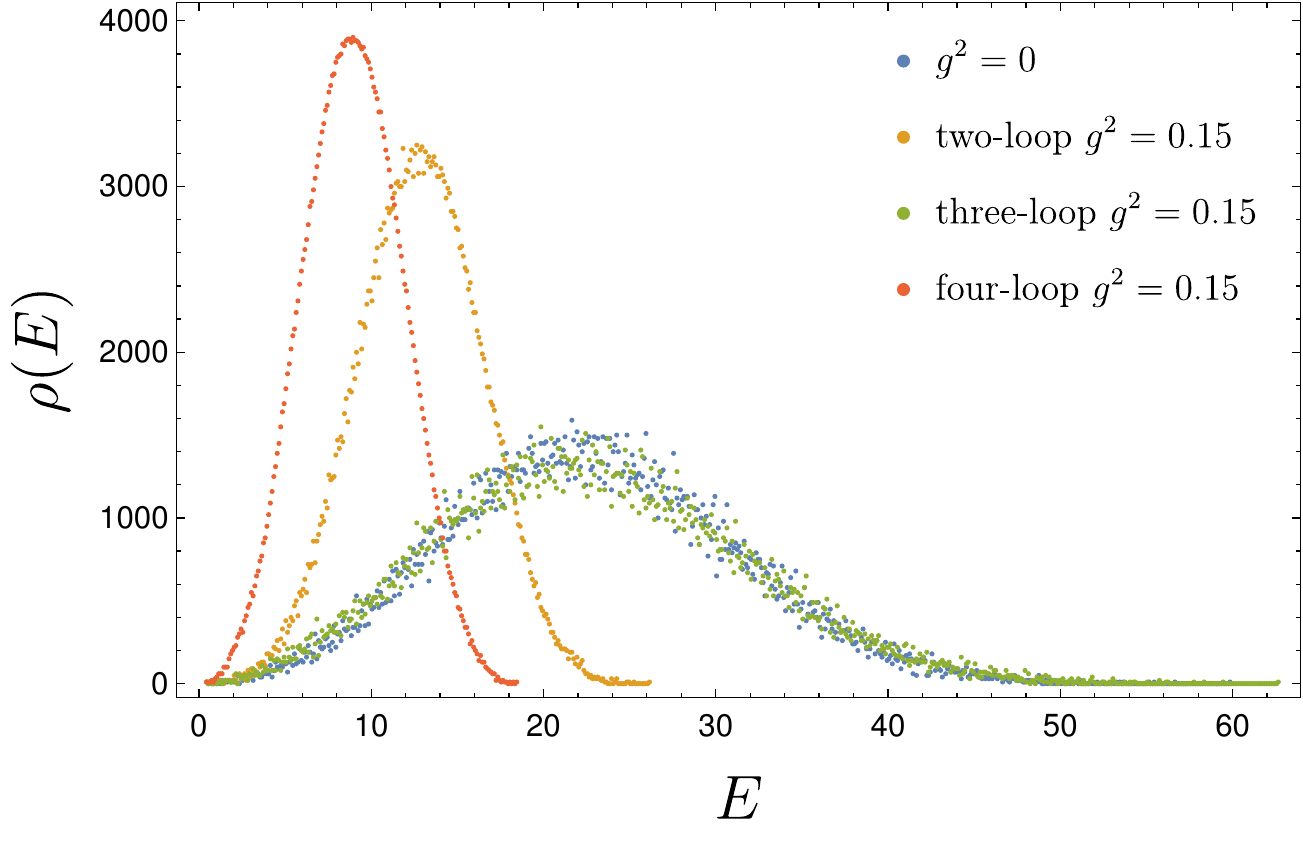}
	\end{center}
	\caption{(Left) Level spacing distribution for the $L=22$, $M=10$ sector with $\beta=1/\sqrt{26}$ for the integrable one-loop spin chain and the two-, three-, and four-loop truncated spin chains at $g^2=0.15$. (Right) Density of states for same sector for the one-, two-, three-  and four-loop spin chain.
    }	\label{fig:LS_comp}
\end{figure}
The nearest-neighbour XXX spin chain is integrable and so the its level-distribution is Poisson. This can be seen in the $g^2=0$ curve in Figure \ref{fig:LS_comp} (Left) were we show the level-spacing distribution for the $L=22$, $M=10$ sector. Including the two-loop correction results in a chaotic spectrum for sufficiently large coupling, as does including up to the four-loop correction, where the distribution is given by the GOE Wigner surmise. Interestingly, if one truncates at the three-loop term the distribution does not become chaotic even at large values of the coupling. This appears to be a consequence of the alternating sign structure of the anomalous dimension of planar $\mathcal{N}=4$ SYM and the fact that the higher-loop coefficients become larger. For example, the $L=5$ $M=2$ operator Tr$(X^2Z^3)+\dots$ to four loops has the dimension \cite{Beisert:2006ez} 
\begin{align}
    \Delta=5+8g^2-24 g^4 +136g^6-16(115/2+8\zeta(3))g^8+\dots ~\,,
\end{align}
and we can see that at $g^2=0.15$ the three-loop term cancels the two-loop contribution to about one part in six. We can see the same effect in a more general sense by examining the full spectrum for the $L=22$, $M=10$ sector. In Figure \ref{fig:LS_comp} (Right) we show a numerical approximation to the density of energy eigenvalues, $\rho(E)$, by plotting the number of states in each energy bin of width $dE=0.1$ for the one-loop chain i.e. $g^2=0$, as well as the two-, three- and four-loop chains with $g^2=0.15$. We can see that the spectrum has an approximately Gaussian distribution, at least away from the edges, however there are significant modifications when the higher-loop terms are included. In particular, the two-loop correction results in a decrease in the mean value. The three-loop term however cancels most of this shift and the density is close to the one-loop result, while the four-loop term again produces a shift of the mean towards zero. 

\begin{figure}
	\vspace{0.3cm}
	\begin{center}
    \includegraphics[height=0.31\linewidth]{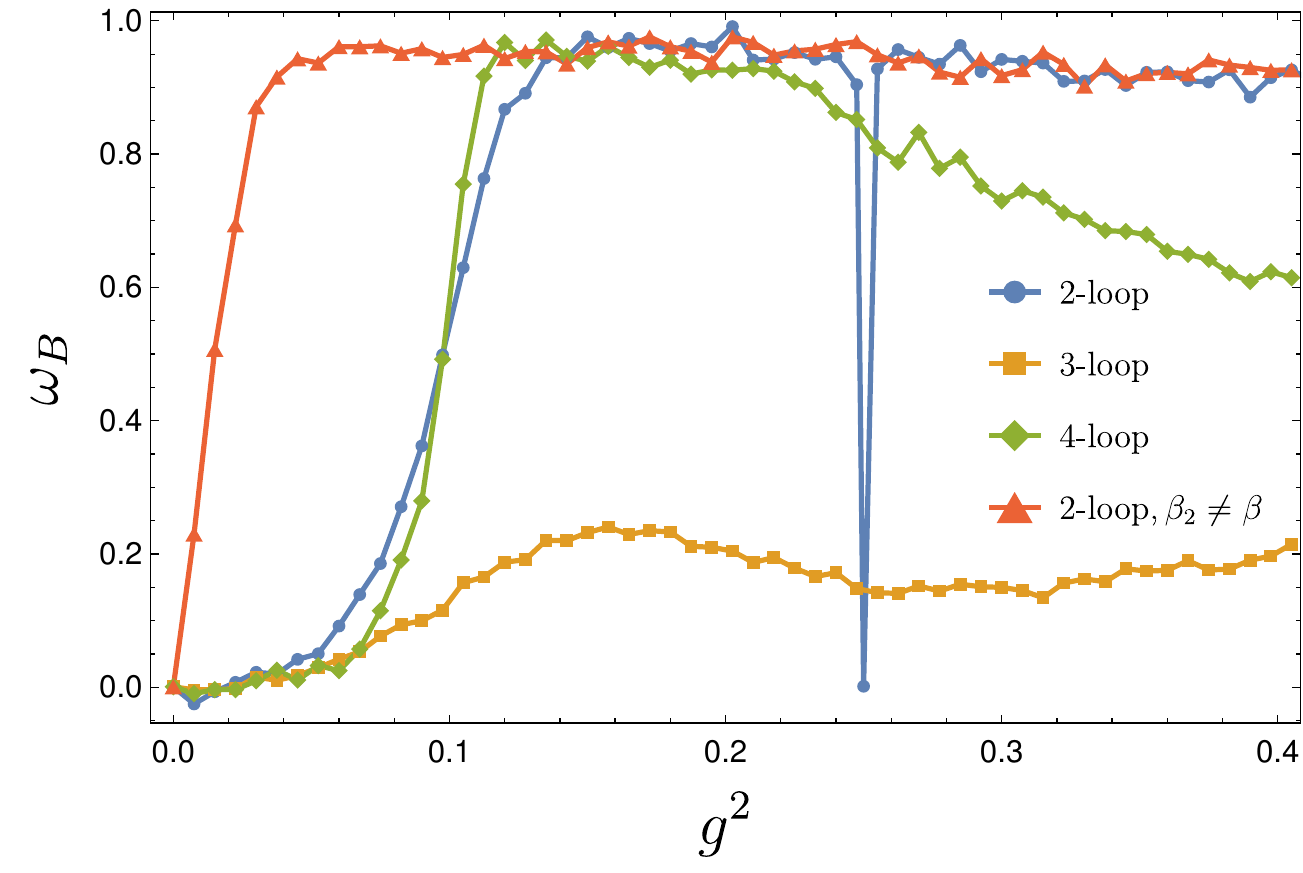}\hspace{3mm}
		\includegraphics[height=0.31\linewidth]{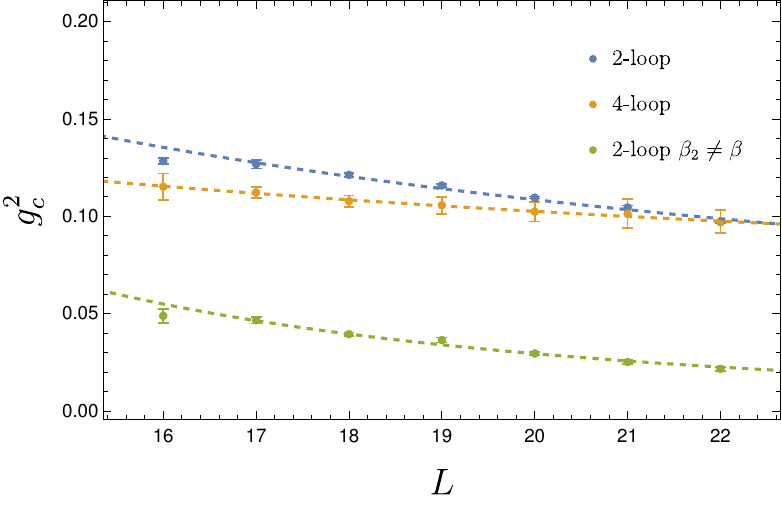}
	\end{center}
	\caption{
    (Left) Brody parameter as a function of coupling for $L=22$ $M=10$ sector for the two- and four-loop spin chain as well as model with different twist beyond one-loop.
    (Right)
    The critical coupling at which $\omega_B=1/2$ for the two- and four-loop spin chains with lengths $L\in [16,22]$. The two-loop chain with $\beta_2\neq \beta$ is shown for reference. Dashed lines are the best power-law fits.}	\label{fig:crit_comp}
\end{figure}

While the level-spacing distribution of the truncated two- and four-loop spin chains are of the usual Wigner-Dyson form for sufficiently large $g$, the transition from integrable to chaotic differs from generic theories and is instead of the form of weak integrability breaking models \cite{Szasz-Schagrin:2021pqg, McLoughlin:2022jyt}. This can be seen by plotting the Brody parameter as a function of the coupling $g^2$, see Figure \ref{fig:crit_comp} (Left). For reference we include the two-loop spin chain where the $\beta$-deformation parameter at one- and two-loops is taken to differ by $\pi/2$, i.e. $\beta_2=\beta+\pi/2$. This model does not arise from a gauge theory dilatation operator and cannot be found by the boost deformation method, rather it is a generic integrability breaking Hamiltonian. For this model, the Brody parameter is initially close to a linear function of $g^2$ and quickly reaches the plateau chaotic value $\omega_B=1$. By comparison, the two-loop spin chain displays a more S-shaped or sigmoid type functional form, where $\omega_B$ remains close to zero for small values of $g$ and the spectrum becomes chaotic only at larger values. For the two-loop case we see the special $g^2=0.25$ point where the Hamiltonian again becomes integrable as the NN interactions cancel and we have two decoupled NN spin chains on even and odd sites. It is interesting to note that the transition between integrable and chaotic behaviours is very rapid at this point which contrasts with the behaviour near $g=0$. The transition is in fact sharper than that of the $\beta_2\neq \beta$ model and to capture it we need to evaluate the spectrum at almost exactly $g^2=0.25$. Beyond $g^2=0.25$, the $\omega_B$ for the two-loop dilatation operator declines slightly though it remains close to one. 

Truncating at three-loops, we see that the spectrum does not become chaotic for $g^2\in [0, 0.4]$\footnote{We plot in step sizes of $0.0075$ and include the special point $0.25$ throughout this work.} while at four-loops the transition is similar to the two-loop case for small values of the coupling. Though the initial difference between the two- and four-loop cases is small, it is clear that for the four-loop case $\omega_B$ remains close to zero for larger values of $g^2$ and then transitions quicker. In this regard the curve approximates more closely a step function.  This leads to the conjecture that including even higher-loop corrections will further increase this resemblance with the all-order spin chain having $\omega_B=0$ for extended values of the coupling and then a sharp step-function at some critical coupling. 
At four-loops there is no special behaviour at $g^2=0.25$ but the decline in $\omega_B$ at larger values of $g$ becomes significant, reaching close to $0.6$ for $g^2\simeq 0.4$. At this intermediate coupling, $g^2\gtrsim 0.2$, the four-loops terms are larger than the one to three-loop terms and the model should be thought of a long-range spin chain perturbed by shorter-range interactions. 

\paragraph{Volume dependence} The onset of chaos depends on the size of the system and we can study this dependence by computing a critical coupling $g_c$, defined by the condition $\omega_B(g_c)=1/2$, at different values of the length, see Figure \ref{fig:crit_comp} (Right). On general grounds, we expect the integrability-breaking terms to produce chaotic behaviour at smaller values of the coupling for larger system sizes. We can fit the data to a power-law functional form:
\begin{align}
    g^2_c=\frac{a}{L^b}\,,
\end{align}
for the two- and four-loop chains as well as the two-loop chain with $\beta_2\neq \beta$. For the latter case we find that the best fit is given by $a_g=123\pm 30$, 
$b_g= 2.8 \pm 0.1$
which is broadly consistent with the expected result of $b=3$ for generic gapless systems \cite{modak2014universal, modak2014finite}~. For the two-loop case we find that $a_{2l}=2.1\pm 0.6$ and $b_{2l}=0.99\pm 0.09$ which reproduces the result of \cite{McLoughlin:2022jyt} \footnote{The numerical differences from \cite{McLoughlin:2022jyt} are due to changes in parameter values and fitting algorithm. While the results are consistent they suggest that the estimated error is slightly too small.}. For the four-loop chain we find that $a_{4l}=0.5\pm 0.04$ and 
\begin{align}
    b_{4l}=0.53\pm 0.03~.
\end{align}
We can see that while the critical coupling is lower for the four-loop chain at smaller lengths it declines more slowly in the thermodynamic limit and eventually the onset of chaos occurs only for larger values of the coupling. This further demonstrates that the inclusion of the higher-loop terms weakens the integrability breaking. 

\subsection{$r$-ratio}\label{sec:r-ratio}
Computing the level spacing distribution requires unfolding the spectrum which in turn necessitates a number of somewhat arbitrary parameter choices,  as highlighted in appendix \ref{sec:unfolding}.  One can mitigate the ambiguities introduced in this process by considering quantities which do not depend on this procedure. In particular, one can evaluate the ratio of adjacent spacings introduced in \cite{Oganesyan_2007}
\begin{equation}\label{eq:rratio}
     r_a=\frac{\min(s_a,s_{a-1})}{\max(s_a,s_{a-1})}\,,
\end{equation}
where here the spacings $s_a=E_{a+1}-E_{a}$ are differences of ordered energy eigenvalues, $E_1<E_2<\dots<E_D$, which have not been unfolded. This observable is dimensionless and independent of the local density of states, hence does not require unfolding. However we still clip 15\% of states from each edge of the spectrum.  Its mean value, $\langle r\rangle$, takes universal values depending on the underlying spectral correlations. For Poisson statistics one finds $\langle r\rangle=2\ln2-1\approx 0.386$, whereas GOE gives $\langle r\rangle\approx 0.5307$ \cite{Atas_2013}.
\begin{figure}
    \centering
    \includegraphics[width=0.5\linewidth]{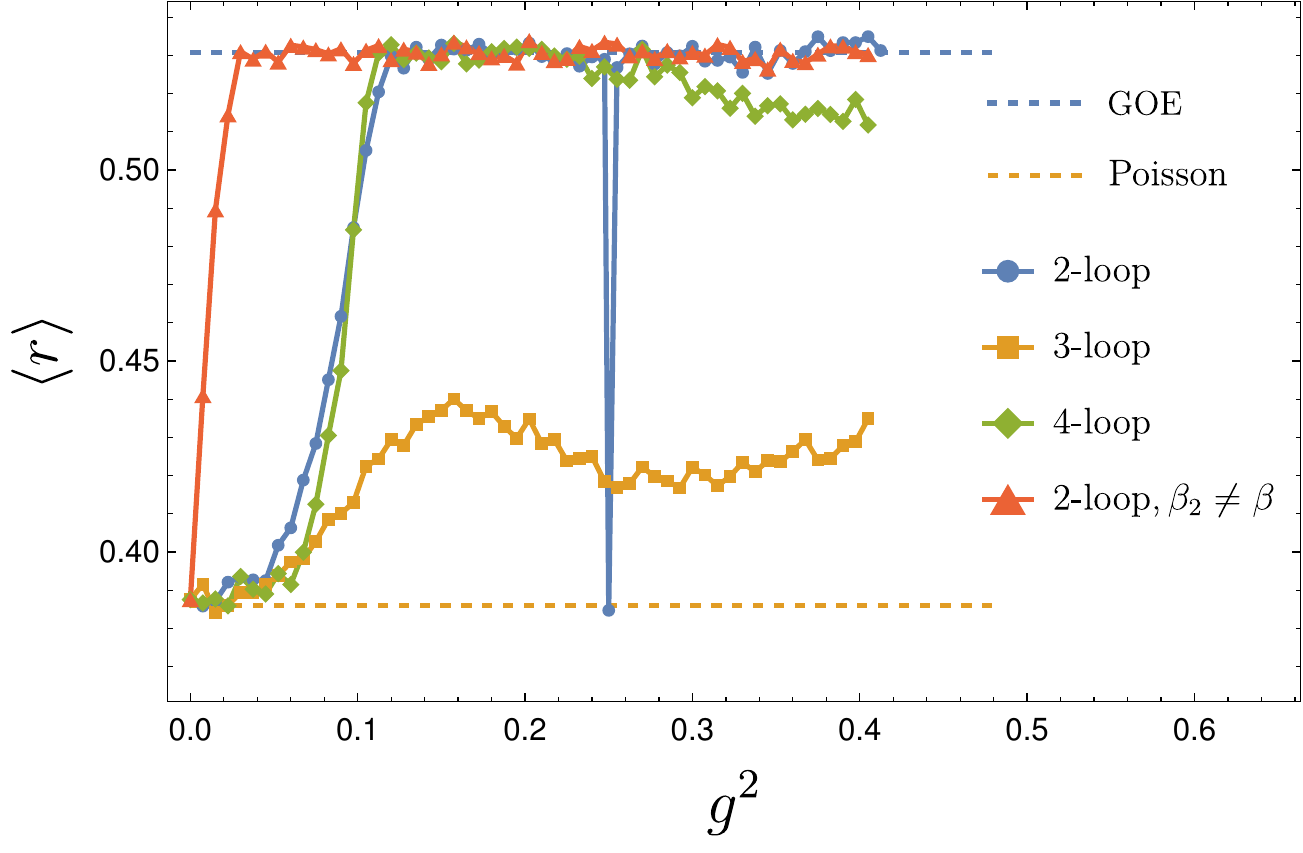}
    \caption{The averaged $r$ ratio for $L=22$, $M=10$ sector with $\beta=1/ \sqrt{26}$ as a function of $g^2$.}
    \label{fig:rRatio}
\end{figure}
We plot $\langle r\rangle$ as a function of the coupling for two-, three- and four-loop truncated Hamiltonians in Figure \ref{fig:rRatio} and compare with the curve for the generic Hamiltonian with unequal twists. The similarity to the plot of the Brody parameter for the unfolded spectrum is quite clear demonstrating that our results are independent of the unfolding process.

\subsection{Spectral form factor}\label{sec:SFF}
The level-spacing distribution and ratio $\langle{r}\rangle$ probe correlations between adjacent energy eigenvalues. Quantities which capture longer range correlations include the Dyson-Mehta statistic $\Delta_3$ and the spectral form factor (SFF). The later has played an important role in a wide range of applications of quantum chaos such as matching RMT predictions with semi-classical results \cite{berry1985semiclassical, sieber2001correlations, muller2009periodic}, understanding black hole physics e.g. \cite{Papadodimas:2015xma,Cotler:2016fpe, Gharibyan:2018jrp} and in quantum many-body physics \cite{Kos:2017zjh}. The SFF is given in terms of the Fourier transform of the two-point function of spectral density or simply
\begin{align}
    |Z(t)|^2=\sum_{a,b}e^{i(E_a-E_b)t}~\,,
\end{align}
though we will use the normalised version
\begin{align}
    g(t)=\frac{\langle |Z(t)|^2\rangle }{|Z(0)|^2}~.
\end{align}
In general $\langle \cdot \rangle$ denotes the ensemble average, however for the spin chain, where we have a deterministic spectrum, we replace this by a local time-average or smoothing in order to remove the late time oscillations
\begin{align}
    \langle F(t)\rangle\equiv\int dt' W_T(t') F(t+t')\,,
\end{align}
where $W_T(t)$ is some window function of width $T$. We also measure time in units of the Heisenberg time $\tau_H$ which we define in terms of the spectrum and Hilbert space dimension $D$
\begin{equation}\label{eq:HeisenbergTime}
    \tau_H := \frac{D}{E_{\text{max}}-E_{\text{min}}} \approx \frac{1}{\langle s \rangle} \quad , \quad \langle s \rangle = \frac{1}{D-1} \sum_{a=1}^{D-1} (E_{a+1}-E_a)~.
\end{equation}

As is well known, the SFF has a characteristic dip-ramp-plateau structure. The early time behaviour, the ``dip", depends on the details of the system but the later time behaviour is universal and for chaotic systems agrees with the predictions of RMT. These predictions depend on the particular ensemble, however in all cases, in the limit of large matrices and for eigenvalues near the centre of the spectrum, there is a ramp region where the SFF increases linearly, starting at the ``ramp" or Thouless time, $t_{\text{th}}$, followed by a plateau at times of order $\tau_H$.
\begin{figure}[t]
	\vspace{0.3cm}
	\begin{center}
		\includegraphics[height=0.31\linewidth]{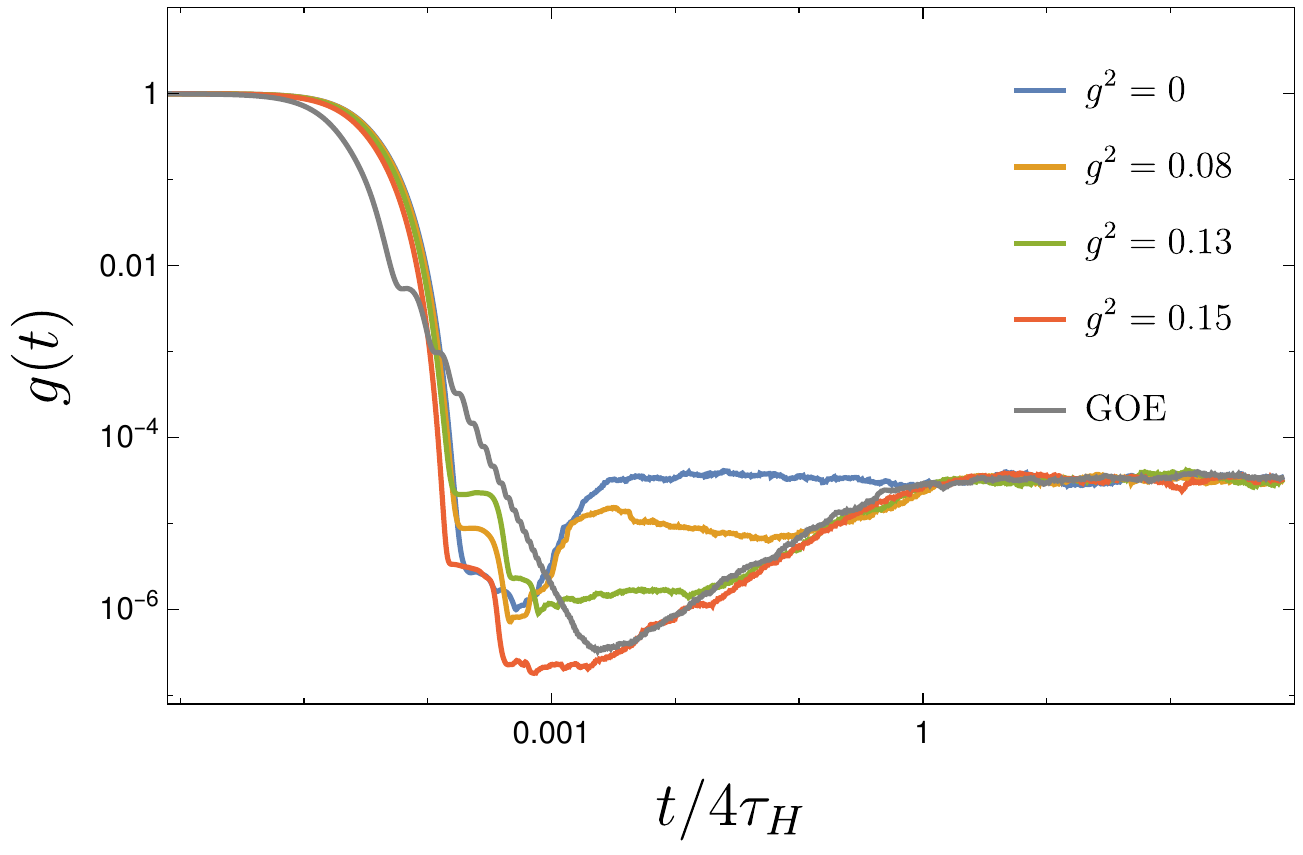}
        \hspace{3mm}
		\includegraphics[height=0.31\linewidth]{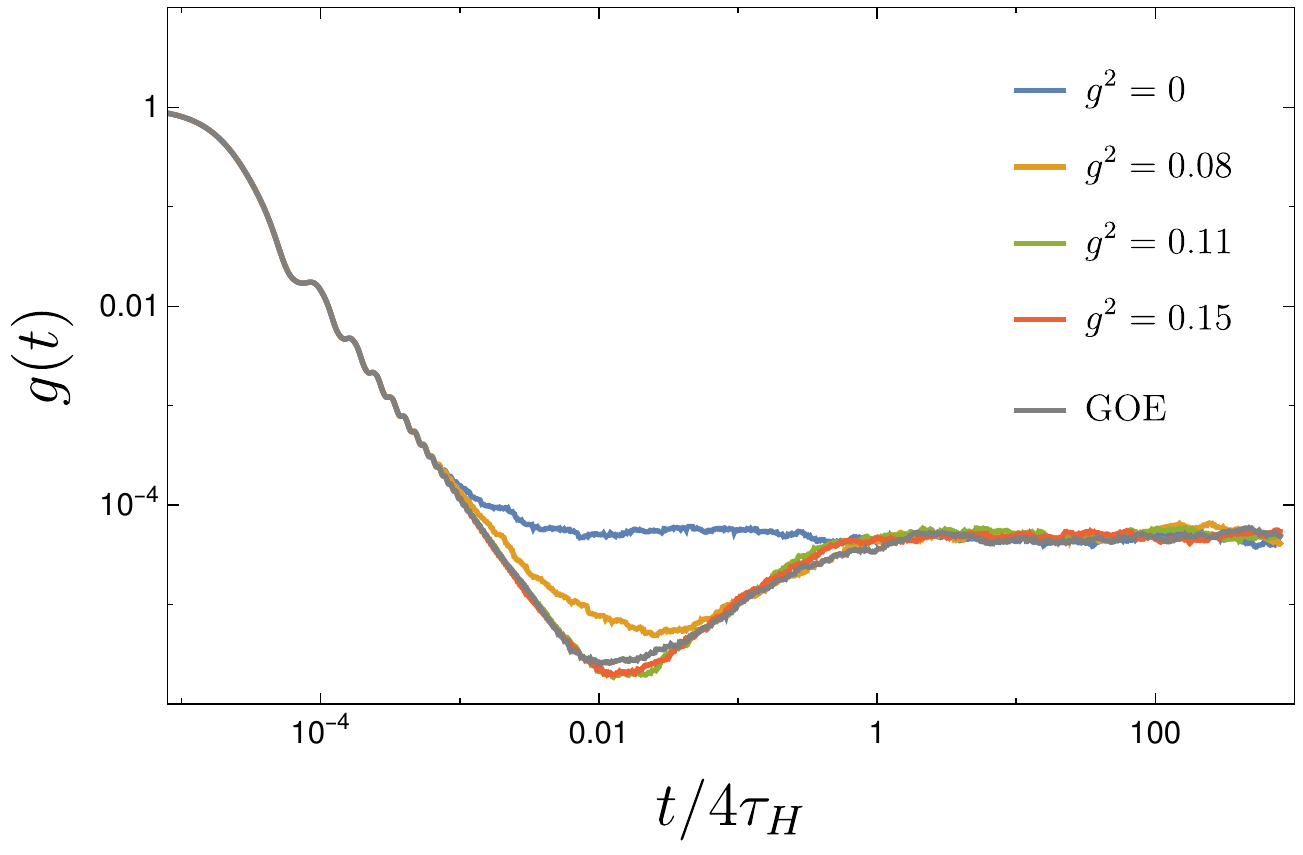}
	\end{center}
	\caption{ Normalised spectral form factor for the $L=22$, $M=10$ sector with $\beta=1/\sqrt{26}$ four-loop truncated spin chains as a function of $g^2$: (Left) Computed using energy eigenvalues (Right) Computed using the unfolded spectrum. Time is measured in units of $\tau_H$, the Heisenberg time, and the same time-averaged curve for one instance of the GOE Hamiltonian is shown for reference. }	\label{fig:SFF}
\end{figure}
We numerically compute the SFF for the $\beta$-deformed   four-loop spin chain in Figure \ref{fig:SFF} for the $L=22$, $M=10$ sector. We consider the SFF using both the energy eigenvalues and their unfolded analogues. As can be seen, the behaviour is somewhat similar though the details differ significantly. For the energy eigenvalues, there is a dip, then a complicated intermediate regime before a plateau. It is possible to distinguish a ramp regime for chaotic  values of the coupling where the SFF joins with the corresponding ramp of the GOE SFF. It can be seen that the beginning of the ramp region occurs earlier for larger values of the coupling. The SFF computed using the unfolded spectrum shows fewer features and lends itself to a straightforward interpretation. For all values of the coupling there is an initial dip to a value of $\sim 10^{-4}$. For the integrable theory, $g^2=0$, and values close to this, the SFF subsequently remains constant. As the coupling is increased and the model becomes chaotic, the bottom of the SFF dip moves lower, reaching close to $10^{-6}$ for $g^2=0.11$ and the ramp appears at progressively earlier times. This is in keeping with the identification of the beginning of the ramp with the Thouless time. 

\section{Eigenvector diagnostics}\label{sec:eigenvecs}
Eigenvalue statistics discussed in the previous section diagnose chaos through correlations in the spectrum. Chaos should however also leave a characteristic imprint on the structure of the eigenvectors. In \cite{berry1977regular}, Berry put forward the conjecture that, in single-particle systems, chaotic eigenfunctions behave as random superpositions of plane waves with random phases and Gaussian-distributed amplitudes. In many-body systems, the analogous expectation is that chaotic eigenvectors should be highly delocalised in a natural many-body basis. This expectation is closely tied to ergodicity and thermalisation through the eigenstate thermalisation hypothesis \cite{Deutsch:1991msp,Srednicki:1994mfb}.

The results presented in section \ref{sec:eigenvals} show that the higher-loop corrections drive the spin chain away from the integrable leading-order contribution to the dilatation operator, or equivalently from the nearest-neighbour XXX spin chain, and lead to chaotic behaviour in the sense of eigenvalue statistics. Notably, the spectral signatures indicate weak and loop-dependent chaos. 

The natural question is therefore whether these higher-loop corrections are also chaotic in the eigenvector sense. This question is interesting from two complementary perspectives. On the one hand, the two-loop, next-to-nearest deformation has already been studied in \cite{McLoughlin:2022jyt} but longer range deformations have not been explored from this point of view. Moreover, we do not study arbitrary long-range perturbations, but the particular combination of terms dictated by the planar dilatation operator. On the other hand, from the holographic point of view, one expects integrability to be restored in the all-loop planar theory. This statement should be treated with some care at larger values of the coupling, beyond the perturbative radius of convergence, where the finite-loop Hamiltonians are better viewed as spin chain models in their own right rather than controlled perturbative approximations to the gauge-theory dilatation operator. It is therefore natural to ask whether this constrained structure is also visible at the level of the eigenstates.

Indeed, it is known that systems that feature GOE-eigenvalue statistics need not approach the random-matrix limit of eigenvector diagnostics uniformly \cite{Backer:2019avi}.  In particular, finite-size fractal dimensions may approach their GOE-reference values only slowly, suggesting weaker ergodic behaviour than one might infer from the eigenvalue statistics alone. The eigenvector diagnostics shown below will probe these two related issues: whether the spectral signatures of chaos are accompanied by genuinely ergodic eigenstates, and whether the loop dependence seen in section \ref{sec:eigenvals} persists.

In order to quantify these questions we compute the delocalisation of the eigenstates with respect to two reference bases. The first is the symmetry-fixed $S^z$ spin-product basis at fixed magnetization. Its elements are 
\begin{equation}\label{eq:SpinProdBasisDef}
     |s_a\rangle=|m_1,m_2,\ldots,m_L\rangle \,, \qquad \text{where} \qquad m_j=\{\uparrow,\downarrow\}\,,
\end{equation}
denotes the spin configuration at site $j$ while $a$ runs from $1,\dots,2^L$, and the total magnetization sector is held fixed. This is the natural local basis of the spin chain, and is directly related to the basis of single-trace words. Delocalisation in this basis therefore measures how broadly an eigenstate spreads over spin configurations. 

The second basis is the eigenbasis of the one-loop dilatation operator, or equivalently of the integrable nearest-neighbour XXX spin chain, in the same symmetry sector. We denote it by
\begin{equation}\label{eq:OneLoopBasis}
        H_1 |E_a^{(1)}\rangle = E_a^{(1)} |E_a^{(1)}\rangle \,, \qquad a=1,\ldots,D \,,
\end{equation}
where $H_1$ is the one-loop Hamiltonian restricted to the fixed $(L,M)$ sector. Expanding the finite-loop eigenstates in this basis measures their delocalisation away from the integrable reference point. The two choices are therefore complementary: the spin-product basis probes ergodicity with respect to the microscopic spin chain degrees of freedom, whereas the one-loop eigenbasis probes how much memory of the integrable starting point is retained by the higher-loop truncations.

Throughout this subsection we work again in the fixed $(L,M)$-symmetry sector with $M/L\cong 7/16$ and we denote the dimension of the desymmetrised Hilbert space at fixed magnetization by $D=\mathrm{dim}\,\mathcal H_{L,M}^{(\mathrm{sym})}$. For each value of the coupling $g^2$, we diagonalise the Hamiltonian numerically and express the eigenvector in the chosen reference basis. In order to reduce edge effects, all averaged eigenvector observables are computed in the bulk of the spectrum, obtained by, as in the previous section, discarding the lowest and highest 15\% of eigenstates in energy, and averaging over the remaining central states. We denote the number of bulk eigenstates by $N_{\mathrm{bulk}}$.

Consider a normalized eigenstate $|\psi_\alpha\rangle$ expanded in the chosen basis,
\begin{equation}\label{eq:EigsinOneLoopBasis}
    |\psi_\alpha\rangle = \sum_a c_a^{(\alpha)} |\varphi_a\rangle\,,
\end{equation}
where $\ket{\varphi_a}$ is chosen to be either $\ket{s_a}$ or $\ket{E^{(1)}_a}$. The coefficients $c_a^{(\alpha)}$ define a probability distribution $p_a^{(\alpha)}=|c_a^{(\alpha)}|^2$ over the reference basis $\{|\varphi_a\rangle\}_a$. Observables constructed from these probabilities allow us to quantify how strongly an eigenstate is concentrated on a small subset of basis states or, conversely, how broadly it spreads across the Hilbert space. One measure of degree of delocalisation of individual eigenvectors is captured by the associated inverse participation ratio
\begin{align}\label{eq:IPR_def}
    \mathrm{IPR}_{q}^{(\alpha)} = \sum_a |c_a^{(\alpha)}|^{2q}\,.
\end{align}
The IPR measures how many reference basis states contribute appreciably to a given eigenstate. A large value of $\mathrm{IPR}_q$ means that the dilatation eigenstate is concentrated on a relatively small set of reference states, while a small value indicates that the state is spread over many basis vectors. In the spin-product basis this corresponds to broad mixing among operators with different spin configurations, while in the one-loop eigenbasis it measures delocalisation away from the integrable reference Hamiltonian.
Thus the same quantity has two complementary interpretations in the two bases: it measures operator mixing in the local spin basis, and remoteness from the integrable one-loop eigenbasis. For a state uniformly distributed over $D$ basis states one has $\mathrm{IPR}_q\sim D^{1-q}$, while for a localised state $\mathrm{IPR}_q$ is parametrically larger. 

It is also convenient to associate to each eigenstate a finite-$D$ fractal dimension
\begin{equation}\label{eq:Dq}
    D_q^{(\alpha)}=-\frac{\log \mathrm{IPR}^{(\alpha)}_q}{(q-1)\log D}\,.
\end{equation}
The fractal dimension characterises the effective support of the eigenstate in the chosen reference basis. Values $D_q \simeq 1$ correspond to eigenstates that are close to ergodic random vectors in the chosen basis, while values $D_q \simeq 0$ correspond to strong localisation. For intermediate values the state is extended but not fully ergodic.

More generally, the moments $\mathrm{IPR}_q$, and hence $D_q$, for different values of $q$ probe different parts of the wavefunction amplitude distribution. Larger values of $q$ emphasise rare large amplitudes, while smaller $q$ are more sensitive to the bulk of the distribution. The full set of exponents $D_q$ characterises the scaling properties of the eigenstate amplitudes. In particular, if the dimensions $D_q$ depend non-trivially on $q$, the wavefunction is said to be multifractal \cite{evers2008anderson}, meaning that different moments of the amplitude distribution scale with different effective dimensions. This behaviour reflects strong fluctuations of the coefficients $c_i^{(\alpha)}$ and indicates that the eigenstate occupies a hierarchically structured subset of the Hilbert space rather than spreading uniformly over it.

A complementary measure of eigenstate structure is the Shannon entropy of the probability distribution $p_i^{(\alpha)}$
\begin{equation}
    S_\alpha=-\sum_a |c_a^{(\alpha)}|^2 \log |c_a^{(\alpha)}|^2\,,\label{ShannonEntropy}
\end{equation}
together with the structural entropy \cite{pipek1992universal}
\begin{equation}\label{eq:str_ent}
    S_{\mathrm{str},\alpha}=S_\alpha+\log \mathrm{IPR}^{(\alpha)}_2\,\,.
\end{equation}
By construction, the structural entropy $S_{\mathrm{str}}$ isolates the component of the Shannon entropy that is not determined by the inverse participation ratio. In particular, it allows one to distinguish states that exhibit similar levels of delocalisation but differ in the detailed distribution of their amplitudes. In what follows we also consider bulk averages such as
\begin{equation}\label{eq:av_Dq_Sstr}
    \langle D_q\rangle=\frac{1}{N_{\mathrm{bulk}}}\sum_{\alpha\in \mathrm{bulk}} D_q^{(\alpha)}\,, \qquad \langle S_{\mathrm{str}}\rangle=\frac{1}{N_{\mathrm{bulk}}}\sum_{\alpha\in \mathrm{bulk}} S_{\mathrm{str},\alpha}\,.
\end{equation}
For GOE random vectors, one has the reference value $S^{\rm GOE}_{\rm str}\approx 0.3646$ \cite{Santos:2010qxi}, which we use as a comparison for the values obtained from the higher-loop dilatation operators.  

We now apply these diagnostics first in the spin-product basis, where they probe ergodicity in the local many-body basis, and then in the one-loop eigenbasis, where they probe the distance from the integrable reference point.

\subsection{Spin-product basis}\label{sec:eigenv_spin_basis}

\begin{figure}
    \hspace{-13pt}
    \includegraphics[width=1.05\linewidth]{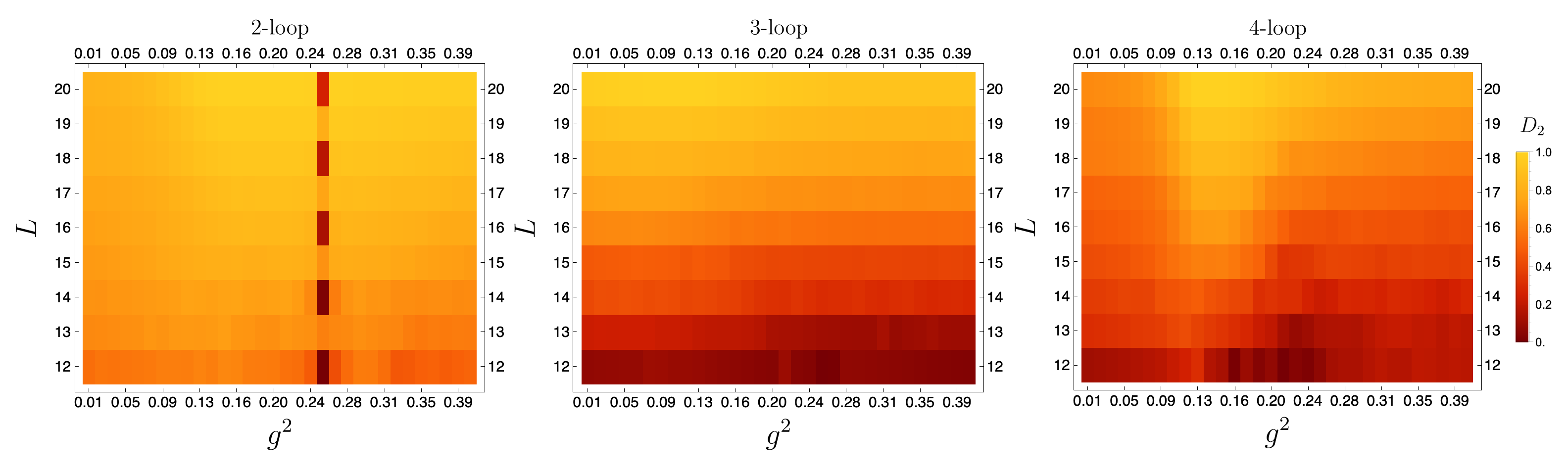}
    \caption{The (averaged) fractal dimension $D_2$ for $L\in \{12,\dots,20\}$ and $g\in (0,0.4]$, for the Hamiltonian at 2-, 3-, and 4-loops. }
    \label{fig:D2vsLandg}
\end{figure}

We first turn to the fractal and entropy diagnostics in the spin-product-basis. Figure \ref{fig:D2vsLandg} shows the fractal dimension $D_2$, see \eqref{eq:av_Dq_Sstr} for $q=2$, for the two-, three- and four-loop spin chains as a function of the coupling $g^2\in(0,0.4]$ and for a range of chain lengths $L\in \{12,\dots,20\}$. Several features are immediately visible. First, away from the special value of the coupling at $g^2=1/4$, the dependence on the spin chain length $L$ weakens as the system size increases, indicating the spectral dimensions $D_2$ asymptote to a fixed value for large $L$. This is consistent with \cite{Backer:2019avi}, where it was shown that, in general, finite-size fractal dimensions approach their asymptotic values only slowly in ergodic many-body system. Second, the values of $D_2$ remain systematically below the ergodic value $D_2=1$ through the range shown. Thus, even where the level statistics are already close to GOE, the eigenvectors are not fully random vectors in Hilbert space and retain some non-trivial fractal structure. Finally, the two-loop truncation shows a sharp feature at $g^2=1/4$, the special point where the nearest-neighbour term cancels. In the spin product basis, the corresponding even/odd splitting is largely an artefact of the finite-size normalisation\footnote{At the special $g^2=1/4$, the two-loop Hamiltonian becomes purely next-to-nearest neighbour. For even $L$, the range-two interaction splits into two disconnected sub-chains, giving two decoupled Heisenberg spin chains. For odd $L$, the step-two map runs through all sites, so one obtains a single relabelled Heisenberg spin chain instead. The even/odd splitting visible in the spin-product in Figure \ref{fig:D2vsLandg} is therefore largely due to a normalisation artefact arising from the different fixed-$M$ Hilbert-space dimension.}.
Taken altogether, the fractal dimension indicates that the onset of chaos in the truncated models is accompanied only by weak ergodicity in the eigenstates.

\begin{figure}[t!]
    \centering
    \includegraphics[width=0.98\linewidth]{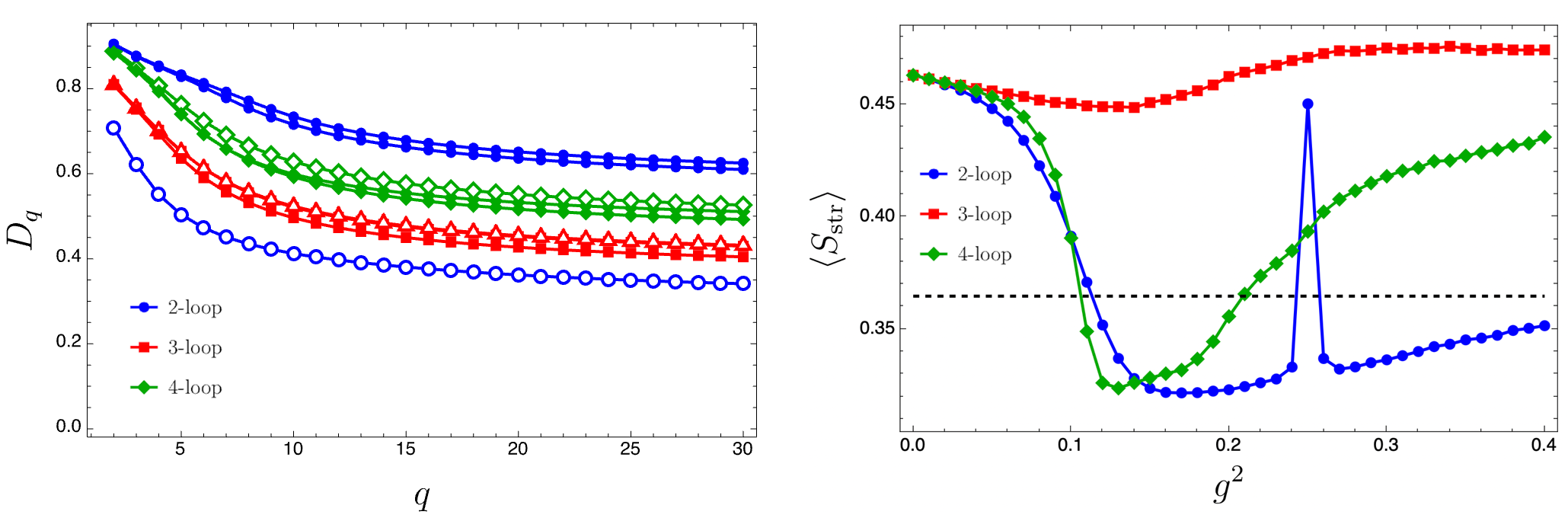}
    \caption{Left: the multi-fractal dimension $D_q$ for $q \in \{1,2,\dots ,30\}$ at $L=20$. The curves indicated by open markers are evaluated at $g^2=1/4$, while solid markers are evaluated at values slightly below and above that value; $g^2=0.2375$ and $g^2=0.2625$. Right: the (averaged) structural entropy for coupling values $g\in (0,0.4]$ at $L=20$. The black dashed line indicates the reference value $S^{\rm GOE}_{\rm str} \approx 0.3646$ for GOE ensembles. }
    \label{fig:S_str_g_Dq_qL}
\end{figure}

Figure \ref{fig:S_str_g_Dq_qL} (left panel) shows the multifractal dimensions $D_q$ at fixed $L=20$ for couplings just below, at, and just above $g^2=1/4$. The monotonic decrease with $q$ is the expected behaviour of multifractal dimensions, since larger $q$ probe the largest wavefunction amplitudes more strongly \cite{Backer:2019avi}. The main effect is in the two-loop chain, where the curves on the two sides of $g^2=1/4$ are clearly separated over the whole range of $q$, showing that this point reorganises the eigenvector structure. By contrast, the corresponding shift is much smaller for the three- and four-loop truncations. Independently of this jump, the large-$q$ hierarchy is also informative: the two-loop chain approaches the largest values of $D_q$, the three-loop chain the smallest, and the four-loop case lies in between. 

Figure \ref{fig:S_str_g_Dq_qL} (right panel) shows the averaged structural entropy with respect to the spin-product basis. Since the structural entropy \eqref{eq:str_ent} is sensitive to the shape of the coefficient distribution but factors out the global degree of delocalisation, it allows one to distinguish between eigenstates which have similar participation ratios but not equally random structure. For the two-loop chain, the averaged structural entropy \eqref{eq:av_Dq_Sstr} decreases rapidly around $g^2\simeq 0.1$ and then remains close to the GOE-reference value over a broad range below $g^2=1/4$. By contrast, the three-loop curve stays well above the GOE-value throughout the coupling range, indicating that although the eigenstates delocalise, their amplitude distribution remains far from the random-matrix behaviour. The four-loop case is intermediate; it begins to decrease near $g^2\simeq 0.1$, but subsequently shows a crossover around $g^2\simeq 0.2$ before going up again. As with the eigenvalue diagnostics in section \ref{sec:eigenvals}, the initial onset of chaos for the four-loop chains happens at a slightly larger value of the coupling compared to the two-loop case and is closer to a step function. These consistently low values imply that, contrary to what one could expect from the GOE-eigenvalue statistics, the eigenvectors retain over a wide range of coupling additional structure, reinforcing the interpretation that the systems are only weakly ergodic. On the other hand, the loop hierarchy in Figure \ref{fig:S_str_g_Dq_qL} closely mirrors the one already seen in the spectral diagnostics of section \ref{sec:eigenvals}, in particular in the $r$-ratio shown in Figure \ref{fig:rRatio}: the two-loop truncation comes closest to chaotic behaviour, the three-loop truncations remains closest to the integrable regime and the four-loop case lies in between.

\begin{figure}[t!]
    \centering
    \includegraphics[width=0.8\linewidth]{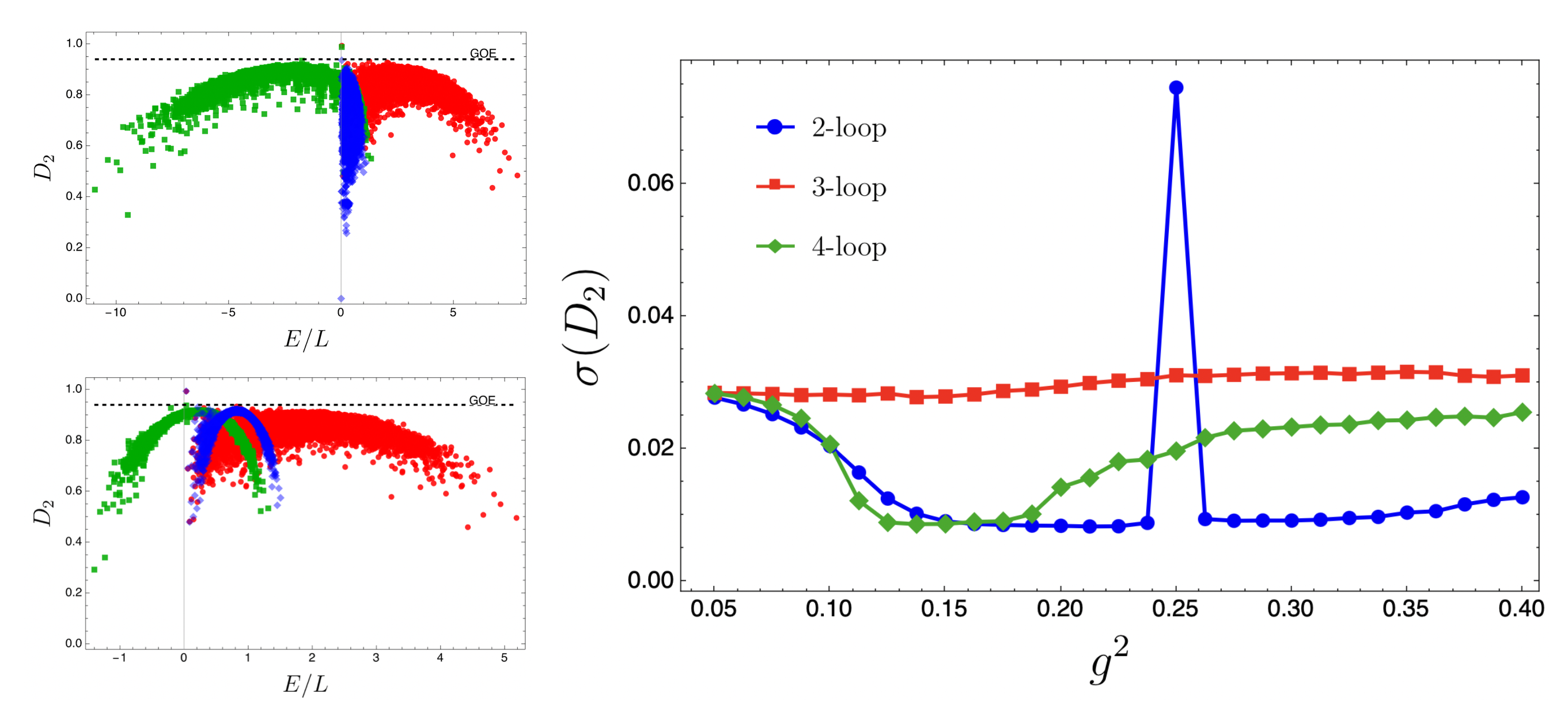}
    \caption{Left, up: Fractal dimensions $D_q$ for $q = 2$ vs. the scaled
eigenenergies at $g^2=0.175$. Left, down: same plot but at  $g^2=0.25$. For both plots the GOE reference line is displayed for $\dim \mathcal H_{(20,9)}^{\mathrm{sym}}=8398$. Right: variance of the fractal dimension $D_2$ as a function of $g^2$. All plots are for a spin chain of length $L=20$ and $M=9$.}
    \label{fig:scatter_st_dev_D2_L18}
\end{figure}

While the previous figures characterise the averaged ergodic behaviour, one could wonder which portion of the eigenstates are ergodic. To understand how these averaged quantities arise, in Figure \ref{fig:scatter_st_dev_D2_L18} we show (on the left) representative scatter plots of the fractal dimension $D_2$ across the spectrum together, and (on the right) the standard derivation $\sigma (D_2)$ as a function of the coupling $g^2$. The scatter plots suggest that, while not all eigenstates are fully ergodic, most states lie in a relatively narrow band of fractal dimensions. This is quantitatively captured by the variance. For the two-loop chain, $\sigma(D_2)$ decreases rapidly as the coupling increases from zero, indicating that the weakly chaotic regime is relatively homogeneous at the level of the eigenstates. At the special point $g^2=1/4$, a large portion of the eigenstates drop further away from the GOE-reference value as can be seen in the scatter plot on the left bottom. This is also reflected in a sharp peak in the variance in the right panel. The three-loop chain interpolates  between these behaviours while showing a mild crossover around $g\simeq 0.2$. This corroborates behaviour already witnessed in the generic next-to-next-neighbour XXZ-spin chain in \cite{Backer:2019avi}, where the chaotic spin chain nonetheless exhibited weak ergodicity with only a fraction of eigenstates being truly ergodic. 

Overall, the eigenvector diagnostics in the spin-product basis show that the higher-loop spin chains dictated by planar $\mathcal N=4$ SYM are only weakly ergodic. Although the eigenvalue diagnostics of section \ref{sec:eigenvals} display the onset of chaotic spectral correlations, the corresponding eigenstates remain systematically less random than GOE vectors in the local many-body basis. This mismatch between eigenvalue chaos and full eigenvector ergodicity is known to occur in generic spin chains \cite{Santos:2010qxi,Backer:2019avi}, including next-to-next-nearest-neighbour models. What is non-trivial here is that the same phenomenon persists for the particular long-range deformations fixed by the gauge theory. At the same time, the loop-dependent hierarchy seen in the eigenvalue statistics remains clearly visible at the level of the eigenvectors: the two-loop truncation comes closest to GOE-like behaviour over an intermediate range of couplings, the three-loop truncation remains closest to the integrable regime, and the four-loop case again lies in between. Thus the finite-loop dilatation operators display weaker ergodicity than one might infer from the spectral data alone, while preserving the same relative pattern of chaos across loop order.

\subsection{Integrable basis}\label{sec:eigenv_intb_basis}
We now repeat the eigenvector analysis in a basis adapted to the integrable one-loop truncation. Instead of expanding the eigenstates of the finite-loop Hamiltonians in the local spin-product basis, we use the eigenbasis of the leading nearest-neighbour Hamiltonian $H_1$ defined in \eqref{eq:123loops}. Thus, in a fixed $(L,M)$ symmetry sector, we first diagonalise, see \eqref{eq:OneLoopBasis}, and then expand the eigenstates of the two-, three- and four-loop truncated Hamiltonians as \eqref{eq:EigsinOneLoopBasis} in the eigenstate basis of the integrable XXX spin chain. The probabilities $|c^{(\alpha)}_a|^2$ therefore measure how strongly the finite-loop eigenstates spread over this basis. This basis is conceptually different from the spin-product basis studied above. In the spin-product basis, delocalisation measures the mixing of microscopic spin configurations, or equivalently of single-trace words. In the present basis, delocalisation measures the loss of memory of the integrable one-loop Hamiltonian. Hence, strong ergodicity in this basis would mean that the higher-loop corrections not only mix local spin configurations, but also randomise the eigenstates with respect to the integrable conserved structure of $H_1$. Conversely, localisation or weak delocalisation in this basis indicates that the finite-loop eigenstates remain organised around the integrable reference problem, even when the spectrum already shows signatures of Wigner-Dyson correlations.

\begin{figure}[t!]
    \hspace{-13pt}
    \includegraphics[width=1.05\linewidth]{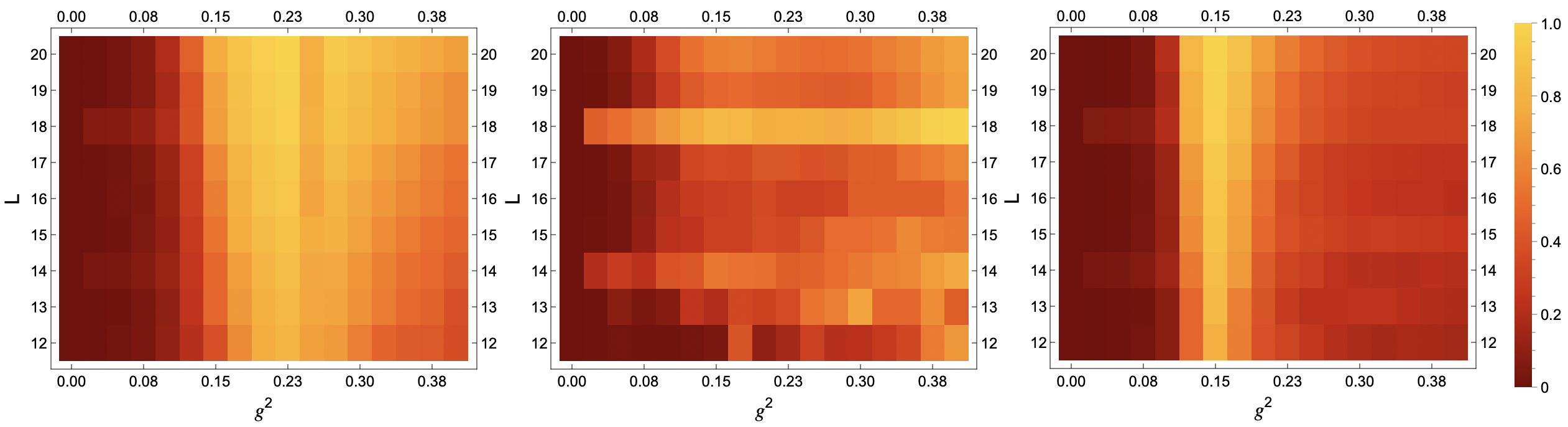}
    \caption{
Fractal dimension $D_2$ of the eigenstates of the finite-loop Hamiltonians in the eigenbasis of the one-loop nearest-neighbour Hamiltonian $H_1$, defined in \eqref{H2g2trunc}. The plots show the two-loop, three-loop and four-loop truncations, from left to right, as a function of the coupling $g^2$ and the spin chain length $L\in\{12,\ldots,20\}$, at fixed magnetization sector with $M/L\simeq 7/16$.}
    \label{fig:eigbasis_heatmap}
\end{figure}

Figure \ref{fig:eigbasis_heatmap} shows the averaged fractal dimension $D_2$ in the $H_1$-eigenbasis for $L$ between $12$ and $20$, and $g^2\in(0,0.4]$. At small coupling, all finite-loop Hamiltonians are close to the one-loop Hamiltonian, and the corresponding eigenstates remain sharply localised in the $H_1$-basis. This is reflected in the small values of $D_2$ near $g^2=0$. As the coupling is increased, the two-loop truncation shows a pronounced delocalisation crossover: the eigenstates spread over a large fraction of the $H_1$-eigenbasis and $D_2$ approaches values close to the ergodic regime. 

This behaviour is qualitatively different from the spin-product basis result of Figure \ref{fig:D2vsLandg}. There, the two-loop eigenstates were already extended over local spin configurations but remained systematically below the GOE random-vector value. In the integrable basis, the same two-loop deformation appears more directly as a loss of memory of the nearest-neighbour XXX eigenstates. The special point $g^2=1/4$, however, again leaves a clear imprint. Around this value the fractal dimension is suppressed, consistent with the cancellation of the nearest-neighbour interaction in the two-loop Hamiltonian and the associated reorganization of the model into a more integrable structure. 

The three-loop truncation behaves very differently. Its fractal dimension remains small throughout the coupling window, showing that the eigenstates remain close to the $H_1$-basis and do not strongly delocalise away from the integrable one-loop problem. This is consistent with the spectral diagnostics, where the three-loop truncation did not display a clear transition to GOE statistics over the same range of couplings. The four-loop truncation lies between these two behaviours. It develops a visible window of delocalisation at intermediate coupling, but the eigenstates do not remain fully extended over the whole range; instead, $D_2$ decreases again at larger coupling. Thus, compared with the spin-product basis, the integrable basis makes the loop-dependent hierarchy sharper: the two-loop chain departs most strongly from the $H_1$-eigenbasis, the three-loop chain remains closest to it, and the four-loop chain displays an intermediate and coupling-dependent loss of integrable-basis structure.

\begin{figure}[t!]
    \centering
     \begin{subfigure}[b]{0.49\textwidth}
     \centering
         \includegraphics[width=\textwidth]{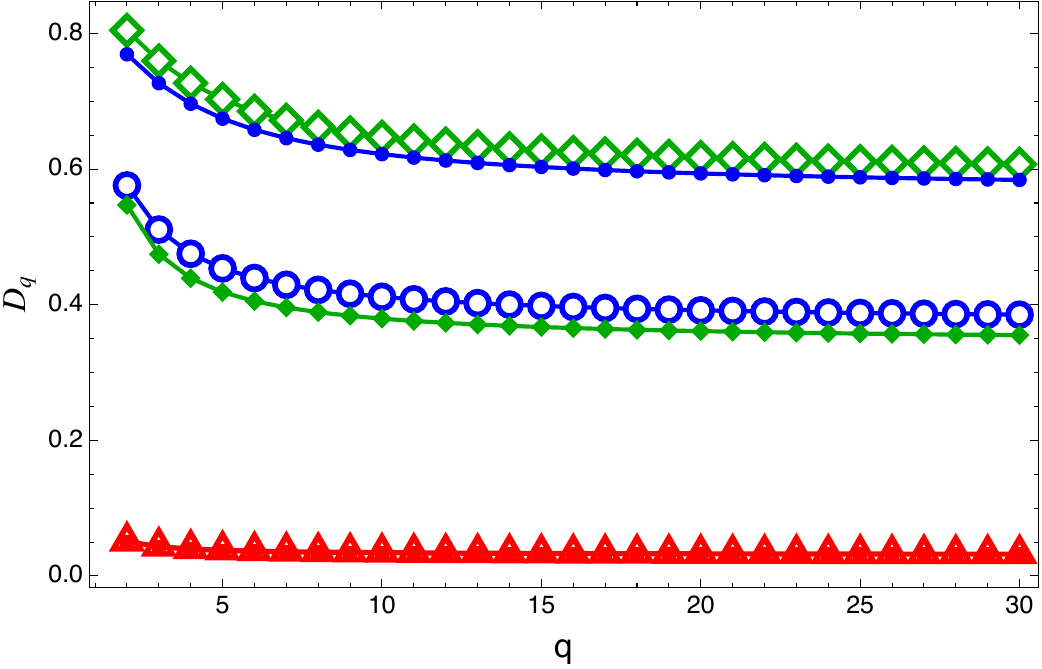}
         \label{fig:dq.eigbasis}
     \end{subfigure}
     \hfill
     \begin{subfigure}[b]{0.49\textwidth}
         \centering
         \includegraphics[width=\textwidth]{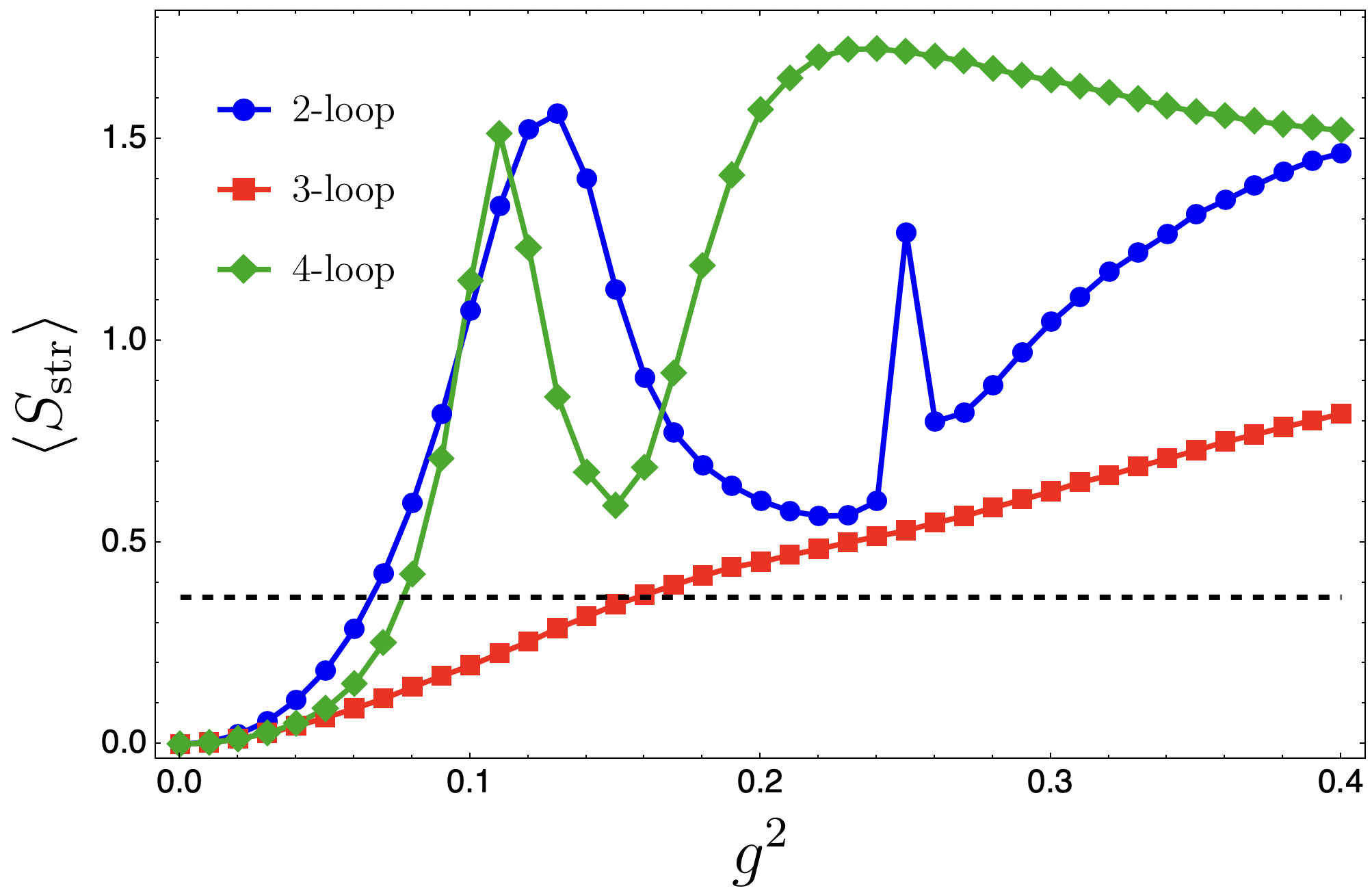}
         \label{fig:F4avggue}
     \end{subfigure}
        \caption{Left: the multi-fractal dimension $D_q$ for $q \in \{1,2,\dots ,30\}$ at $L=20$. The curves indicated by open markers are evaluated at $g^2=0.15$, while solid markers are evaluated at $g^2=0.2$. Right: the averaged structural entropy for coupling values $g\in (0,0.4]$ at $L=20$. The red dashed line indicates the reference value $S^{\rm GOE}_{\rm str} \approx 0.3646$ for GOE ensembles.}
        \label{fig:eigb}
\end{figure}

The multifractal dimensions in Figure \ref{fig:eigb} provide a more refined version of this statement. The left panel shows $D_q$ for $q\in\{1,2,\ldots,30\}$ at $L=20$, comparing representative values of the coupling. For all loop orders, $D_q$ decreases as $q$ is increased, as expected for multifractal diagnostics: larger values of $q$ give more weight to the largest amplitudes in the wavefunction and are therefore more sensitive to residual localisation. The hierarchy between loop orders is, however, different in the $H_1$-basis than in the spin-product basis. The two-loop chain has the largest $D_q$ over the full range of $q$, indicating the broadest spreading over the integrable eigenstates. The four-loop chain is also extended but less strongly so, while the three-loop chain remains close to localised in this basis, with $D_q$ staying small and only weakly dependent on $q$. 

The right panel of Figure \ref{fig:eigb} shows the averaged structural entropy in the same basis. Unlike in the spin-product basis, where the GOE reference value $S_{\rm str}^{\rm GOE}\simeq 0.3646$ provides a direct benchmark for the shape of the local-basis amplitude distribution, the structural entropy in the integrable basis grows to much larger values in the delocalised regimes. This should not be interpreted simply as convergence to GOE eigenvectors. Instead, it reflects the fact that, when the finite-loop eigenstates spread over the highly non-local eigenbasis of $H_1$, their coefficient distributions acquire additional structure not captured by the inverse participation ratio alone. The two-loop curve rises as the coupling is turned on, reaches the maximum in the same region where the eigenstates are most delocalised in the $H_1$-basis, and then shows a visible jump near $g^2=1/4$. The four-loop curve displays a similar but shifted crossover: it rises at intermediate coupling and subsequently decreases, indicating that the strongest departure from the integrable eigenbasis occurs only in a finite coupling window. The three-loop curve remains much smaller throughout, confirming that the corresponding eigenstates stay comparatively close to the one-loop integrable basis. 
\begin{figure}[t!]
    \centering
    \includegraphics[width=0.9\linewidth]{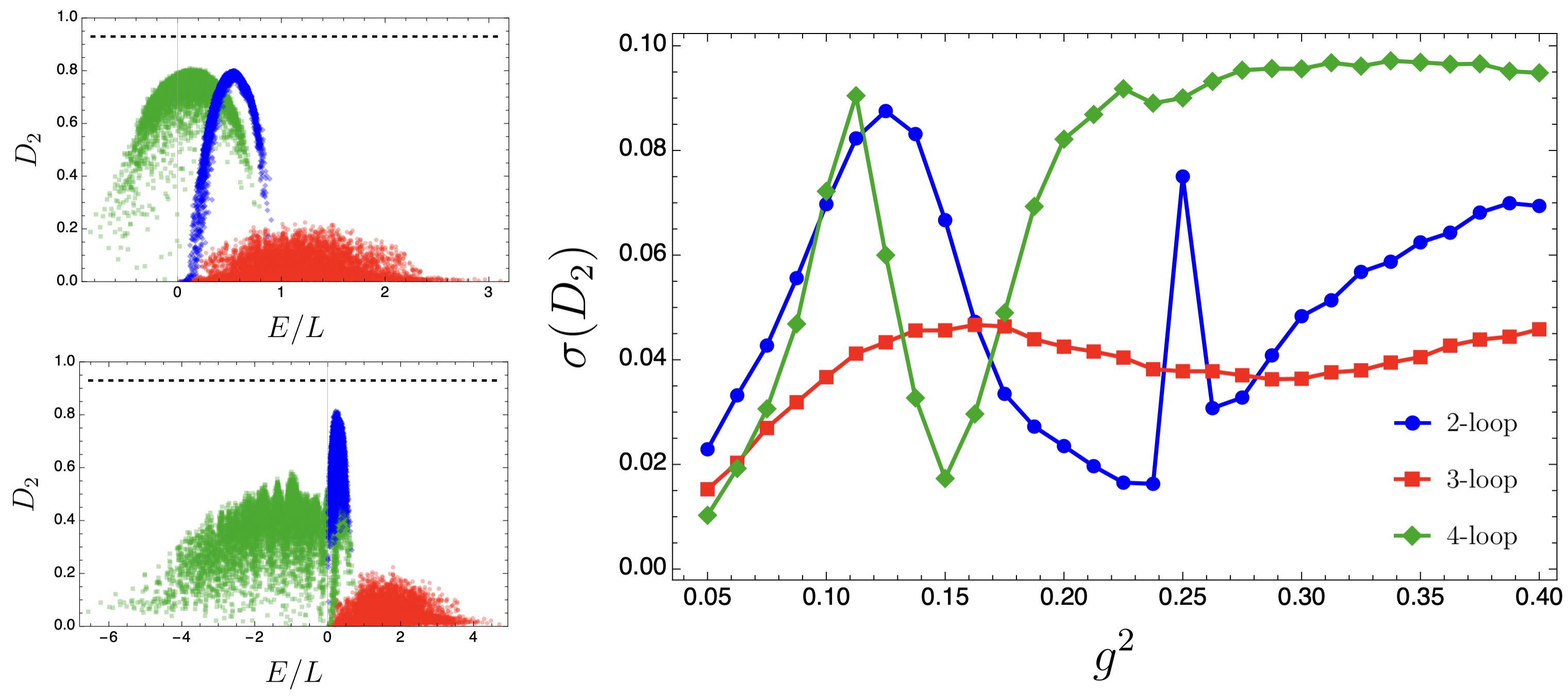}
    \caption{
Distribution and fluctuations of the fractal dimension $D_2$ in the eigenbasis of the one-loop nearest-neighbour Hamiltonian $H_1$ for a spin chain of $L=20$ and $M=9$. Left panels: scatter plots of $D_2$ against the rescaled eigenenergy $E/L$. The dashed horizontal line indicates the GOE reference value for random vectors in the corresponding Hilbert-space dimension. Right panel: standard deviation $\sigma(D_2)$ of the fractal dimension as a function of $g^2$, computed over the bulk of the spectrum. Enhanced values of $\sigma(D_2)$ mark coupling regions where different eigenstates delocalise at different rates.
}\label{fig:scatter_L20}
\end{figure}

Figure \ref{fig:scatter_L20} shows how the averaged quantities arise from the distribution of eigenstates across the spectrum. The scatter plots of $D_2$ against the rescaled eigenenergy indicate that the delocalisation in the $H_1$-basis is not uniform. In the two-loop and four-loop cases, a substantial fraction of the bulk eigenstates becomes extended over the integrable basis in the chaotic crossover regime, but the distribution remains structured and energy-dependent. The eigenstates therefore do not behave as uniformly random vectors in the $H_1$-basis. The three-loop data, by contrast, remain concentrated at much smaller values of $D_2$, again showing that most eigenstates preserve a strong overlap with only a limited number of $H_1$-eigenstates. The right panel of Figure \ref{fig:scatter_L20} quantifies this through the standard deviation $\sigma(D_2)$. The two-loop and four-loop curves display enhanced fluctuations in precisely the regions where the averaged fractal dimension changes most rapidly. This indicates that the crossover away from the integrable basis does not occur uniformly across the spectrum; rather, different eigenstates delocalise at different rates. The sharp feature in the two-loop curve near $g^2=1/4$ is the integrable-basis analogue of the reorganization already visible in the spin-product basis. The four-loop curve has a broader region of enhanced fluctuations, consistent with a more gradual and less complete departure from the $H_1$-basis. The three-loop variance remains comparatively small, in agreement with its weak delocalisation and the absence of a strong spectral-chaos crossover.

 Overall, the eigenvector diagnostics in the one-loop basis confirms that the finite-loop Hamiltonians remain only weakly ergodic. Although the eigenvalue statistics indicate the onset of chaotic spectral correlations, the eigenstates do not become fully random in the basis adapted to the integrable point. The loop-dependent hierarchy on the other hand is again clear: the two-loop truncation departs most strongly from the integrable basis, the three-loop truncation remains closest to it, and the four-loop case is intermediate.

\section{Krylov basis diagnostics}\label{sec:Krylov}
In this section we use Krylov-space tools and diagnostics to obtain a complementary perspective on the transition to quantum chaos in $\mathcal{N}=4$ SYM and in the long-range spin chains discussed above.
\subsection{Spread complexity}
Spread complexity \cite{Balasubramanian_2022} is a recently introduced measure of quantum complexity built upon the Krylov basis and the recursion method \cite{ViswanathMuller1994}. It quantifies the extent to which an initial quantum state spreads through Hilbert space under unitary evolution generated by a time-independent Hamiltonian (or, more generally, by an arbitrary generator of dynamics). Because spread complexity depends on both the generator and the chosen initial state, it is sensitive to both the eigenvalue chaos discussed in section \ref{sec:eigenvals} as well as the eigenvector chaos examined in section \ref{sec:eigenvecs}. In this section, we review the definition and physical interpretation of spread complexity, and then discuss the specific features that encode various signatures of quantum chaos. For comprehensive reviews, see \cite{ViswanathMuller1994,Nandy_2025,Baiguera_2026,Rabinovici:2025otw}. 

Given a time-independent Hamiltonian $H$ and an initial state $\ket{\psi(0)}$ we expand the time-evolved state as follows
\begin{equation}
    \ket{\psi(t)}=e^{-iHt}\ket{\psi(0)}=\sum^{\mathcal{K}-1}_{n=0}\psi_n(t)\ket{K_n}\,,
\end{equation}
where $\ket{K_n}$ is the Krylov basis of dimension $\mathcal{K}$ constructed iteratively using the Lanczos algorithm (see appendix \ref{LanczosAlg})  through repeated actions of the Hamiltonian on the initial state and applications of the Gram-Schmidt procedure. As shown in \cite{Balasubramanian_2022}, the Krylov basis parametrizes the minimal subspace in which the evolution of the initial state by $H$ takes place, with the index $n$ of the basis element related to the time elapsed. The main goal is then to determine the coefficients of this expansion $\psi_n(t)=\langle K_n|\psi(t)\rangle$. 

Fortunately, since the Lanczos algorithm constructs a basis in which the Hamiltonian acts tri-diagonally
 \begin{equation}
    H \ket{K_n} = a_n \ket{K_n}+b_n \ket{K_{n-1}} + b_{n+1}\ket{K_{n+1}}\,, 
\end{equation}
we can show that the coefficients (amplitudes) satisfy a discrete Schroedinger equation fully determined by Lanczos coefficients $a_n$ and $b_n$
\begin{equation}
i\partial_t\psi_n(t)=a_n\psi_n(t)+b_n\psi_{n-1}(t)+b_{n+1}\psi_{n+1}(t)\,,
\end{equation}
with initial condition $\psi_n(0)=\delta_{n,0}$.

Effectively, this representation maps quantum dynamics into a problem of a particle hopping on a 1D chain (known as the Krylov chain) which starts from the origin at $n=0$ and, as time progresses, moves to the right. The Lanczos coefficients $a_n$ determine the amplitude of staying on site $n$, and $b_n$ of moving to the left or right. The unitarity of the evolution ensures that the amplitudes $\psi_n(t)$ define a normalized probability distribution on the Krylov chain: $p_n(t)=|\psi_n(t)|^2$. The spread complexity is then defined as  the average position of the effective particle on the chain
\begin{equation}
    C(t) := \sum_{n=0}^{\mathcal{K}-1} n |\psi_n(t)|^2 \,.
\end{equation}
This will be our main tool in the following part of this article.

A generic time evolution of spread complexity, shows an initial period of universal quadratic growth, followed by linear growth which can become exponential if the Krylov basis dimension is infinite, or can saturate for finite-dimensional Krylov basis. 
The saturation phase can start with a peak, which for random matrix Hamiltonians is directly associated with the energy level-repulsion. In fact it was observed in models with controlled integrability breaking that the presence of the peak and its value can be a reliable diagnostic of chaos. Finally, the complexity plateaus at late times to a constant estimated from complete delocalisation of the wave function on the Krylov chain. Namely, we expect that at late times $p_n(t)=const.=1/\mathcal{K}$ so that
\begin{equation}\label{eq:SaturationOfKrylov}
    \overline{C(t)} = \frac{1}{\mathcal{K}}\sum^{\mathcal{K}-1}_{n=0}n=\frac{\mathcal{K}-1}{2}\sim \frac{\mathcal{K}}{2}\,.
\end{equation}
In integrable models, we can observe additional fluctuations around the plateau value or generic revivals for lower-dimensions $\mathcal{K}$. Moreover, after studying various examples of chaotic and integrable spin chain Hamiltonians, \cite{Balasubramanian:2023kwd} concluded that the variance in the Lanczos coefficients, as a measure of disorder on the chain, can also reliably track the spectral integrability/chaos of the model. We will return to this probe in the following sections.

Before proceeding, a comment is in order. Recall that spectral unfolding and clipping play a central role in conventional diagnostics of quantum chaos. Signatures of eigenvalue chaos are typically extracted from the unfolded spectrum, whereas no analogous or universally accepted unfolding procedure exists for eigenvectors. In addition, one often removes a fraction of the highest- and lowest-energy states, since these edge states are generally not expected to exhibit RMT behavior, even in paradigmatic chaotic systems.
Both procedures require knowledge of the complete set of eigenvalues and eigenvectors. For an initial state specified by its overlaps with the energy eigenbasis, clipping is implemented by setting to zero the overlaps associated with the discarded eigenstates. By contrast, unfolding leaves the overlap distribution unchanged, as it only modifies the eigenvalue spectrum. For states whose coefficients explicitly depend on the energies, such as the thermofield double (TFD) state, the energies entering the overlaps are replaced by their unfolded counterparts.

The remainder of this section is organised as follows. We compute the Krylov data for the case of an initial infinite temperature coherent Gibbs state (CGS) \cite{Carabba_2022} (defined in \eqref{eq:DefCGS}) for fixed $(L,M)$ sectors of the various loop truncated dilatation operators at various values of the coupling. In section \ref{sec:KrylovPeak} the peak of spread complexity is compared to the $r$ ratio. The disorder in Lanczos coefficients is formalised in section \ref{sec:KrylovDisorder} and compared to the $r$ ratio once again. Finally in section \ref{sec:VecandKrylov} we discuss whether and how the analysis of section \ref{sec:eigenvecs} on eigenvector chaos can be correlated with choices of initial states in spread complexity.
\subsection{Lanczos coefficients and spread complexity in $\mathcal{N}=4$ SYM}\label{sec:LanczosSpreadN4}
The purpose of this section is to introduce our setups and to compare the Lanczos coefficients and spread complexity computed from both the raw and the unfolded spectra. In the following sections, we will analyze these quantities in greater detail and connect them to the diagnostics discussed in sections \ref{sec:eigenvals} and \ref{sec:eigenvecs}. This provides a bridge between the Krylov-basis approach and more conventional probes of the emergence of quantum chaos in $\mathcal{N}=4$ SYM. 

More precisely, we consider examples of both the spread complexity and the Lanczos coefficients obtained from the recursion method based on the Hamiltonian given by the dilatation operator\footnote{As in previous sections, we consider the $\beta$ deformation to be fixed at $\beta=1/\sqrt{26}$, which should capture the qualitative behaviour for any irrational $\beta$.}, for a given loop truncation ($l$) and coupling strength ($g^2$), see \eqref{eq:HamiltonianExpansion}. Then, we chose the initial state of the evolution as
\begin{equation}
\ket{\psi(0)}= \frac{1}{\sqrt{D}} \sum_{a=1}^D \ket{E^{(l)}_a(g^2)}\,.\label{InState0}
\end{equation}
This state is equivalent to the infinite temperature coherent Gibbs state (CGS) \cite{Carabba_2022}, which is an analogue of the thermofield double in the single Hilbert space
\begin{equation}\label{eq:DefCGS}
   \ket{\text{CGS}(T)} := \frac{1}{\sqrt{\mathcal{Z}(T)}} \sum_{a=1}^D e^{-\frac{1}{2T} E_a} \ket{E_a}\,, \qquad \mathcal{Z}(T)=\sum_{a=1}^D e^{-\frac{1}{T}E_a}\,, 
\end{equation}
where $\mathcal{Z}(T)$ is the thermal partition function at temperature $T$ and $\ket{E_a}$ are the eigenstates of the evolving Hamiltonian. For our purposes the CGS will also have labels for the loop truncation ($l$) and coupling strength ($g^2$): $\ket{\text{CGS}^{(l)}(T;g^2)}$.

It should be noted that the results that follow are not just valid for the mixing matrix of anomalous dimensions but for the full dilatation operator, including the tree level contribution ($H^{(l)}\to H^{(0)}+g^2H^{(l)}$, $H^{(0)}=L\mathbbm{1}$). This follows from the well-understood effect of shifts and rescaling of the evolving Hamiltonian on dynamics and the Krylov data \cite{ViswanathMuller1994,Muck:2024fpb,sanchezgarrido2024krylovcomplexity}. Namely, under linear transformation of the Hamiltonian (energies)
\begin{equation}
    H \to \alpha H + \delta \mathbbm{1}\,, \qquad \alpha,\delta \in \mathbb{R}\,,
\end{equation}
the Krylov dimension, $\mathcal{K}$, is invariant, while the Lanczos coefficients are modified to
\begin{equation}
    a_n \to \alpha a_n + \delta , \quad b_n \to |\alpha|b_n, \label{eq:LinearTfromOnLanczos}
\end{equation}
and the Krylov/spread complexity changes in a simple manner
\begin{equation}
    C(t) \to C(\alpha t)\,.
\end{equation}

\begin{figure}[b!]
    \centering
    \includegraphics[width=0.49\linewidth]{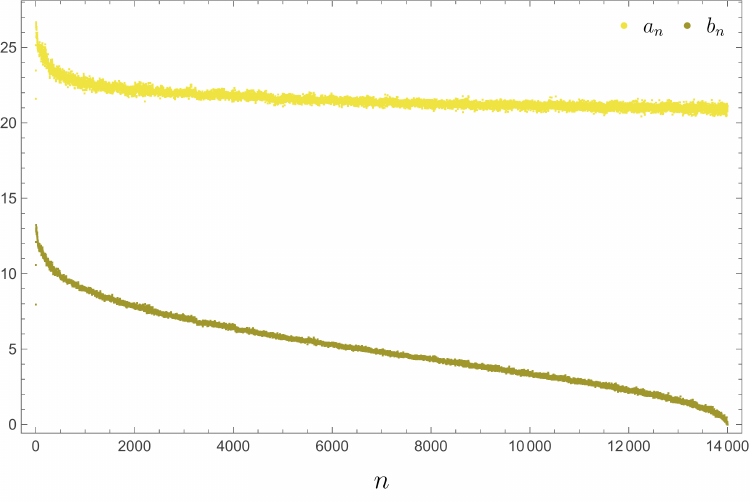}
    \hspace{0.01\linewidth}\includegraphics[width=0.49\linewidth]{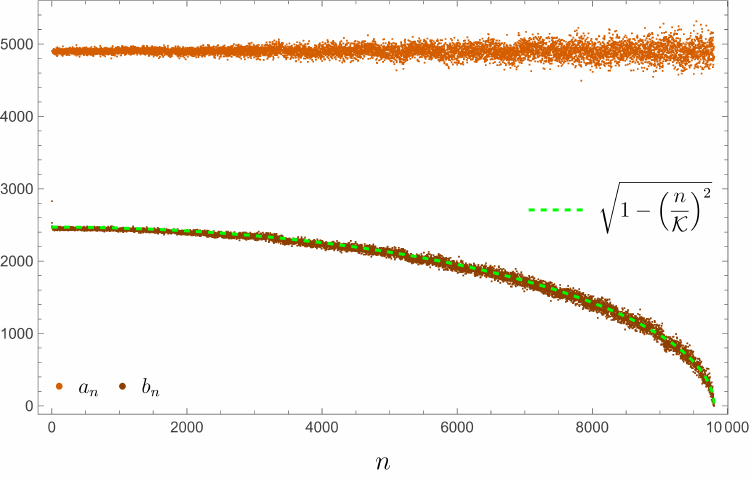}
    \caption{Lanczos coefficients for the one-loop dilatation operator's (left) raw and (right) unfolded spectrum in the $L=21$, $M=9$ sector with an initial state \eqref{InState0}.}
    \label{fig:loop1LanCoef}
\end{figure}

With this setup and known ambiguities in mind, we present the Lanczos coefficients and spread complexity from one- to four-loop truncation sequentially for three representative values of $g^2$. To start, we plot the Lanczos coefficients for the one-loop truncated dilatation operator in Figure \ref{fig:loop1LanCoef}.

For the raw spectrum, plotted on the left, the Lanczos coefficients are similar to those found in other integrable spin chain models such as the Ising model \cite{Camargo_2024}. The diagonal Lanczos coefficients, $a_n$, change significantly at small $n$ but vary slowly at larger $n$. Recalling that $a_n$ is the average energy in the $n^{\text{th}}$ Krylov basis element, the difference in behaviour at small and large $n$ indicates that the extremes of the spectrum are eliminated early on the in Lanczos procedure. There is a very small fraction of the off-diagonal Lanczos coefficients, $b_n$, which are linear \cite{Parker_2019}, however they saturate quickly.

The results for the unfolded spectrum for the same initial state are plotted on the right of Figure \ref{fig:loop1LanCoef}. The diagonal coefficients $a_n$ oscillate around the centre of the spectrum, since the CGS has maximal Krylov dimension ($\mathcal{K}=D$) it is convenient to note that the unfolded spectrum runs from $1$ to $\mathcal{K}$. We find that $b_n\sim\sqrt{1-(\frac{n}{\mathcal{K}})^2}$ (plotted in green), which is structurally similar to the continuum limit of the RMT bulk prediction \cite{Balasubramanian:2022dnj}. In fact, as shown in \cite{Balasubramanian:2022dnj}, for an ensemble of $\mathcal{K}\times \mathcal{K}$ matrices, the ensemble average of the Lanczos coefficients in the large-$\mathcal{K}$ limit, $b(x)$ with $x=n/\mathcal{K}$, can be related to the mean density of states $\rho(E)$. For the semi-circle density of RMT, $\rho=\tfrac{1}{2\pi} \sqrt{4-E^2}$, one has $b(x)=\sqrt{1-x}$ while for a constant $\rho$ one finds the result for the unfolded spectrum $b(x)=\sqrt{1-x^2}$.

In this work, we do not interpret the overall functional dependence of the Lanczos coefficients as a diagnostic of integrability or chaos. Instead, as discussed above, our primary measure will be the degree of disorder in the Lanczos coefficients themselves. As this model is known to be integrable, it is unsurprising that there are significant fluctuations between neighbouring Lanczos coefficients, as reflected in the substantial width of the bands formed by the data points. We will contrast this behaviour with the higher-loop results later in this section and quantify it more systematically in subsequent sections. For now, it is worth noting that, after unfolding, the level of disorder appears to increase with $n$, whereas in the raw data it remains approximately uniform across the Krylov chain. 

\vspace{0.5cm}

\begin{figure}[hbtp]
    \centering
    \includegraphics[width=0.49\linewidth]{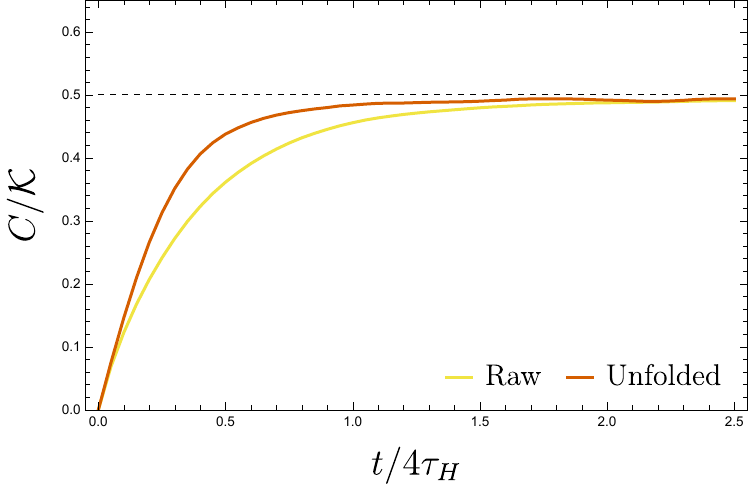}
    \caption{Spread complexity for the one-loop dilatation operator's raw and unfolded spectrum in the $L=21$, $M=9$ sector with an initial state \eqref{InState0}. }
    \label{fig:loop1KCExamples}
\end{figure}

The corresponding spread complexities as functions of time are plotted in Figure\,\ref{fig:loop1KCExamples}. We observe that the Heisenberg time  \eqref{eq:HeisenbergTime} is a good scale for the saturation of the complexity for both the raw and unfolded spectrum, signalling that it shares the same origin as the saturation of the SFF \cite{Erdmenger:2023wjg} ($t_{\text{sat}}=8\tau_H$ \cite{Gharibyan:2018jrp}). The late time plateau of the spread complexity is consistent with the value $\mathcal{K}/2$, implying that the wave function becomes fully delocalised on the Krylov chain at late times. Finally, the effect of unfolding plays a role only at intermediate times, increasing slightly the value of spread complexity. Nevertheless, without the appropriate scales for $C$ and $t$ ($\mathcal{K}$ and $\tau_H$) the discrepancy would be substantial, highlighting the importance of a consistent normalisation when comparing the raw and unfolded setups. 

\vspace{0.5cm}

\begin{figure}[t!]
    \centering
   \includegraphics[width=0.485\linewidth]{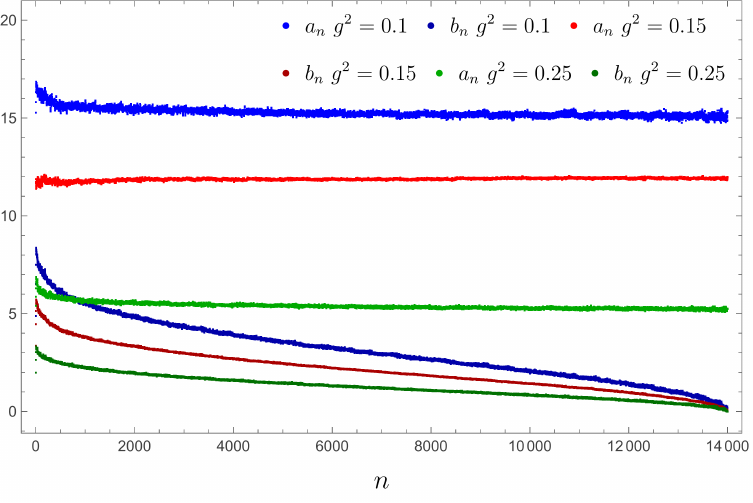}
\hspace{0.01\linewidth}\includegraphics[width=0.49\linewidth]{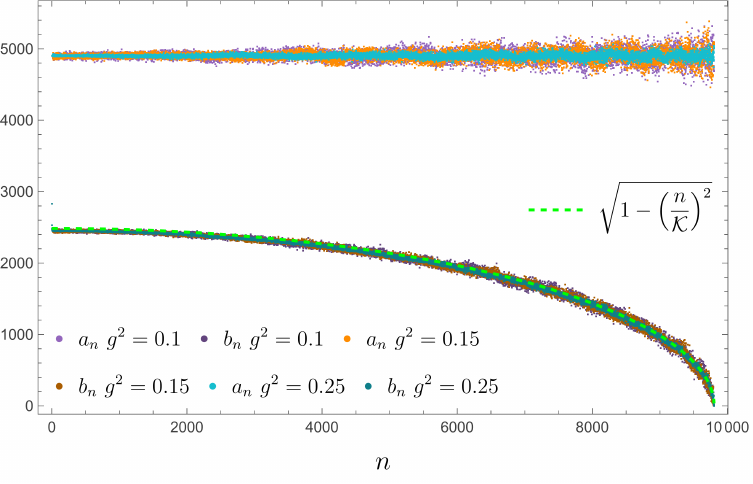}
    \caption{Lanczos coefficients for the two-loop dilatation operator's (left) raw and (right) unfolded spectrum in the $L=21$, $M=9$ sector with an initial state \eqref{InState0}.}
    \label{fig:loop2LanCoef}
\end{figure}

For the two-loop truncation, signatures of chaos can be observed by increasing $g^2$ as shown in previous sections. Here, in Figure\,\ref{fig:loop2LanCoef} we plot the Lanczos coefficients for three representative values of $g^2$. At $g^2=0.1$, the average $r$-ratio \eqref{eq:rratio}, $\langle r \rangle$, is approximately halfway between the predicted integrable and chaotic value, thus is a representative of intermediate behaviour. For $g^2=0.15$, $\langle r \rangle$ matches the chaotic value, and for $g^2=0.25$ it matches the integrable value for this special point as discussed previously. Once again, the raw spectrum Lanczos coefficients have typical functional form from integrable and chaotic spin chains. Furthermore, the unfolded spectrum produces $a_n$ that oscillate around the centre of the spectrum and $b_n$ that are functionally well described by $\sqrt{1-(\frac{n}{\mathcal{K}})^2}$.

\vspace{0.5cm}

For the raw spectrum, variations in the degree of disorder are sufficiently small that they are not readily visible. After unfolding, however, the behaviour at larger $n$ suggests an apparent hierarchy of disorder, with increasingly chaotic spectra exhibiting progressively larger fluctuations in the Lanczos coefficients.

\begin{figure}[h!]
    \centering
    \includegraphics[width=0.49\linewidth]{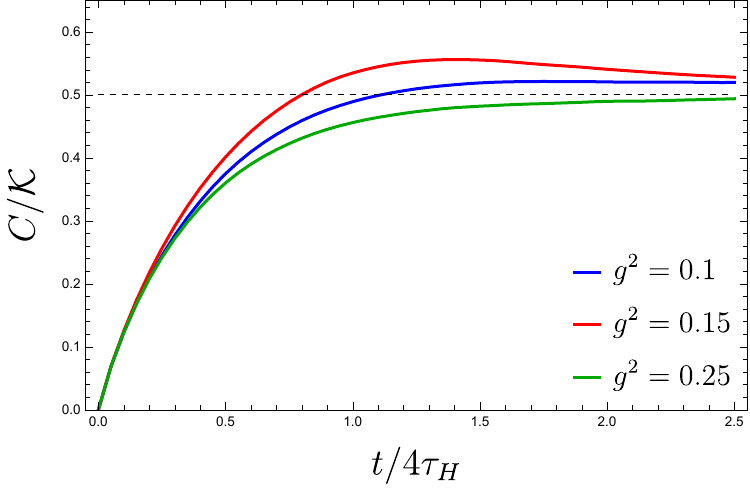}\hspace{0.01\linewidth}
    \includegraphics[width=0.49\linewidth]{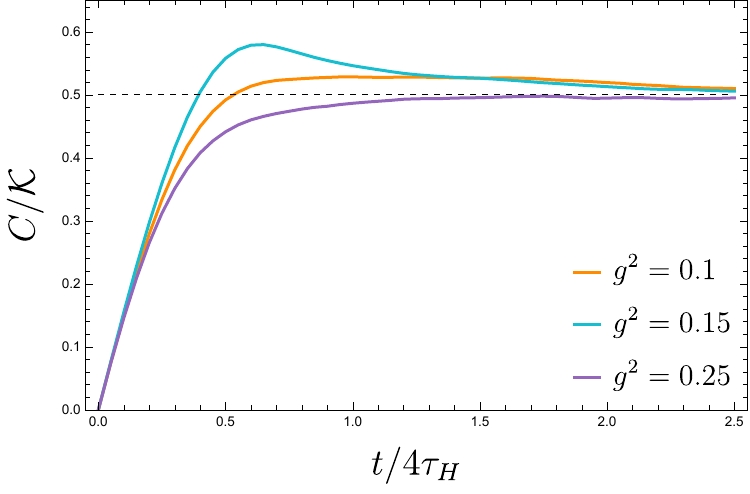}
    \caption{Spread complexity for the two-loop dilatation operator's (left) raw and (right) unfolded spectrum in the $L=21$, $M=9$ sector with an initial state \eqref{InState0}.}
    \label{fig:loop2KCExamples}
\end{figure}

The associated spread complexities are plotted in  Figure\,\ref{fig:loop2KCExamples} for the raw and unfolded spectra, which both develop a peak as $g^2$ moves towards chaotic values. After unfolding, this peak is enhanced in maximum magnitude and shortened in duration\footnote{For the raw spectrum, the time plotted is not long enough to show full relaxation to the saturation value of $\mathcal{K}/2$, in addition this gives the peak a more stretched appearance.}. This is akin to other eigenvalue chaos measures where unfolding enhances the agreement with the Wigner surmise or RMT predicted SFF. It is important to note again that the unfolding procedure does not lead to a development of a peak for the integrable case of $g^2=0.25$. 

\begin{figure}[hbtp]
    \centering
    \includegraphics[width=0.485\linewidth]{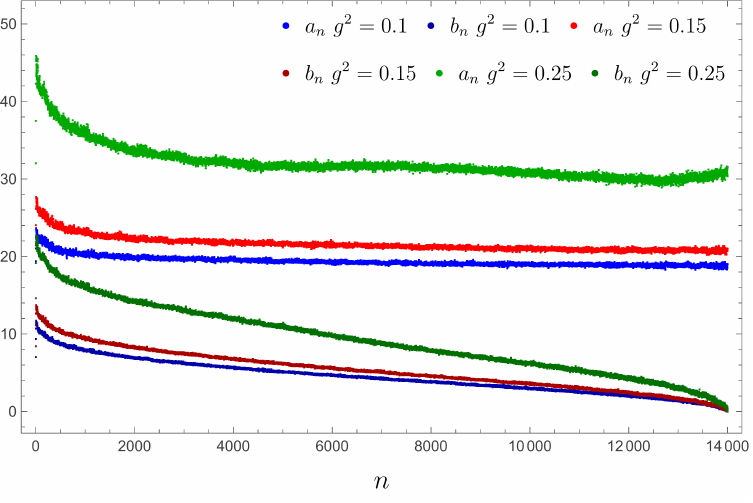}
    \hspace{0.01\linewidth}\includegraphics[width=0.49\linewidth]{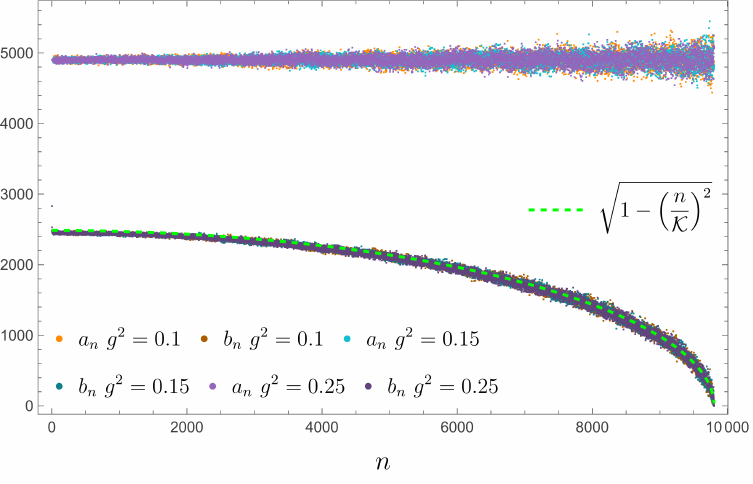}\\
    \caption{Lanczos coefficients for the three-loop dilatation operator's (left) raw and (right) unfolded spectrum in the $L=21$, $M=9$ sector with an initial state \eqref{InState0}.}
    \label{fig:loop3LanCoef}
\end{figure}

\begin{figure}[hbtp]
    \centering
    \includegraphics[width=0.49\linewidth]{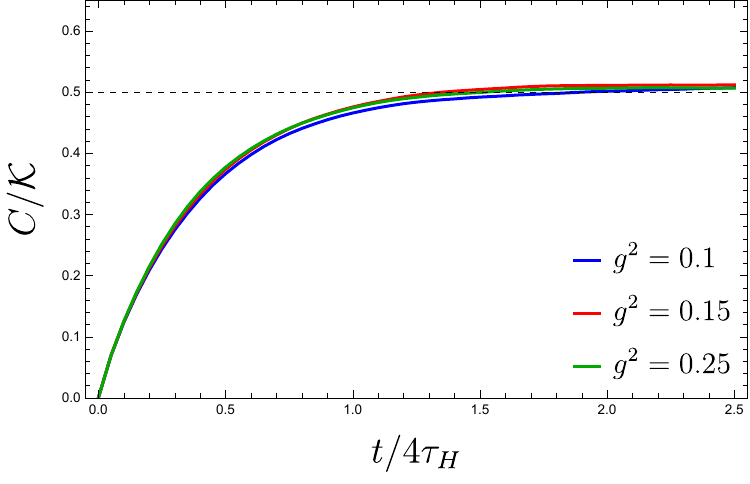}
    \hspace{0.01\linewidth}\includegraphics[width=0.49\linewidth]{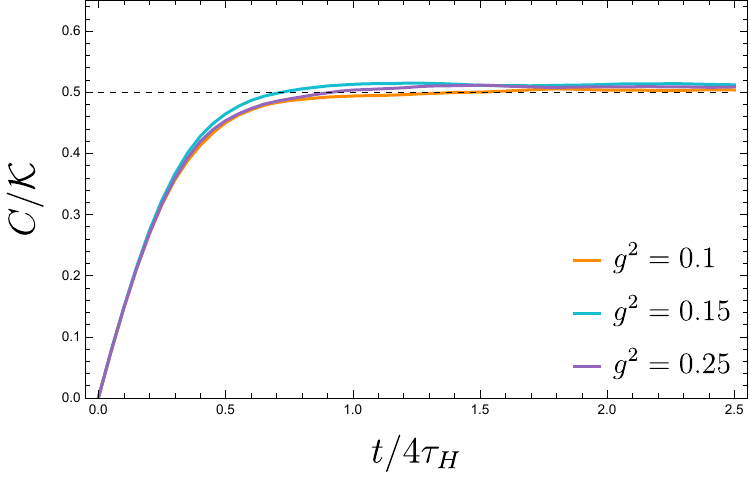}
    \caption{Spread complexity for the three-loop dilatation operator's (left) raw and (right) unfolded spectrum in the $L=21$, $M=9$ sector with an initial state \eqref{InState0}.}
    \label{fig:loop3KCExamples}
\end{figure}

For the three loop truncation, we previously observed that there was almost no change in the $r$ statistics across $g^2$. This is again reflected in both, the Lanczos coefficients remaining similar, in  Figure\,\ref{fig:loop3LanCoef}, and the spread complexity's lack of any peak in  Figure\,\ref{fig:loop3KCExamples}.

\begin{figure}[hbtp]
    \centering
    \includegraphics[width=0.485\linewidth]{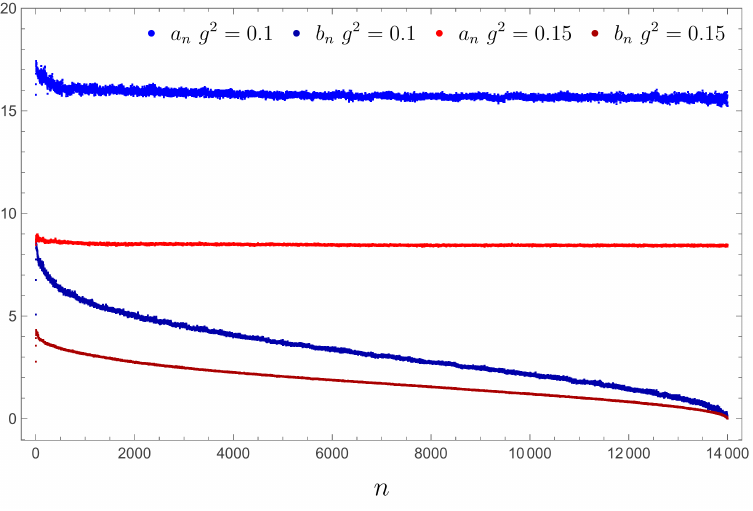} 
    \hspace{0.01\linewidth}\includegraphics[width=0.49\linewidth]{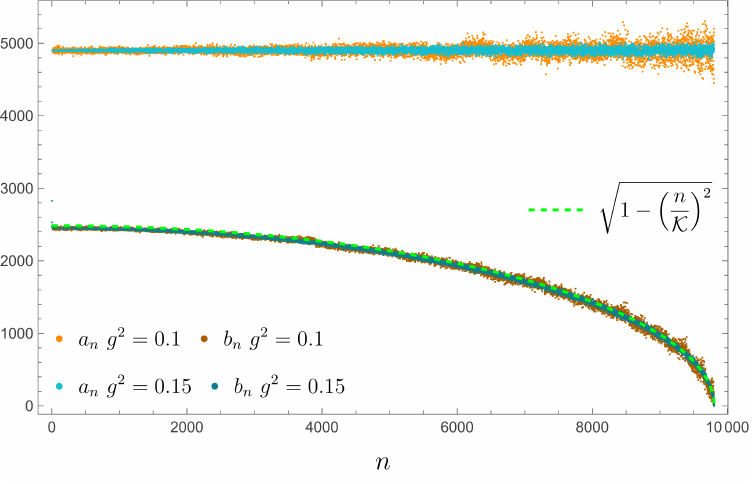}
    \caption{Lanczos coefficients for the four-loop dilatation operator's (left) raw and (right) unfolded spectrum in the $L=21$, $M=9$ sector with an initial state \eqref{InState0}.}
    \label{fig:loop4LanCoefF}
\end{figure}

\begin{figure}[hbtp]
    \centering
    \includegraphics[width=0.49\linewidth]{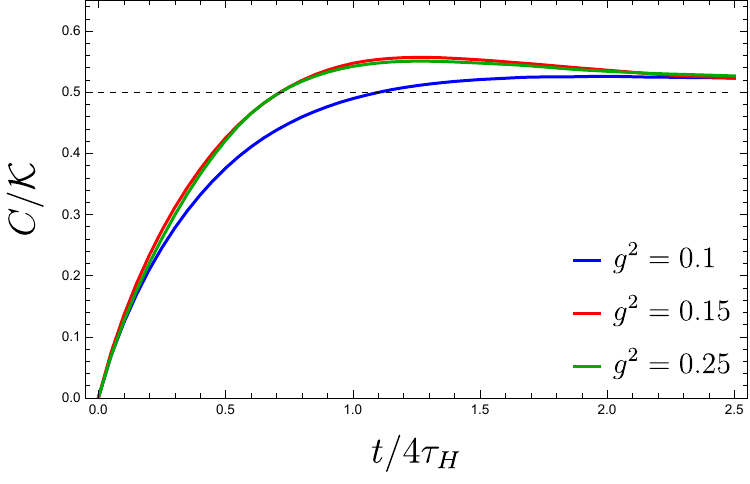}
    \hspace{0.01\linewidth}\includegraphics[width=0.49\linewidth]{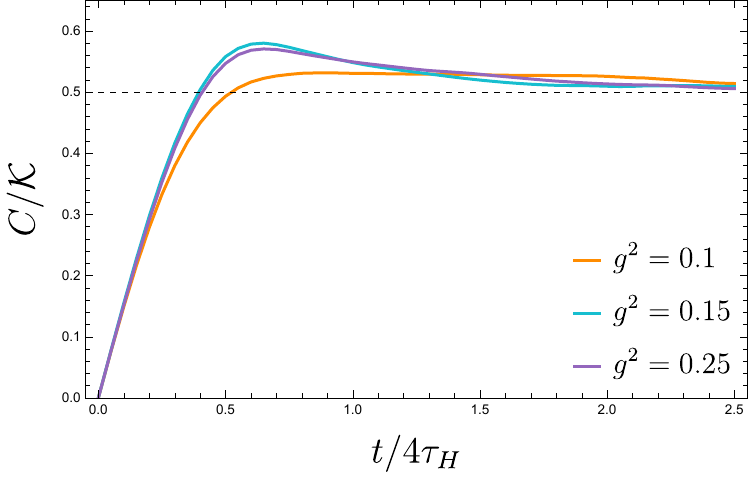}
    \caption{Spread complexity for the four-loop dilatation operator's (left) raw and (right) unfolded spectrum in the $L=21$, $M=9$ sector with an initial state \eqref{InState0}.}
    \label{fig:loop4KCExamples}
\end{figure}

Finally, in  Figure\,\ref{fig:loop4LanCoefF}, we plot the Lanczos coefficients at four loop truncation for $g^2=0.1$ and $0.15$. At four loops, $g^2=0.1$ is still in between a chaotic and integrable model, while at $g^2=0.15$ it is fully chaotic. Observe, again at large $n$, that the disorder is smaller for the more chaotic model. In Figure\,\ref{fig:loop4KCExamples} we restore the $g^2=0.25$ case and find similar results to the two loop case, however, now $0.25$ is no longer an integrable point and as such develops a peak in the spread complexity. 

\subsection{$r$-ratio and the peak of spread complexity}\label{sec:KrylovPeak}
We now compare the Krylov-space signatures of the integrable-chaotic crossover with the average $r$-ratio, $\langle r \rangle$, as discussed in section~\ref{sec:r-ratio}. As mentioned previously, the diagnostic will be the presence of a peak in the spread complexity. More explicitly, we define the late time plateau by $\overline{C}$, and the peak of spread complexity as 
\vspace{0.05cm}
 \begin{equation}\label{eq:PeakofKrylov}
   C_{peak} := \text{max} [ C(t) ] - \overline{C} \,,
\end{equation}
where $\overline{C}$ is also the saturation/late time value estimated in \eqref{eq:SaturationOfKrylov}. The saturation time of approximately $8\tau_H$ (where the Heisenberg time is defined in \eqref{eq:HeisenbergTime}), is estimated by the RMT SFF saturation time \cite{Gharibyan:2018jrp} as well as numerical evidence in the previous sections. By definition, the peak should occur before the saturation. Moreover, in cases where there is no chaos, small, sporadic/numerical ``peaks" may occur after the saturation time but should be disregarded.

\vspace{0.5cm}

\begin{figure}[hbtp]
    \centering
    \includegraphics[width=0.475\linewidth]{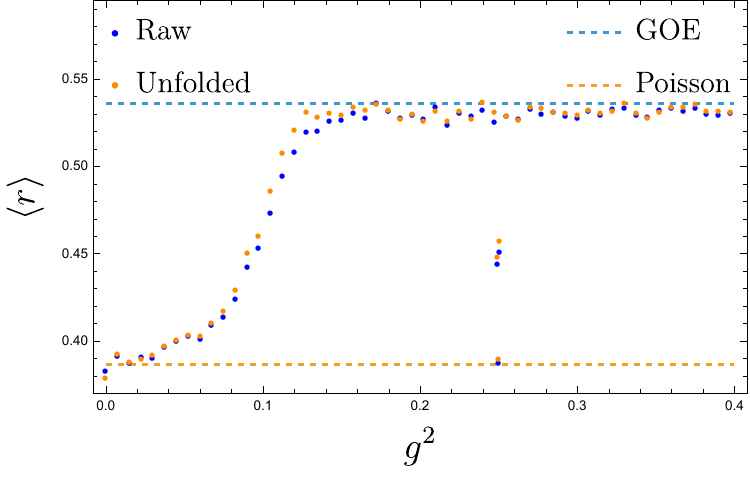} \hspace{0.01\linewidth}
    \includegraphics[width=0.475\linewidth]{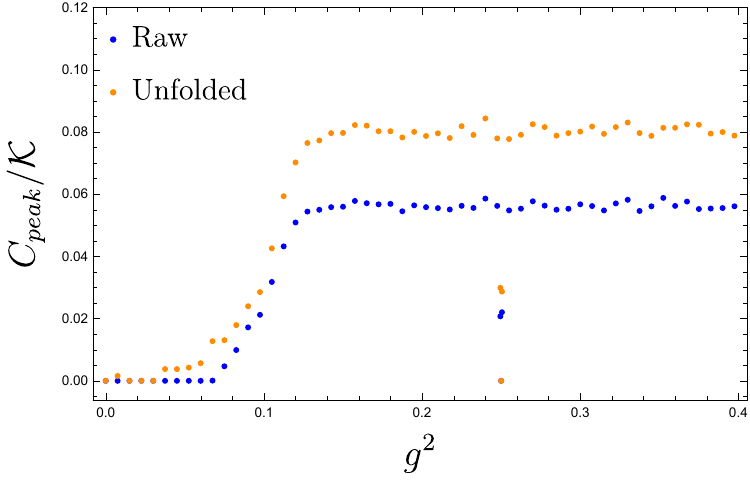}\\
    \includegraphics[width=0.475\linewidth]{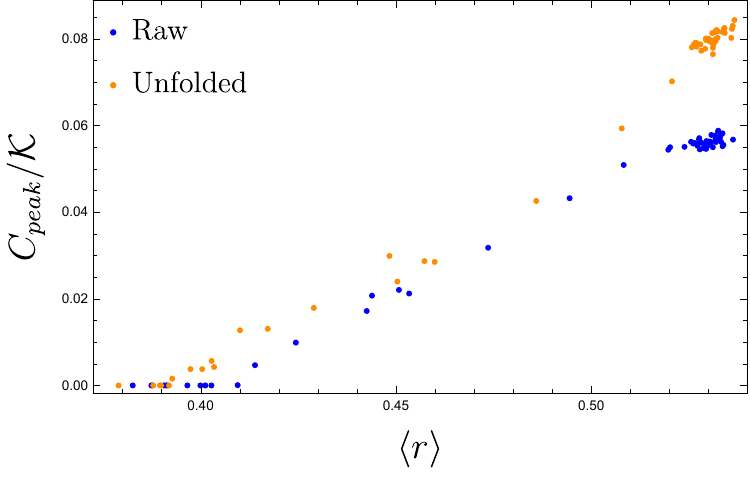}
    \caption{(Top left) $\langle r \rangle$ and (top right) peak of spread complexity \eqref{eq:PeakofKrylov} for the raw and unfolded spectrum in the $L=21$ and $M=9$ sector with $\beta=1/\sqrt{26}$ at $2$-loop truncation as a function of $g^2$ (steps of $0.0075$). (Bottom) Peak of spread complexity vs $\langle r \rangle$.}
    \label{fig:PeakvsrPlot1}
\end{figure}

In Figure\,\ref{fig:PeakvsrPlot1} (top left) we plot $\langle r \rangle$ for the raw and unfolded spectrum for the two-loop Hamiltonian in the $L=21$, $M=9$ sector, demonstrating that it is unaffected by unfolding, and compare it to the corresponding peak values. We confirm our prior observations from the example plots that the unfolding procedure enhances the height of the peak (top right) and that the peak follows the same functional form as $\langle r \rangle$ in $g^2$. To further illustrate this, we plot the peak values as a function of $\langle r \rangle$ (bottom) and find approximately linear dependence. However, for the raw case the higher values of $r$ deviate from this linear behaviour, which is restored by unfolding.

\vspace{0.5cm}

\begin{figure}[h]
    \centering
    \includegraphics[width=0.475\linewidth]{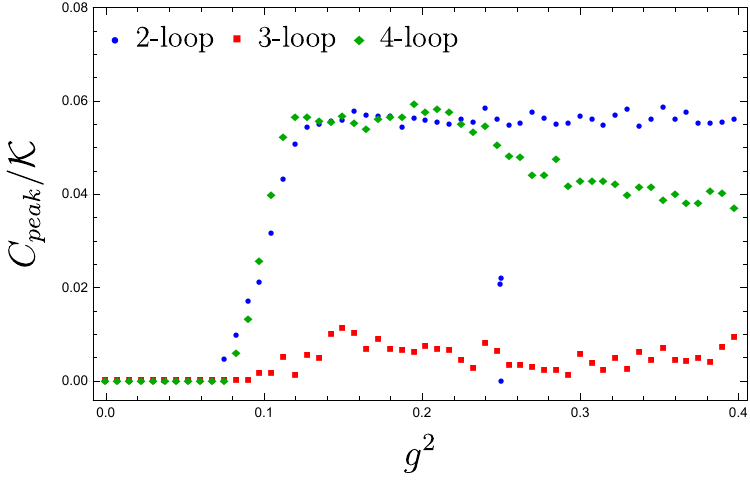}\hspace{0.01\linewidth}\includegraphics[width=0.475\linewidth]{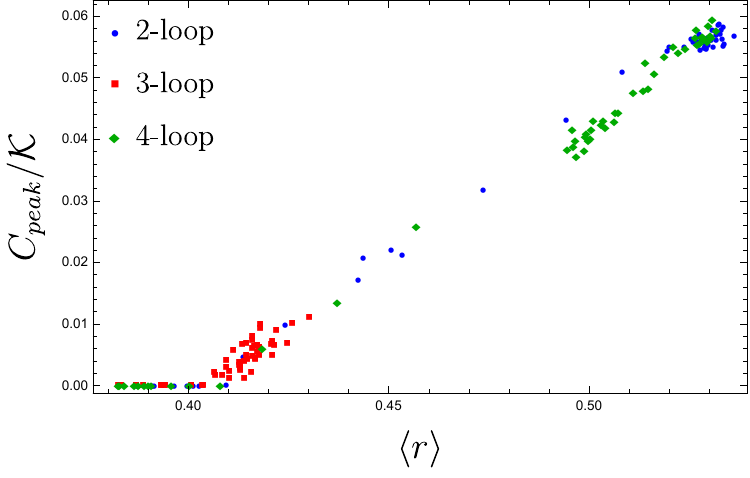}
    \caption{The peak of spread complexity \eqref{eq:PeakofKrylov} as a function of (left) $g^2$ and (right) $\langle r \rangle$ for $L=21$, $M=9$ for $\beta=1/\sqrt{26}$, for Hamiltonian truncations from two to four loops.}
    \label{fig:PeakvsrPlotAllLoops}
\end{figure}

To close out this section we present the peak values of spread complexity, for loop truncations two to four, as a function of $g^2$ and $\langle r \rangle$ in  Figure\,\ref{fig:PeakvsrPlotAllLoops}. We again observe a linear dependence with some deviations at large values of $\langle r \rangle$ due to the folded nature of the spectrum. 

%

\subsection{Eigenvalue chaos through the variance in Lanczos coefficients}\label{sec:KrylovDisorder}

\begin{figure}[h!]
    \centering
    \includegraphics[width=0.475\linewidth]{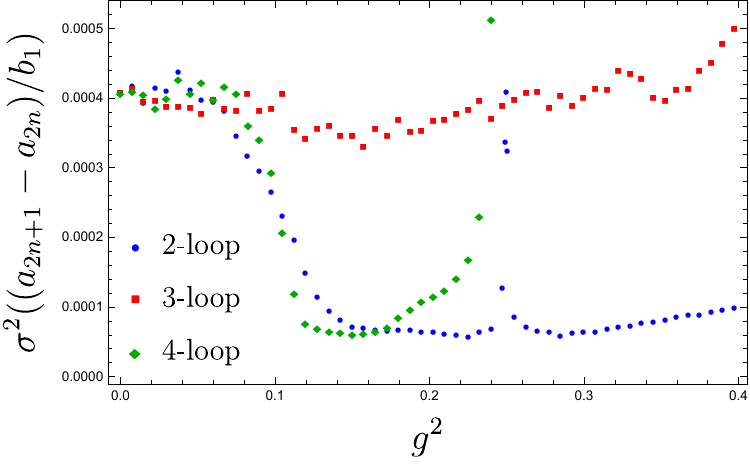}\hspace{0.01\linewidth}\includegraphics[width=0.475\linewidth]{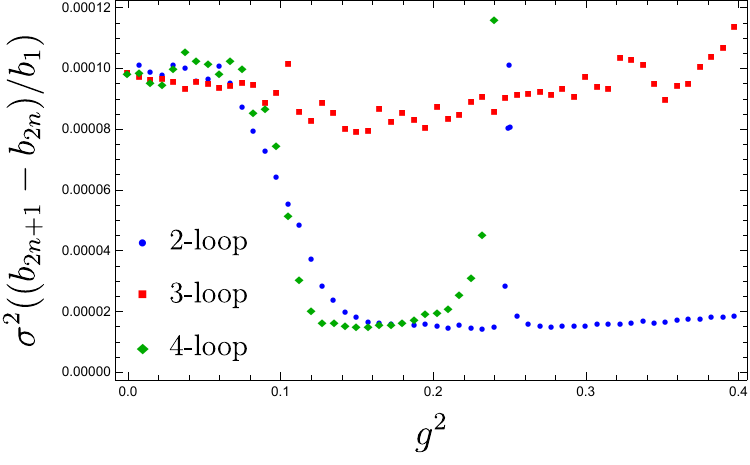}
    \caption{Variance of (left) $(a_{2n+1}-a_{2n})/b_1$ and (right) $(b_{2n+1}-b_{2n})/b_1$ for the raw spectrum of $L=21$ $M=9$ sector with $\beta=1/\sqrt{26}$ as a function of $g^2$.}
    \label{fig:RawDiffVariance}
\end{figure}

Phenomenologically, the absence of the peak is caused by an underlying Anderson localisation-like effect \cite{Rabinovici_2022} that occurs on the Krylov chain. The localisation arises from the local disorder in the hopping (Lanczos) coefficients, the local variance can quantify this effect. With the ambiguities of the Lanczos coefficients under linear transformations of the spectrum \eqref{eq:LinearTfromOnLanczos} we must form some invariant quantities. We choose the following example from \cite{Balasubramanian:2024ghv}
\begin{equation}
    \sigma^2 \left( \frac{a_{2n+1}-a_{2n}}{b_1} \right)\,, \quad  \quad \sigma^2 \left( \frac{b_{2n+1}-b_{2n}}{b_1} \right)\,, \label{eq:RathVar} 
\end{equation}
where the variance is $\sigma^2(x)\equiv \langle (x-\langle x\rangle)^2\rangle$. Moreover, taking the difference of adjacent coefficients removes the shift but not the scaling symmetry, as such we propose to set the scale by dividing by $b_1$. Other choices also exist in the literature, in particular the logarithm of the ratio of successive coefficients has been studied in \cite{Rabinovici_2022}. However, after closer examination, \eqref{eq:RathVar} seems to produce the most stable results. Figure \ref{fig:RawDiffVariance} shows our results for this metric, noting that we have removed $g^2$ values beyond $0.25$ at four-loops, where the spectrum becomes problematic. 

\begin{figure}[h!]
    \centering
    \includegraphics[width=0.475\linewidth]{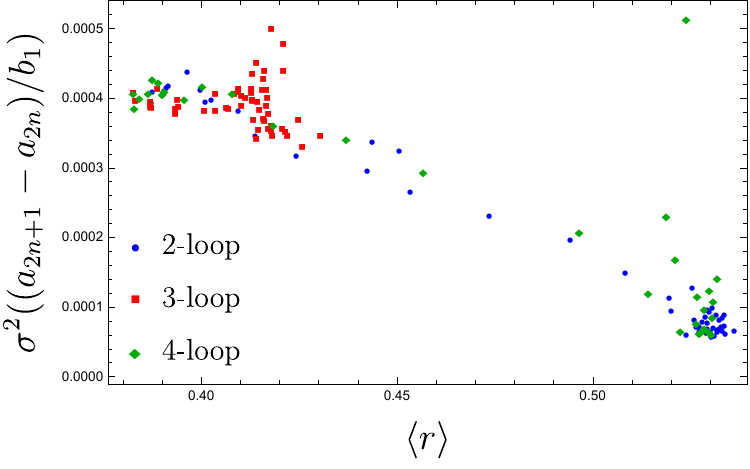}\hspace{0.01\linewidth}\includegraphics[width=0.475\linewidth]{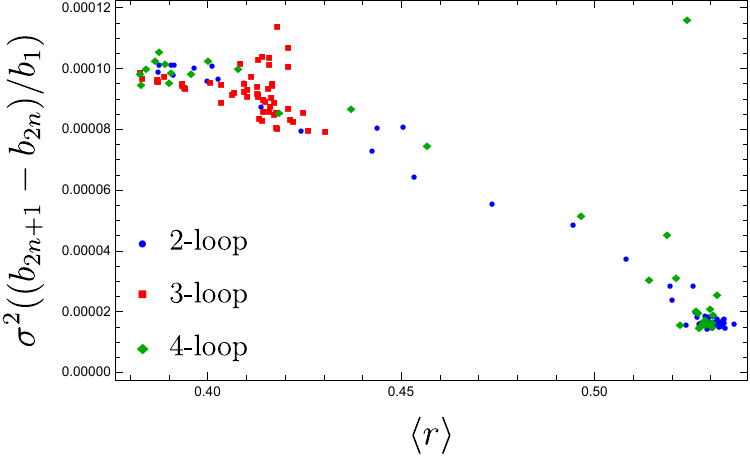}
    \caption{Variance of (left) $(a_{2n+1}-a_{2n})/b_1$ and (right) $(b_{2n+1}-b_{2n})/b_1$ compared to the $r$ ratio at equal $g^2$. }
    \label{fig:RawDiffVariancevsr}
\end{figure}

It can be seen that the variance follows a similar pattern to $\langle r \rangle$, here less variance is related to chaos and as such the trend is inverted. To demonstrate the agreement, in Figure\, \ref{fig:RawDiffVariancevsr} we plot the disorders with their corresponding $\langle r \rangle$ at each value of $g^2$ and find linear agreement for the bulk values.

\begin{figure}[h!]
    \centering
    \includegraphics[width=0.475\linewidth]{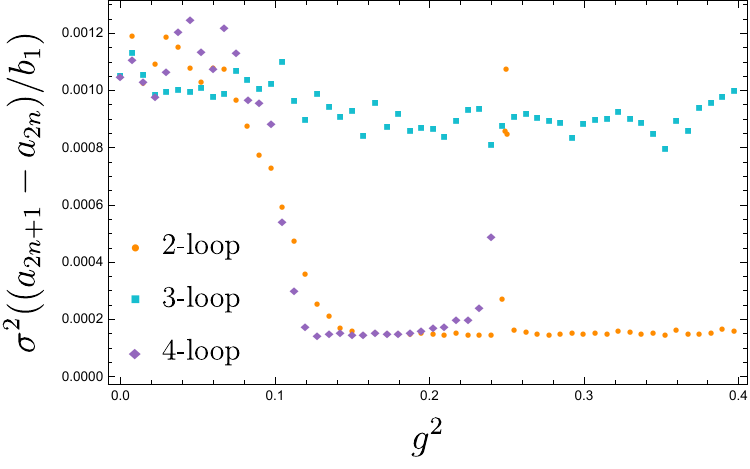}\hspace{0.01\linewidth}\includegraphics[width=0.475\linewidth]{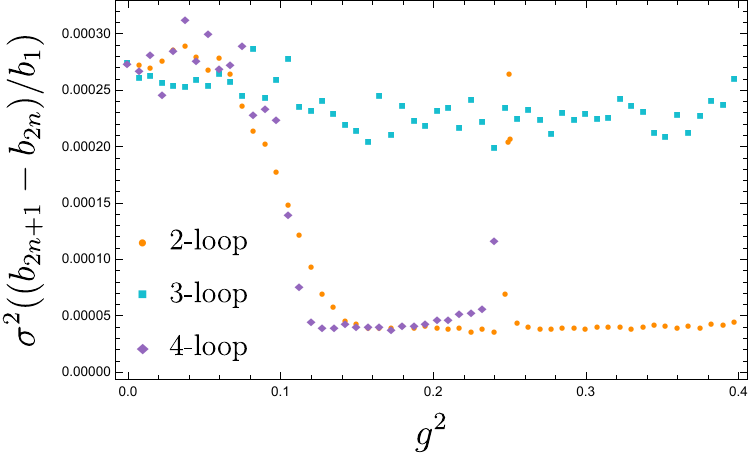}
    \caption{Variance of (left) $(a_{2n+1}-a_{2n})/b_1$ and (right) $(b_{2n+1}-b_{2n})/b_1$ for the unfolded spectrum of $L=21$ $M=9$ with $\beta=1/\sqrt{26}$ as a function of $g^2$.}
    \label{fig:UnfDiffVariance}
\end{figure}

\begin{figure}[h!]
    \centering
    \includegraphics[width=0.475\linewidth]{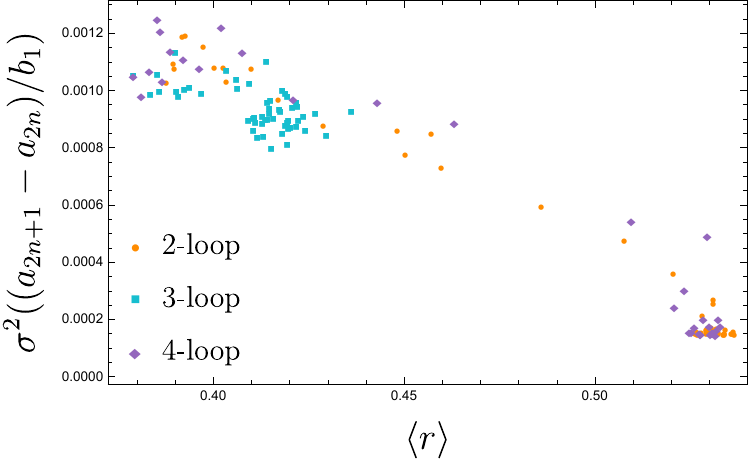}\hspace{0.01\linewidth}\includegraphics[width=0.475\linewidth]{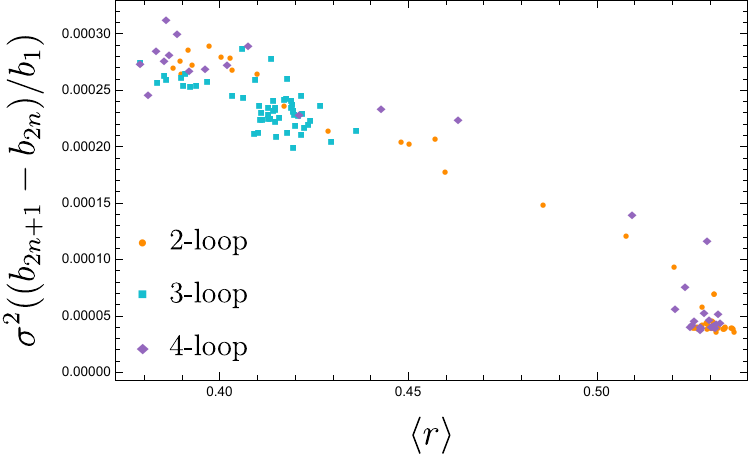}
    \caption{Variance of (left) $(a_{2n+1}-a_{2n})/b_1$ and (right) $(b_{2n+1}-b_{2n})/b_1$ compared to the averaged $r$ ratio at equal $g^2$.}
    \label{fig:UnfDiffVariancevsr}
\end{figure}

Repeating the analysis using the unfolded spectrum we plot the same measures for the same choice of parameters in Figure \ref{fig:UnfDiffVariance}. There is a slight improvement in the results, notably in the high $g^2$ behaviour across the different loop truncations. This is further demonstrated by Figure \ref{fig:UnfDiffVariancevsr}, where the disorder measures are plotted against their corresponding $\langle r \rangle$ values at equal $g^2$ again. Observe that the three loop truncation is lower compared to the raw spectrum results, which aligns with the other loop truncations for low $\langle r \rangle$.

\subsection{Eigenvector chaos through initial state dependence}\label{sec:VecandKrylov}

In this section, we discuss how changing the initial state can be used to capture eigenvector chaos. Previously, in section \ref{sec:eigenvecs}, we studied the decomposition of eigenstates of the two- to four-loop truncated dilatation operator at various values of $g^2$ into two reference bases; the spin product basis (gauge invariant words) $\{ \ket{s_a}\}$ \eqref{eq:SpinProdBasisDef} and the one-loop truncation eigenbasis $\{ \ket{E_a^{(1)}} \}$ \eqref{eq:OneLoopBasis}. For Krylov and spread complexity, as outlined several times in the literature \cite{Espa_ol_2023,PhysRevE.109.054209,Baggioli_2025,PG_2025}, the initial state's decomposition in the eigenbasis of the generator of the dynamics greatly affects the complexity and associated features; namely the saturation value, $\overline{C}$, and sometimes the peak, $C_{peak}$ \cite{Alishahiha_2025}. 

To connect with the results of section \ref{sec:eigenvecs}, we can choose the initial state to be a basis element from one of the two above-mentioned bases, and compute the Krylov data using the two- to four-loop truncations of the dilatation operator as the generator of dynamics. This way, the Krylov data will depend on the coefficients of the chosen basis element in the eigenbasis of the truncated dilatation operator, $c_\alpha^{(a)}$ in the notation of \eqref{eq:EigsinOneLoopBasis}. We will reuse $D_2$ \eqref{eq:Dq} to quantify this decomposition. When averaged over all basis elements\footnote{This procedure is analogous to \cite{Craps_2025} and we observe that the choice of basis to average over can be correlated with features of the eigenvector chaos.}, we expect that the Krylov data would capture some of the features of results in section \ref{sec:eigenvecs}. For example, the average IPR$_2$ of the eigenstates in the reference basis is equivalent to the average IPR$_2$ of the reference basis states in the eigenbasis (suppressing the $g^2$ dependence): 
\begin{equation}\label{eq:EquivofAvg}
    \frac{1}{D}\sum_{a=1}^D \text{IPR}_2(\{\bra{s_b}\}, \ket{E_a^{(l)}}) = \frac{1}{D} \sum_{a,b=1}^D |\bra{E^{(l)}_a} s_b \rangle|^4 = \frac{1}{D} \sum_{b=1}^D \text{IPR}_2(\{\bra{E_a^{(l)}}\}, \ket{s_b} ) \,.
\end{equation}

However, repeating this for the Krylov data by computing for all initial basis states and then taking an average would be too computationally expensive. One might hope that taking the average state before computing the Krylov data would be equivalent. This is not the case, as can be seen e.g. for an average over one-loop truncated eigenstates (the infinite temperature CGS defined in \eqref{eq:DefCGS}) and considering the IPR$_2$ (again we suppress the $g^2$ dependence, which need only be restored for the higher loop truncation):
\begin{eqnarray} 
    \text{IPR}_2(\{\bra{E_b^{(l)}}\},\ket{CGS^{(1)}(\infty)}) &=& \sum_{b=1}^D |\bra{E_b^{(l)}} CGS^{(1)}(\infty) \rangle |^4 = \frac{1}{D^2} \sum_{b=1}^D \bigg| \sum_{a=1}^D \bra{E^{(l)}_b} E_a^{(1)} \rangle  \bigg|^4\nonumber \\
    &\neq& \frac{1}{D} \sum_{a=1}^D \text{IPR}_2(\{\bra{E_b^{(l)}}\},\ket{ E^{(1)}_a }).
\end{eqnarray} 
Similarly, we do not expect that the Krylov data for the infinite temperature CGS state to correspond to an average over initial basis states and so will not directly match with the results of sections \ref{sec:eigenv_spin_basis} and \ref{sec:eigenv_intb_basis}. Nonetheless, it still provides a representative of an important class of initial states and we will see how it affects the spread complexity, in particular the saturation $\overline{C}$.
\subsubsection{Spin product basis}
\begin{figure}[hbtp]
    \centering
    \includegraphics[width=0.49\linewidth]{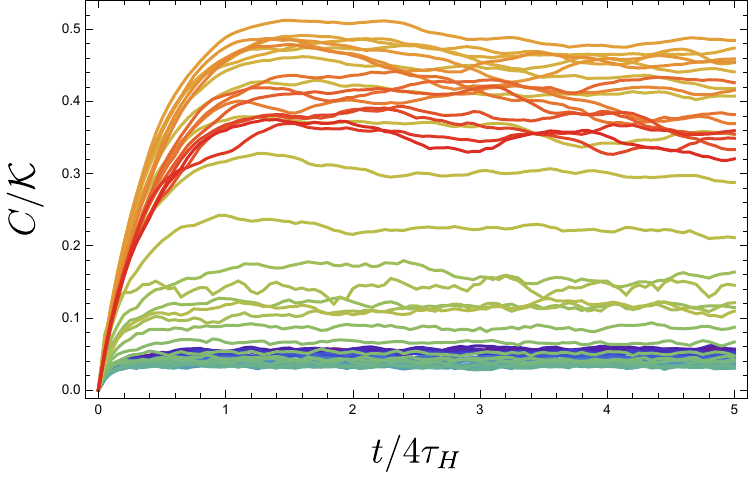}\hspace{0.01\linewidth}\includegraphics[width=0.49\linewidth]{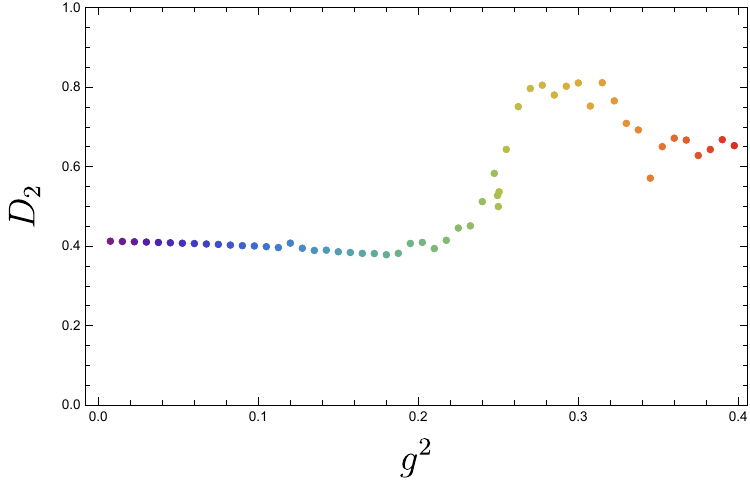}\\
    \includegraphics[width=0.49\linewidth]{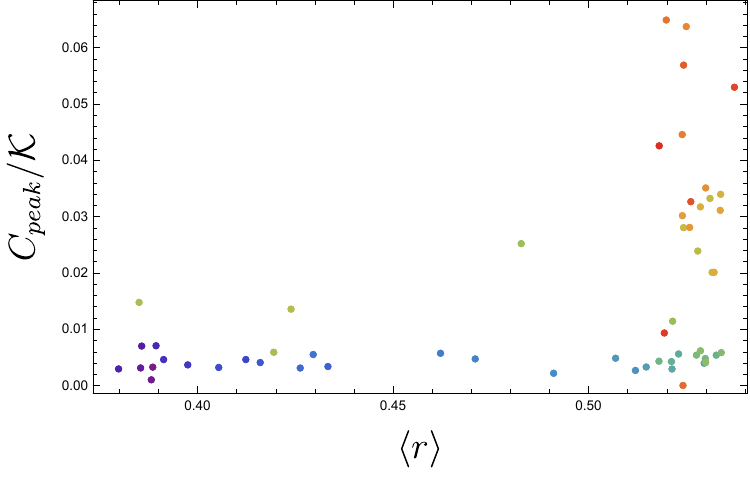}\hspace{0.01\linewidth}\includegraphics[width=0.49\linewidth]{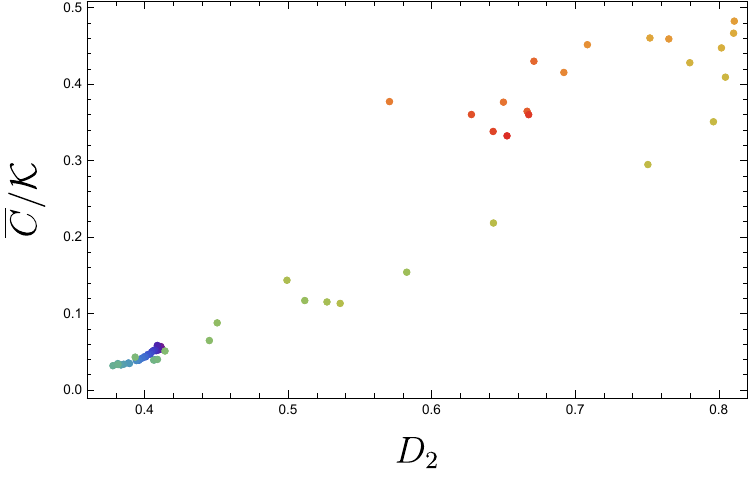}
    \caption{The (top left) spread complexity, (top right) $D_2$ in the energy eigenbasis, (bottom left) the peak of spread complexity and corresponding $\langle r \rangle$, (bottom right) saturation of spread complexity and $D_2$ for the domain wall state Tr$[X^MZ^{L-M}]$ at two loop truncation (raw spectrum) for $g^2\in (0,0.4]$.}
    \label{fig:SP1Krylov}
\end{figure}

To begin, we take a simple example of a state from the spin product basis, namely a ``domain wall"-like state; Tr$[X^M Z^{L-M}]$ in the language of gauge invariant words. In Figure \ref{fig:SP1Krylov}, we plot the results for the domain wall state in the $L=18$, $M=8$ sector at two loop truncation for all values of $g^2$ previously tested. Each colour corresponds to an individual value of $g^2$, see the top right panel for $D_2$ as a function of $g^2$ for example. The top left panel contains all the spread complexities, and we see significant changes in the saturation value as a function of $g^2$. Notably, some of the individual values seem to even show a slight peak. This change in the saturation value can be seen as a consequence of the profile of the domain wall states' $D_2$ in the eigenbasis as shown in the upper right plot. The state becomes delocalised after the integrable point, recall that $D_2\in[0,1]$.

In the bottom set of panels we analyse the eigenvalue and eigenvector chaos measures in the Krylov framework, $C_{peak}$ and $\overline{C}$, with their counterparts $\langle r \rangle$ and $D_2$. In the right panel, the saturation value and $D_2$ seem linearly correlated with quite a large variance\footnote{One might try to account for this with higher $D_q$ or more coarse-grained information}. The impact of the initial state on the peak is expected but still surprising, we see that for high $D_2$ states (orange and red) the peak value is close to that of the CGS tests in Figure \ref{fig:PeakvsrPlotAllLoops}, however, for lower $D_2$ the linear trend is eliminated.

\begin{figure}[hbtp]
    \centering
    \includegraphics[width=0.475\linewidth]{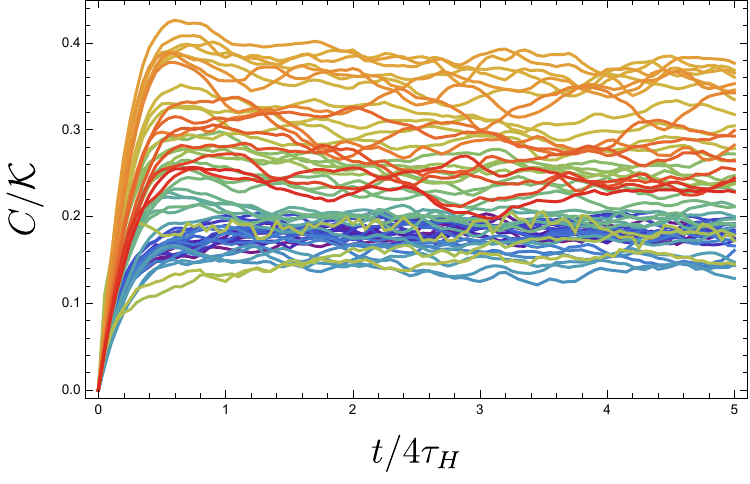}\hspace{0.01\linewidth}\includegraphics[width=0.475\linewidth]{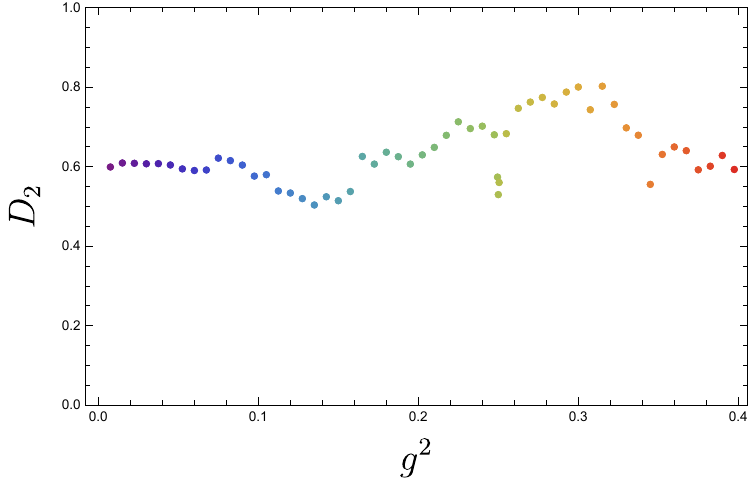}\\
    \includegraphics[width=0.475\linewidth]{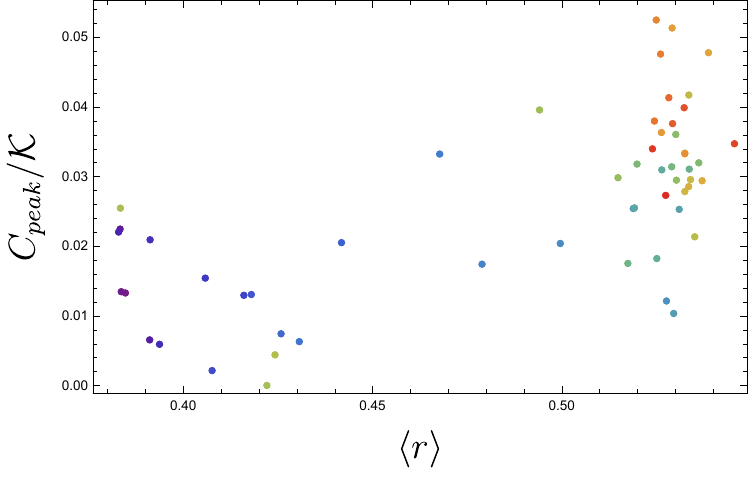}\hspace{0.01\linewidth}\includegraphics[width=0.475\linewidth]{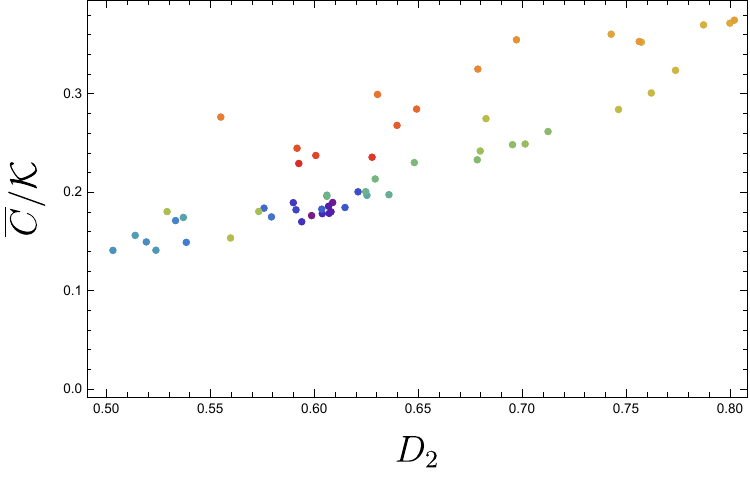}
    \caption{The (top left) spread complexity, (top right) $D_2$ in the energy eigenbasis, (bottom left) the peak of spread complexity and corresponding $\langle r \rangle$, (bottom right) saturation of spread complexity and $D_2$ for the ``domain wall" state Tr$[X^MZ^{L-M}]$ at two loop truncation (unfolded spectrum) for $g^2\in (0,0.4]$.} 
    \label{fig:SP1KrylovUnf}
\end{figure}

To demonstrate that unfolding can also help identify chaos in this case, the analysis is repeated in Figure \ref{fig:SP1KrylovUnf} using the unfolded and clipped spectrum. For most values of $g^2$ the clipping of the spectrum increases the $D_2$ of the domain wall state, indicating that the overlaps of the state with the extremes of the spectrum were minor. This may have been expected as the extremes of the spectrum are not as ergodic in the spin product basis, see Figure \ref{fig:scatter_st_dev_D2_L18}. Correspondingly, the saturation values of the spread complexity increase and there is still a clear relationship with $D_2$. As expected the peaks become more defined and the approximate linear trend emerges, but still has quite large variance. The emergence of some peaks for low values of $\langle r \rangle$ is most likely due to very erratic random fluctuations and the predicted saturation time being inaccurate for these cases.

\subsubsection{One-loop basis}
To compare with section \ref{sec:eigenv_intb_basis} the infinite temperature CGS at one-loop \eqref{InState0} is taken to be the initial state and we present the results for the time evolution with the two loop truncated dilatation operator at various $g^2$. This is the example of averaging the initial state that we discussed at the beginning of this section, as the infinite temperature CGS is exactly the average over all one-loop eigenstates. As we will see, the average state does not fully reproduce the results of section \ref{sec:eigenv_intb_basis}, but it may still serve as a good indication for how the features of the initial state are correlated with eigenvector chaos.

\begin{figure}[h]
    \centering
    \includegraphics[width=0.49\linewidth]{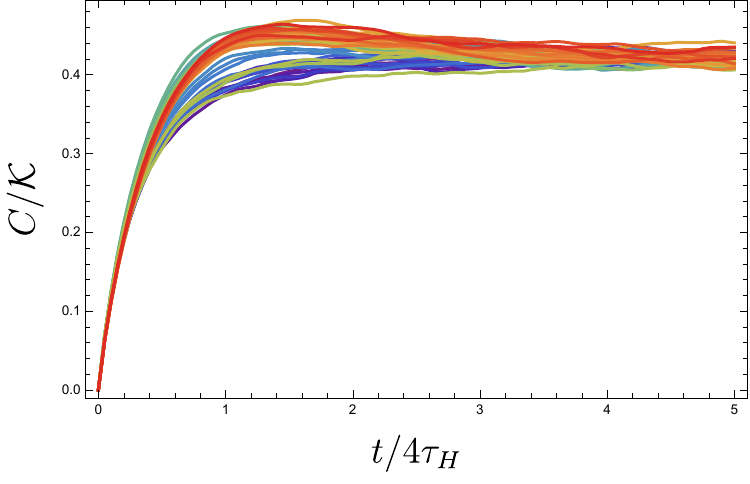}\hspace{0.01\linewidth}\includegraphics[width=0.49\linewidth]{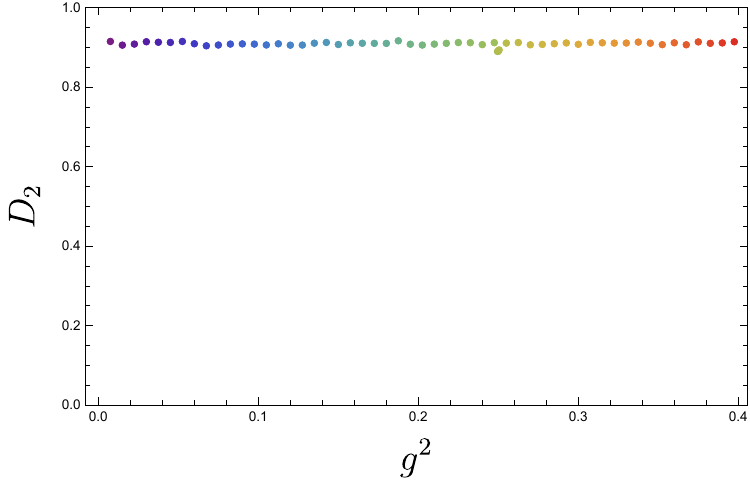}\\
    \includegraphics[width=0.49\linewidth]{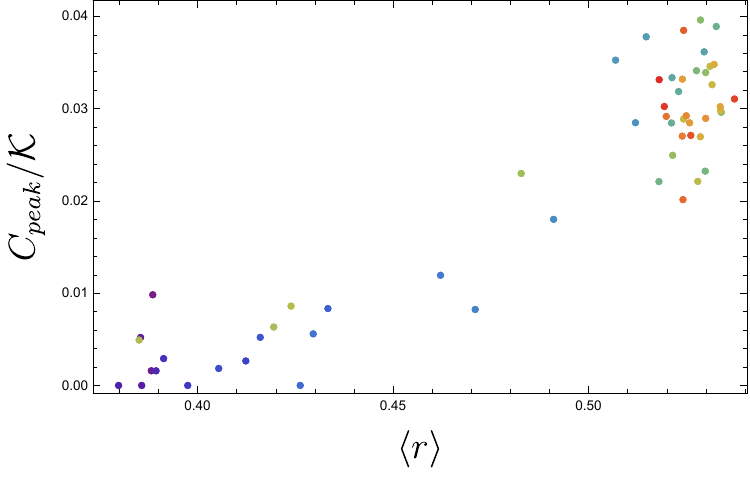}\hspace{0.01\linewidth}\includegraphics[width=0.49\linewidth]{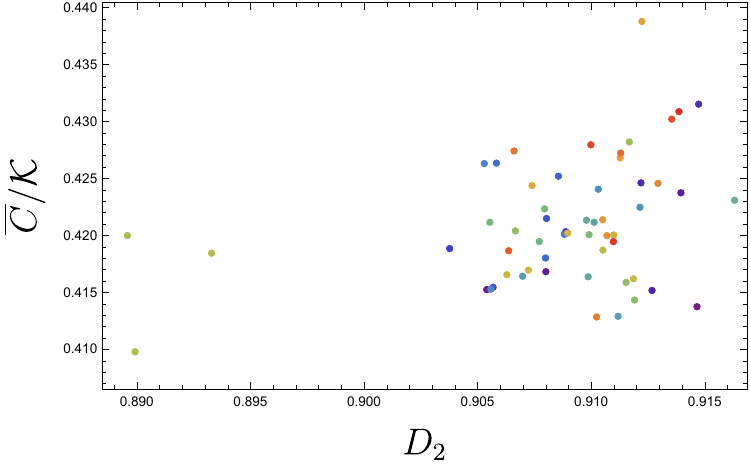}
    \caption{The (top left) spread complexity, (top right) $D_2$ in the energy eigenbasis, (bottom left) the peak of spread complexity and corresponding $\langle r \rangle$, (bottom right) saturation of spread complexity and $D_2$ for the one-loop truncated, infinite temperature, CGS at two loop truncation (raw spectrum) for $g^2\in (0,0.4]$. }
    \label{fig:OneLoopCGS}
\end{figure}

In Figure \ref{fig:OneLoopCGS} we repeat the analysis of the previous section. The infinite temperature one-loop CGS is an ergodic state in the two loop basis for all $g^2$ as seen in the upper right panel by $D_2$ remaining constant in $g^2$. It is not a surprise that there is an emergence of a peak as $g^2$ is increased and the model becomes chaotic. The approximate linear relation in the peak and $\langle r \rangle$ is almost recovered in the bottom left panel. The saturation value does not reach $0.5$ for any value of $g^2$, however, the variation with $D_2$ is approximately proportional as shown in the bottom right plot. 

\begin{figure}[t!]
    \centering
    \includegraphics[width=0.49\linewidth]{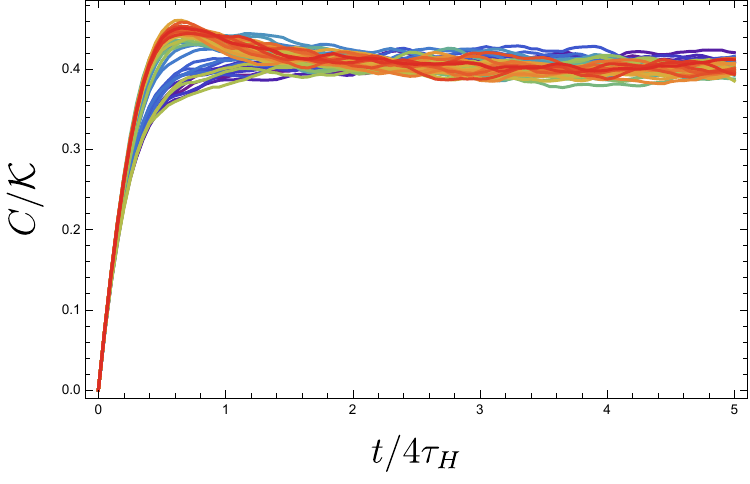}\hspace{0.01\linewidth}\includegraphics[width=0.49\linewidth]{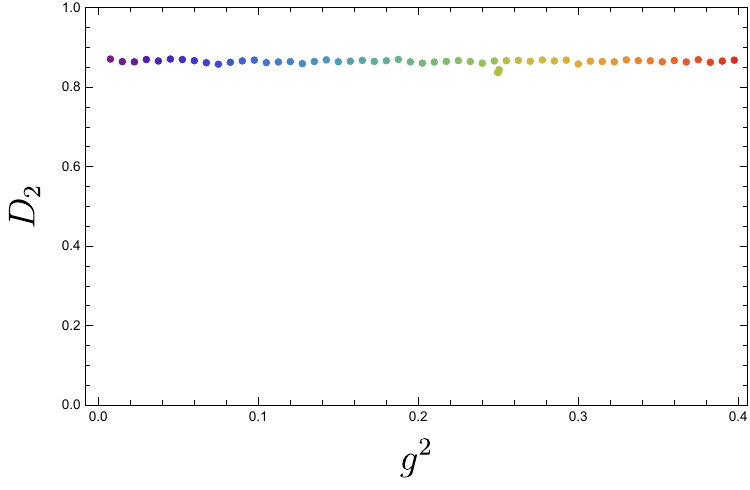}\\
    \includegraphics[width=0.49\linewidth]{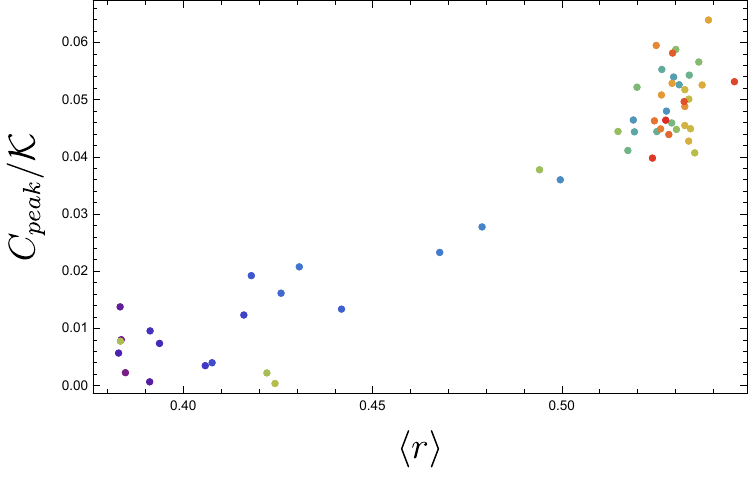}\hspace{0.01\linewidth}\includegraphics[width=0.49\linewidth]{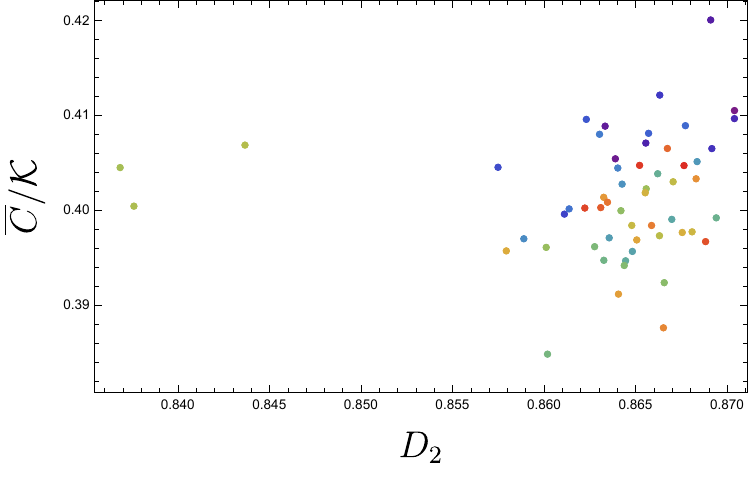}
    \caption{The (top left) spread complexity, (top right) $D_2$ in the energy eigenbasis, (bottom left) the peak of spread complexity and corresponding $\langle r \rangle$, (bottom right) saturation of spread complexity and $D_2$ for the one-loop truncated, infinite temperature, CGS at two loop truncation (unfolded spectrum) for $g^2\in (0,0.4]$.}
    \label{fig:OneLoopCGSUnfolded}
\end{figure}

Turning to the unfolded spectrum, as the one-loop infinite temperature CGS was already ergodic for all values of $g^2$ the clipping of the spectrum does not dramatically alter $D_2$, and thus the saturation of spread complexity, as seen in the right side of Figure \ref{fig:OneLoopCGSUnfolded}. As seen in section \ref{sec:KrylovPeak}, the peak's height is enhanced and its duration shortened when using the unfolded spectrum. This provides further evidence that the peak of spread complexity is controlled by the eigenvalue chaos, but we have learned in the previous section that the initial state may also obscure a peak. 

We note again that this does not capture, for example, the results of Figure \ref{fig:eigbasis_heatmap}. One expects that the average $D_2$ of the eigenstates is low for small $g^2$, however from Figure \ref{fig:OneLoopCGS} we see that the average of the eigenstates' $D_2$ (note the difference in when the averaging occurs) is featureless, remaining consistently close to maximum throughout $g^2$. Similar statements can be made for the spin product basis if one takes the average over all spin basis states as the initial state. However, one might still hope to extract some physical information from particular states. For instance the domain wall state might be physically relevant as a state that can be consistently prepared in an experimental set-up, or the (in)finite temperature CGS (more generally its mixed state analogue) may be seen as an interesting state in a quench protocol where the higher-loop truncations are turned on. We hope to return to more detailed analysis of this dependence in the future.
%
\section{Summary and Discussion}\label{sec:disc}

In this paper we studied finite-loop truncations of the planar dilatation operator in the $\mathfrak{su}(2)$-sector as long-range deformations of the nearest neighbour XXX-spin chain. The one-loop Hamiltonian is integrable, while the all-loop planar theory is expected to be again integrable. Intermediate orders are not. 
These finite-loop truncations therefore provide a useful setting in which to study how chaotic behaviour appears between these two integrable limits. We addressed this question using spectral statistics, eigenvector diagnostics and Krylov basis diagnostics.

At the level of the spectral statistics, we found that the two- and four-loop truncations develop GOE-like level correlations for sufficiently large values of the coupling. This was seen both from the level-spacing distribution and from the ratio of adjacent level spacings. The transition from Poisson to Wigner-Dyson statistics is however slower than for a generic integrability-breaking perturbation. This is particularly clear when comparing with the model in which the twist parameters are chosen differently for the terms occurring at different loop order.
The latter rapidly approaches the chaotic value of the Brody parameter and average $r$-ratio, while the holographic truncations remain close to the integrable point over a larger range of the coupling. 

In addition, our analysis shows that the spectral diagnostics are non-monotonic in the loop order. The two-loop truncation reproduces the expected weak crossover and has a special point at $g^2=1/4$, where the nearest-neighbour interaction cancels and the model becomes integrable again. The three-loop truncation never fully reaches GOE statistics and deviates from the Poisson predictions at much larger values of the coupling compared to the two-loop truncation. This seems connected to the alternating sign structure of the gauge theory perturbative series. At four-loops the GOE predicted values are attained, but the model breaks from the Poisson prediction at coupling strengths between that of the two- and three-loop truncations. This suggests that the higher-loop terms do not simply increase the amount of chaos but instead, in an alternating fashion, show convergence towards the integrable all-loop behaviour, even at very low truncation order. The four-loop terms also display weaker integrability breaking as measured by the volume scaling of the critical coupling, $g^2_c\propto L^b$. The two-loop volume exponent, $b_{\text{2-loop}}\sim 1$ \cite{McLoughlin:2022jyt}, which is smaller than the generic value $b=3$ and the weakly broken XXZ chain $b\sim 2$ \cite{Szasz-Schagrin:2021pqg}, is further reduced in the four-loop truncation to $b_{\text{4-loop}}\sim 0.5$.

The eigenvector diagnostics give a complementary picture. Although the two- and four-loop truncations display chaotic spectral correlations, their eigenstates do not feature the ergodicity of GOE random vectors. We tested this by computing fractal dimensions and structural entropies in the spin-product basis and in the eigenbasis of the integrable one-loop Hamiltonian. The former probes delocalisation in the local many-body basis of single-trace words, while the latter measures memory of the nearest-neighbour XXX chain. In the spin-product basis, the fractal dimensions remain below their ergodic values and the structural entropy does not uniformly approach the GOE reference value; in the integrable one-loop basis, the eigenstates likewise remain only partially delocalised. Thus the onset of spectral chaos is accompanied only by weak eigenvector ergodicity. Such a mismatch between eigenvalue chaos and full eigenvector ergodicity is known to occur in generic spin chains \cite{Santos:2010qxi,Backer:2019avi}, including next-to-next-nearest-neighbour models, and here persists for the particular long-range deformations fixed by the planar $\mathcal N=4$ SYM dilatation operator. 

Moreover, the loop-dependent hierarchy seen in the eigenvalue statistics is reproduced in both eigenvector diagnostics: the two-loop truncation comes closest to GOE-like behaviour at sufficiently large coupling, the three-loop truncation remains closest to the integrable regime, and the four-loop truncation lies between them. Thus the finite-loop dilatation operators are chaotic in the spectral sense, but remain only weakly ergodic in their eigenstate structure.

In this work we subsequently used Krylov complexity tools as a way of bridging eigenvalue and eigenvector diagnostics of chaos. 
For an infinite-temperature coherent Gibbs state, the peak of the spread complexity provides a Krylov-space signature of the spectral crossover: $C_{\rm peak}$ follows the same dependence on $g^2$ as $\langle r\rangle$. The two are linearly correlated but, unlike $\langle r \rangle$, the peak value is sensitive to unfolding. The same eigenvalue-chaos transition is also visible in the Lanczos coefficients. In particular, the variance of appropriately combined adjacent Lanczos coefficients gives a measure of disorder on the Krylov chain, closely related to the Anderson-localisation interpretation of integrability in Krylov dynamics \cite{Rabinovici_2022}. These results show that spread complexity tracks standard spectral diagnostics, such as $\langle r\rangle$, level statistics and the spectral form factor, but now through a dynamical probe of the onset of chaos.

While previous works relating chaos and Krylov methods have largely focused on eigenvalue chaos, we also studied the relation to eigenvector chaos.  
Our results suggest that this dependence is a useful feature. The saturation value of the spread complexity is sensitive to the distribution of the initial state in the Hamiltonian eigenbasis, and changes in parallel with the initial state's delocalisation, as measured for example by $D_2$. At the same time, this initial state dependence means that spectral chaos need not always be visible in a given Krylov evolution: weak eigenvector localisation can obscure the peak associated with chaotic spectral correlations. Thus Krylov dynamics provide not only another diagnostic of eigenvalue chaos, but also a promising way to probe eigenvector chaos through the choice of initial state. While a direct comparison with averaged eigenvector diagnostics remains computationally difficult, the Krylov saturation value appears to encode complementary information about the degree of eigenstate ergodicity. 

Taken together, these results show that finite-loop truncations of the planar dilatation operator are not generic long-range spin chains. Their couplings are fixed by the gauge theory, and the resulting Hamiltonians are therefore not generic long-range spin chains. Instead, the loop dependence suggests a staggered approach towards the all-loop integrable structure, with the finite-loop truncations displaying only intermediate and progressively constrained forms of chaos. This is visible in the delayed onset of Wigner-Dyson statistics, in the incomplete randomization of the eigenvectors, and in the explicit initial state dependence of the Krylov dynamics. From this perspective, the finite-loop truncations provide a concrete realization of a specific form of weak chaos in planar holography.

\bigskip

\noindent
It would be interesting to extend the analysis presented here in the following directions:

\vspace{-4pt}

\paragraph{Higher-loop truncations and larger system sizes.} A natural next step is to extend the present analysis to higher-loop truncations and to larger spin chain lengths. Within the context of this work the higher-loop Hamiltonians are not arbitrary long-range deformations: they are generated from higher loop correction to the one-loop dilatation operator and are closely related to spin chain $T\bar T$-type deformations. Our analysis shows that up to four-loops the resulting chaos weakens and ergodicity properties are weak, and it would be important to test how far this persists at higher-loops. In particular, our results suggest two related possibilities. First, the critical coupling for the onset of random-matrix behaviour may continue to drift to larger values, with a progressively weaker volume scaling, as expected if the all-loop planar Hamiltonian restores finite-coupling integrability. Second, the alternating pattern seen between the two-, three-, and four-loop truncations may reflect the persistent alternative asymptotic behaviour towards restoration of integrability, at least in the region of convergence $g^2<1/16\pi$, rather than generic long-range chaos. Distinguishing these possibilities will require both higher-loop data and larger system sizes. Beyond the connection to gauge theory, such long-range chains play an important role as models of many-body systems such as cold atoms. Much of the previous work on weak integrability-breaking in this context has focussed on models with NNN interactions e.g \cite{Jung_2006, Kurlov_2022, Surace:2023wqq, Vanovac:2024tkj} and it would be interesting to understand how the longer-range interactions affect the thermodynamic properties of the spin chain such as thermalisation time.

\paragraph{Spread complexity further analytics.} 
It would be useful to understand more thoroughly the relation between Krylov diagnostics and different measures of chaos. Our results show that spread complexity is sensitive both to spectral correlations and to the structure of the initial state in the Hamiltonian eigenbasis. This makes it a natural tool for connecting eigenvalue and eigenvector chaos, but it also raises several questions.

First, the relevant time scales should be disentangled systematically: the peak time, the saturation time, and their relation to the Thouless and Heisenberg times in the context of the SFF \cite{Tan2025ScalingSFFandKrylovAtFiniteTemp}. In particular, while the peak of the spread complexity tracks the same integrability-to-chaos crossover as $\langle r\rangle$, it should be a direct complexity realisation of the SFF ramp.  
Its sensitivity to unfolding, together with its proximity to the shared saturation time of the spread complexity and SFF, suggest that the peak arises due to late time chaos. It will be important to make this statement sharper in controlled settings, including RMT. 

Second, the role of the initial state deserves a more systematic treatment. The saturation value of spread complexity is sensitive to the distribution of the initial state in the Hamiltonian eigenbasis. We have argued that one can capture the eigenvector chaos through this by averaging the results from all basis elements as initial states, however computationally expensive it may be. As noted previously, there are still interesting cases to be studied using particular initial states, not just averaging. In addition to those mentioned previously, it would be interesting to see applications towards holography and being able to distinguish black hole microstates from other states in the gauge theory.  

Finally, the precise role of unfolding in Krylov diagnostics remains largely open. We have demonstrated across multiple scenarios that the unfolding and clipping of the spectrum and initial state, make the emergence of the peak before the saturation time more apparent by increasing its maximum height and shortening its duration. These procedures are useful in numerical studies, but lack a physical or operational meaning and progress in this direction may improve our understanding of quantum chaos from the Krylov basis perspective.

\acknowledgments
PC is supported by the NCN Sonata Bis 9 2019/34/E/ST2/00123 grant. PC and BC are supported by the ERC Consolidator grant (number: 101125449/acronym: QComplexity). Views and opinions expressed are however those of the authors only and do not necessarily reflect those of the European Union or the European Research Council. Neither the European Union nor the granting authority can be held responsible for them.  RND's work leading to this publication was supported by the PRIME programme of the German Academic Exchange Service (DAAD) with funds from the German Federal Ministry of Research, Technology and Space (BMFTR). RND is also supported by Germany's Excellence Strategy through the W\"urzburg‐Dresden Cluster of Excellence on Complexity, Topology and Dynamics in Quantum Matter ‐ ctd.qmat (EXC 2147, project‐id 390858490). RND gratefully acknowledges the COST Action CA22113 ``THEORY-CHALLENGES'' for supporting a scientific visit to CUNEF Universidad during which part of this work was carried out. RND and SD would like to acknowledge the Mainz Institute for Theoretical Physics (MITP) of the Cluster of Excellence PRISMA+ (Project ID 390831469), for its hospitality and support.
%

\appendix
\section{Hamiltonians in terms of Pauli operators}\label{app:Ham_Pauli}
Using the interaction-symbol notation
$\{a,b,c,\ldots\}=\sum_{p=1}^{L}P_{p+a}P_{p+b}P_{p+c}\cdots$,
we rewrite the loop contributions to the spin chain Hamiltonian in \eqref{eq:123loops}
in terms of Pauli matrices, making the interaction range and operator content more
transparent than in the compact symbolic form used in the main text.
All site labels are understood modulo $L$. The nearest-neighbour permutation
operator is related to the Pauli matrices by,
$P_{a,b}
=
\frac12\left(\mathbf 1_{a,b}
+
\vec\sigma_a\cdot\vec\sigma_b
\right)$.
Therefore, for the nearest-neighbour permutation operator appearing in the draft,
\begin{equation}
P_p=P_{p,p+1}
=
\frac12
\left(
1+\vec\sigma_p\cdot\vec\sigma_{p+1}
\right)~~~\text{and}~~~\{\}=\sum_{p=1}^{L}1=L.\label{eq:a3}
\end{equation}
For a general interaction symbol,
\begin{equation}
\{a_1,a_2,\ldots,a_n\}
=
\frac{1}{2^n}
\sum_{p=1}^{L}
\left(
1+\vec\sigma_{p+a_1}\cdot\vec\sigma_{p+a_1+1}
\right)
\left(
1+\vec\sigma_{p+a_2}\cdot\vec\sigma_{p+a_2+1}
\right)
\cdots
\left(
1+\vec\sigma_{p+a_n}\cdot\vec\sigma_{p+a_n+1}
\right).\label{eq:a4}
\end{equation}
All site labels are understood modulo $L$. For $H_1$, we start from $H_1$ in \eqref{eq:123loops}.
Using \eqref{eq:a3} and $\{1\}
=
\sum_{p=1}^{L}P_{p+1}
=
\frac12
\sum_{p=1}^{L}
\left(
1+\vec\sigma_{p+1}\cdot\vec\sigma_{p+2}
\right)$, one obtains,
\begin{equation}
H_1
=
\sum_{i=1}^{L}
\left(
1-\vec\sigma_i\cdot\vec\sigma_{i+1}
\right).
\end{equation}
For $H_2$ in \eqref{eq:123loops}, noting that,
\begin{align}
\{1,2\}
&=
\frac14
\sum_{i=1}^{L}
\left(
1+\vec\sigma_i\cdot\vec\sigma_{i+1}
\right)
\left(
1+\vec\sigma_{i+1}\cdot\vec\sigma_{i+2}
\right),\\
\{2,1\}
&=
\frac14
\sum_{i=1}^{L}
\left(
1+\vec\sigma_{i+1}\cdot\vec\sigma_{i+2}
\right)
\left(
1+\vec\sigma_i\cdot\vec\sigma_{i+1}
\right)
\end{align}
and using the Pauli algebra
\begin{equation}
\sigma_i^a\sigma_i^b=\delta^{ab}+i\epsilon^{abc}\sigma_i^c.
\end{equation}
one gets
\begin{equation}
H_2
=
\sum_{i=1}^{L}
\left(
-3
+
4\,\vec\sigma_i\cdot\vec\sigma_{i+1}
-
\vec\sigma_i\cdot\vec\sigma_{i+2}
\right).
\end{equation}
Similarly, for $H_3$ in \eqref{eq:123loops}, using \eqref{eq:a3} and \eqref{eq:a4}, we obtain
\begin{align}
\begin{aligned}
H_3
=&
\sum_{i=1}^{L}
\Big[
20
-29\,\vec\sigma_i\cdot\vec\sigma_{i+1}
+10\,\vec\sigma_i\cdot\vec\sigma_{i+2}
\Big] -\frac12
\sum_{i=1}^{L}
\Big[
\left(\vec\sigma_i\cdot\vec\sigma_{i+1}\right)
\left(\vec\sigma_{i+1}\cdot\vec\sigma_{i+2}\right)
\left(\vec\sigma_{i+2}\cdot\vec\sigma_{i+3}\right)
\\&
+
\left(\vec\sigma_{i+2}\cdot\vec\sigma_{i+3}\right)
\left(\vec\sigma_{i+1}\cdot\vec\sigma_{i+2}\right)
\left(\vec\sigma_i\cdot\vec\sigma_{i+1}\right)
\Big] +\frac{\epsilon_2}{2}
\sum_{i=1}^{L}
\Big[
\left(\vec\sigma_{i+1}\cdot\vec\sigma_{i+2}\right)
\left(\vec\sigma_i\cdot\vec\sigma_{i+1}\right)
\left(\vec\sigma_{i+2}\cdot\vec\sigma_{i+3}\right)
\\
&
-
\left(\vec\sigma_i\cdot\vec\sigma_{i+1}\right)
\left(\vec\sigma_{i+2}\cdot\vec\sigma_{i+3}\right)
\left(\vec\sigma_{i+1}\cdot\vec\sigma_{i+2}\right)
\Big].
\end{aligned}
\end{align}
The four-loop contribution, $H_4$ of \eqref{eq:4-loops} can be similarly obtained in a straightforward manner.
\section{Unfolding}\label{sec:unfolding}
  The BGS conjecture \cite{Bohigas:1983er} characterises quantum chaotic spectra as those whose fluctuations, that is deviations from uniformity, match those of RMT. In order to compute many of the standard indicators of the onset of chaos, it is necessary to extract the local properties of the spectrum. This is  known as unfolding and corresponds to rescaling the energy eigenvalues such that they have a uniform density. That is, given a spectrum of eigenvalues for a $D$-dimensional space, $E_1, E_2, \dots, E_D$, with a density 
\begin{align}
    \rho(E)=\sum_{i=1}^D \delta(E-E_i)\,,
\end{align}
we split it into an average, smooth, component and a fluctuation part
\begin{align}
\rho(E)=\rho_{\text{av}}(E)+\rho_{\text{fluc}}(E)\,,
\end{align}
where $\rho_{\text{av}}$ is defined by an appropriate smoothing procedure. For example, in the case where the model is defined by a statistical ensemble this can be done by taking the expectation value $\rho_{\text{av}}=\mathbb{E}(\rho)$ which for RMT gives the usual semi-circle formula $\rho_{\text{av}}(E)\propto \sqrt{4-E^2}$. An alternative method is to use resolvent smoothing 
\begin{align}
\rho_{\text{av}, \eta}(E) =\frac{1}{ \pi}\text{Im~Tr}\frac{1}{H-E-i \eta}\,,
\end{align}
where the parameter $\eta$ controls the degree of smoothing. More practically, in our numerical implementations, we define an averaged cumulative number density $n_{\text{av}}(E)$
by fitting a high-order polynomial to the numerically computed staircase function $n(E)=\sum_i \Theta(E-E_i)$ such that $\rho_{\text{av}}=dn_{\text{av}}/dE$. Given $\rho_{\text{av}}$ we then define the unfolded spectrum 
\begin{align}
    \xi_i=f(E_i)~,~~~\text{where}~~~f(e)=\int_{-\infty}^E \rho_{\text{av}}(E') dE'~.
\end{align}
It immediately follows that the $\xi$'s have uniform density and so carry only the information regarding the fluctuations which can be compared to the predictions of RMT such as energy level correlation functions. 

We can also think of the unfolding as an operation on the Hamiltonian, 
\[
H=U \text{diag}(E_1, \dots, E_D) U^{-1}~,
\]
and so define a new, unfolded, Hamiltonian, 
\begin{align}
    H_{\text{unf}}=f(H)\equiv U\text{diag}(f(E_1),\dots, f(E_D)) U^{-1}~
\end{align}
which has an unfolded spectrum but the same eigenvectors. For chaotic systems, we expect dynamical observables computed with this unfolded Hamiltonian that go beyond the spectrum, for example spread complexity, to demonstrate an increased universality across systems. 
\section{Lanczos algorithm}\label{LanczosAlg}
For completeness, we define the Lanczos algorithm which we use in this work. 
For a given Hamiltonian $H$ and normalized initial state $\ket{\psi(0)}$ the algorithm is defined as:
\begin{align}
    \text{Define: }&   \ket{K_0}:= \ket{\psi(0)} \, , \,  a_0 := \bra{K_0} H \ket{K_0} \, , \, b_0:=0\,,  \nn\\
    \text{Iterate for $n>0$ : }& \ket{A_{n+1}} := (H-a_n) \ket{K_n} - b_{n-1} \ket{K_{n-1}}\,,\nn \\
    &b_{n+1}:= \sqrt{\bra{A_{n+1}}A_{n+1}\rangle}\,, \nn\\
    &\text{If $b_{n+1}=0$ Stop.} \nn\\
    \text{Else: }& \ket{K_{n+1}}:=b^{-1}_{n+1} \ket{A_{n+1}} \, , \, a_{n+1}:= \bra{K_{n+1}} H \ket{K_{n+1}}\,.
\end{align}

Upon completion of the algorithm, the Lanczos coefficients ($\{ a_n,b_n \}_{n=0,\dots,\mathcal{K}-1})$, the Krylov basis ($\{ \ket{K_n} \}_{n=0,\dots,\mathcal{K}-1}$), and the Krylov dimension $\mathcal{K}$, collectively referred to as the Krylov data, are extracted.

\addcontentsline{toc}{section}{References}
\bibliography{wc}
\bibliographystyle{JHEP}

\end{document}